\documentclass[aps]{revtex4}
\usepackage{amsfonts}
\usepackage{amsmath}
\usepackage{amssymb,epsf}

\begin{document}

\title{Charged Black Hole Solutions in Gauss-Bonnet-Massive Gravity}
\author{S. H. Hendi$^{1,2}$\footnote{%
email address: hendi@shirazu.ac.ir}, S. Panahiyan$^{1}$\footnote{
email address: sh.panahiyan@gmail.com} and B. Eslam
Panah$^{1}$\footnote{ email address: behzad.eslampanah@gmail.com}}
\affiliation{$^1$ Physics Department and Biruni Observatory,
College of Sciences, Shiraz
University, Shiraz 71454, Iran\\
$^{2}$ Research Institute for Astronomy and Astrophysics of Maragha (RIAAM),
P.O. Box 55134-441, Maragha, Iran}

\begin{abstract}
Motivated by high interest in the close relation between string theory and
black hole solutions, in this paper, we take into account the
Einstein-Gauss-Bonnet Lagrangian in the context of massive gravity. We
examine the possibility of black hole in this regard, and discuss the types
of horizons. Next, we calculate conserved and thermodynamic quantities and
check the validity of the first law of thermodynamics. In addition, we
investigate the stability of these black holes in context of canonical
ensemble. We show that number, type and place of phase transition points may
be significantly affected by different parameters. Next, by considering
cosmological constant as thermodynamical pressure, we will extend phase
space and calculate critical values. Then, we construct thermodynamical
spacetime by considering mass as thermodynamical potential. We study
geometrical thermodynamics of these black holes in context of heat capacity
and extended phase space. We show that studying heat capacity, geometrical
thermodynamics and critical behavior in extended phase space lead to
consistent results. Finally, we will employ a new method for obtaining
critical values and show that the results of this method are consistent with
those of other methods.
\end{abstract}

\maketitle

\section{Introduction}

Cosmological observations show that about $95\%$ of the universe is made of
dark energy and dark matter. Although the basic nature of the mentioned dark
sector of the universe is an open question, in order to interpret their
effects, many authors have proposed modified gravity models such as;
scalar-tensor theories \cite{Brans} (or their conformally related; $F(R)$
gravity theories \cite{Akbar,Cognola,Hendi}), brane world cosmology \cite%
{Gergely} and Lovelock gravity \cite{Lovelock}. This is one of the main
motivations for considering Lovelock theory.

On the other hand, regarding higher dimensional gravity, the reasonable
effects of higher curvature terms in higher derivative gravity theories have
been investigated. One of the well-known theories of higher derivative
gravity is Lovelock theory which is a natural generalization of Einstein
theory in higher dimensions. Taking into account the first additional term
of Einstein gravity in the context of Lovelock theory (Gauss-Bonnet (GB)
gravity), it is believed that GB gravity can solve some of the shortcomings
of Einstein theory \cite{shortcomings}. GB gravity consists
curvature-squared terms which, interestingly, is free of ghosts and the
corresponding field equations contain no more than second derivatives of the
metric \cite{secondorder,ghostfree}. Another interesting aspect of GB theory
is that it can be arisen from the low-energy limit of heterotic string
theory \cite{string,string2}. It was shown that the low-energy expansion of
a heterotic (closed) string theory effective action has GB term as well as a
scalar field \cite{string2}. The functional form of scalar field remains a
fascinating and challenging topic of observational research and some authors
have stipulated a trivial constant value for the scalar field, and
therefore, ignore its effect on the solutions. Considering GB gravity
context, black hole solutions and brane-world model and their interesting
behavior have been investigated in literature \cite{blackholesGB}. Recently,
it was shown that conserved charges of Einstein-Gauss-Bonnet AdS gravity are
presented in electric part of the Weyl tensor which gives a generalization
to conformal mass definition \cite{Jatkar}. Holographic thermalization in GB
gravity through the Wilson loop and the holographic entanglement entropy has
been investigated in Ref. \cite{thermalization}. In addition, regarding $p$%
-wave phase transition, it was shown that although magnetic field has a
positive contribution to superconductor phase transition, the effect of GB
parameter is negative in case of this phase transition \cite{pwave}. In the
context of AdS/CFT correspondence, there has been several studies with the
subject of GB gravity \cite{GBAdSCFT}. In Ref. \cite{PureGB}, Stuckelberg
scalar field has been employed to study phase transitions in a dual quantum
field theory by considering pure GB gravity in asymptotically AdS space. On
the other hand, in Ref. \cite{Jose}, it was shown that restriction due to
causality violation in GB gravity points to the need to complete the theory
with an infinite tower of massive higher-spin states, as it happens in
string theory \cite{veneziano}. Motivated by the recent results mentioned
above, we study thermodynamic behavior of black holes in GB-massive gravity.

Before we proceed, let us provide a motivation for considering massive
gravity. One of the interesting open question of gravitational field is
graviton and its properties. Although graviton is massless in the Einstein
gravity, a natural question is whether one can build a self-consistent
gravity theory if graviton is massive. Therefore, the Einstein gravity may
be modified to satisfy this point, where massive gravity is one such
modification. From the theoretical perspective, the shear difficulty of
constructing a consistent theory of massive gravity makes the subject more
interesting. Therefore, there are limited considerable works in the massive
gravity context. The development of the ghost-free theory with massive
gravitons which are non-interactive in flat background was done in Ref. \cite%
{Fierz}. It was shown that generalization of this theory to curved
background leads to existence of the ghost instabilities \cite{Boulware}.
The effects of quantum interactions of massive gravity and a nonlinear
theory of massive gravity in absence of ghost field \cite{Hassan} were
investigated in Refs. \cite{Minjoon,Rham}. In addition, a special class of
charged massive black holes has been investigated \cite{Saridakis} (see \cite%
{Babichev} for more details regarding considering massive gravity). A new
class of nontrivial massive black holes in AdS spacetime was investigated in
\cite{Vegh,Hassan2011}. In this theory of the massive gravity, graviton has
similar behavior as lattice in holographic conductor model. In other words,
due to exhibiting a Drude peak which approaches to delta function in limit
of massless gravity, graviton in this theory plays the role of lattice.
Thermodynamical behavior, $P-V$ criticality and geometrothermodynamics (GTD)
of the Vegh's massive gravity has been, recently, investigated \cite%
{Cai2015,Xu2015,HendiPEM}. Some holographic consequences of graviton mass
have been investigated in Ref. \cite{Davison}. In this paper, we intend to
generalize Einstein gravity action by adding GB Lagrangian and obtain exact
charged solutions of GB-Massive gravity. In general, higher derivative
gravity terms (such as GB) are corrections that become relevant at
high--energy regime while the mass terms (of massive gravity) are relevant
at low-energy. Here, we do not regard the energy regime, directly, and we
look for the weight of GB and mass terms contributions into the
thermodynamical behavior of a typical black hole.

Obtaining the solutions of GB-massive gravity, we should check their
stability. In general, one may categorize the stability criteria into two
classes; dynamical stability and thermodynamical one. In the present paper,
we relax the dynamical stability and instead, we focus on thermodynamic
behavior. Thermodynamical aspects of the black holes are among the popular
subjects for studying black hole's behavior. Thermal stability of the black
hole is of great importance and has been widely investigated in literature
\cite{Myung}. This is due to the fact that instability of black holes puts
restriction (conditions) for validation of the physical solutions. One of
the approaches for studying thermal stability is through canonical ensemble
with investigation of the heat capacity. On the other hand, one can look for
the phase transition points of solutions by calculating root(s) and
divergence point(s) of the heat capacity. Also, the thermal stability and
its corresponding conditions can be employed in studies conducted in context
of AdS/CFT correspondence.

Recently, there has been a renewed interpretation regarding cosmological
constant as a thermodynamical variable. It was pointed out that considering
cosmological constant as a thermodynamical variable may lead to removing
ensemble dependency in studying stability in context of canonical (heat
capacity) and grand canonical (Hessian matrix) ensembles for BTZ black holes
\cite{Mamasani}. In addition, in context of AdS/CFT correspondence, it was
shown that variation of cosmological constant in black holes corresponds to
variation of the number of colors in Yang-Mills theory residing in the
boundary of the spacetime \cite{yang}. There has been a consideration of the
cosmological constant as a state-dependant parameter in two dimensional
dilaton gravity \cite{Grumiller}. In an interesting paper, Hawking and Page
showed an interpretation of confinement/deconfinement phase transition in
the dual strongly coupled gauge theory for phase transition between the
stable large black hole and thermal gas in AdS space \cite{Hawking}. On the
other hand, it was shown that there is a similarity between phase transition
in black holes and liquid/gas phase transition in a Van der Waals system
\cite{Dolan}. This consideration has been investigated in literature for
different types of black holes \cite{P-V}.

Another method for studying critical behavior of the system is by
constructing a spacetime by considering a thermodynamical potential and a
set of extensive parameters as components of that thermodynamical spacetime.
The calculated Ricci scalar of constructed thermodynamical spacetime should
diverges in place of phase transition points that were mentioned in studying
heat capacity. There are several approaches for this method that to name a
few, one can say: Weinhold in which mass is considered as thermodynamical
potential with other parameters (entropy, electric charge and etc.) as
extensive parameters \cite{Weinhold}, Ruppeiner approach that is built up
based on entropy as thermodynamical potential \cite{Ruppeiner}, it should be
pointed out that it was shown that these two metrics are conformably related
to each other by a conformal factor (temperature) \cite{Conformal}. The
other method is the one that Quevedo introduced which has mass as
thermodynamical potential like Weinhold but with different structure \cite%
{Quevedo}. There are several types for Quevedo metric. Recently, it was
shown that mentioned approaches in case of some black holes, will not yield
consisting results with studying heat capacity. In order to overcome this
problem, a new metric was introduced in Ref. \cite{HendiPEM,HPEM}. The
denominator of the Ricci scalar of HPEM metric \cite{HendiPEM,HPEM} only
contains expressions of numerator and denominator of the heat capacity.

The outline of the paper will be as follows. In Sec. \ref{GBMassive}, we
introduce action regarding charged massive gravity in presence of GB gravity
and obtain charged black hole solutions in this gravity. We calculate
conserved and thermodynamic quantities related to obtained solutions and
check the validation of the first law of thermodynamics in Sec. \ref{Thermo}%
. In Sec. \ref{Stability} we study thermal stability of the massive charged
GB black hole solutions in canonical ensemble using heat capacity. In the
next section, considering cosmological constant as thermodynamical pressure,
we study phase transition of black holes in context of $P-V$ criticality and
phase diagrams. In Sec. \ref{GTD} we study thermodynamical behavior of these
solutions through geometrical thermodynamics. Sec. \ref{HC} is devoted to
calculation of critical values in extended phase space by using heat
capacity. Last section will be closing remarks.

\section{Charged Black hole solutions in GB-Massive gravity \label{GBMassive}%
}

The $d$-dimensional action of GB-massive gravity with a cosmological
constant ($\Lambda$) can be written as
\begin{equation}
\mathcal{I}=-\frac{1}{16\pi }\int d^{d}x\sqrt{-g}\left[ \mathcal{R}-2\Lambda
+\alpha L_{GB}-\mathcal{F}+m^{2}\sum_{i}^{4}c_{i}\mathcal{U}_{i}(g,\Psi )%
\right] ,  \label{Action}
\end{equation}
where $m$ and $\mathcal{R}$ are, respectively, the massive parameter and the
scalar curvature and $\mathcal{F}=F_{\mu \nu }F^{\mu \nu }$ is the Maxwell
invariant. In addition, $\alpha $ and $L_{GB}$ are, respectively, the
coefficient and the Lagrangian of GB gravity, and $\Psi $ is a fixed
symmetric tensor. In Eq. (\ref{Action}), $c_{i}$ are constants and $\mathcal{%
U}_{i}$ are symmetric polynomials of the eigenvalues of the $d\times d$
matrix $\mathcal{K}_{\nu }^{\mu }=\sqrt{g^{\mu \alpha }\Psi _{\alpha \nu }}$
\begin{eqnarray}
\mathcal{U}_{1} &=&\left[ \mathcal{K}\right] ,  \notag \\
\mathcal{U}_{2} &=&\left[ \mathcal{K}\right] ^{2}-\left[ \mathcal{K}^{2}%
\right] ,  \notag \\
\mathcal{U}_{3} &=&\left[ \mathcal{K}\right] ^{3}-3\left[ \mathcal{K}\right] %
\left[ \mathcal{K}^{2}\right] +2\left[ \mathcal{K}^{3}\right] ,  \notag \\
\mathcal{U}_{4} &=&\left[ \mathcal{K}\right] ^{4}-6\left[ \mathcal{K}^{2}%
\right] \left[ \mathcal{K}\right] ^{2}+8\left[ \mathcal{K}^{3}\right] \left[
\mathcal{K}\right] +3\left[ \mathcal{K}^{2}\right] ^{2}-6\left[ \mathcal{K}%
^{4}\right] ,  \notag
\end{eqnarray}%
and $L_{GB}=R_{\mu \nu \gamma \delta }R^{\mu \nu \gamma \delta }-4R_{\mu \nu
}R^{\mu \nu }+R^{2}$ where $R_{\mu \nu }$ and $R_{\mu \nu \gamma \delta }$
are Ricci and Riemann tensors. Variation of the action (\ref{Action}) with
respect to the metric tensor $g_{\mu \nu }$ and the Faraday tensor $F_{\mu
\nu }$, leads to
\begin{equation}
G_{\mu \nu }+\Lambda g_{\mu \nu }+H_{\mu \nu }-\left[ 2F_{\mu \lambda
}F_{\nu }^{\lambda }-\frac{1}{2}g_{\mu \nu }\mathcal{F}\right] +m^{2}\chi
_{\mu \nu }=0,  \label{Field equation}
\end{equation}%
\begin{equation}
\nabla _{\mu }F^{\mu \nu }=0,  \label{Maxwell equation}
\end{equation}%
where $G_{\mu \nu }$ is the Einstein tensor, $H_{\mu \nu }$ and $\chi _{\mu
\nu }$ are
\begin{eqnarray}
H_{\mu \nu } &=&-\frac{\alpha }{2}\left( 8R^{\rho \sigma }R_{\mu \rho \nu
\sigma }-4R_{\mu }^{\rho \sigma \lambda }R_{\nu \rho \sigma \lambda
}-4RR_{\mu \nu }+8R_{\mu \lambda }R_{\nu }^{\lambda }+g_{\mu \nu
}L_{GB}\right) , \\
\chi _{\mu \nu } &=&-\frac{c_{1}}{2}\left( \mathcal{U}_{1}g_{\mu \nu }-%
\mathcal{K}_{\mu \nu }\right) -\frac{c_{2}}{2}\left( \mathcal{U}_{2}g_{\mu
\nu }-2\mathcal{U}_{1}\mathcal{K}_{\mu \nu }+2\mathcal{K}_{\mu \nu
}^{2}\right) -\frac{c_{3}}{2}(\mathcal{U}_{3}g_{\mu \nu }-3\mathcal{U}_{2}%
\mathcal{K}_{\mu \nu }  \notag \\
&&+6\mathcal{U}_{1}\mathcal{K}_{\mu \nu }^{2}-6\mathcal{K}_{\mu \nu }^{3})-%
\frac{c_{4}}{2}(\mathcal{U}_{4}g_{\mu \nu }-4\mathcal{U}_{3}\mathcal{K}_{\mu
\nu }+12\mathcal{U}_{2}\mathcal{K}_{\mu \nu }^{2}-24\mathcal{U}_{1}\mathcal{K%
}_{\mu \nu }^{3}+24\mathcal{K}_{\mu \nu }^{4}).
\end{eqnarray}

We consider the metric of $d$-dimensional metric with the following form
\begin{equation}
ds^{2}=-f(r)dt^{2}+f^{-1}(r)dr^{2}+r^{2}h_{ij}dx_{i}dx_{j},\
i,j=1,2,3,...,n~,  \label{Metric}
\end{equation}%
where $h_{ij}dx_{i}dx_{j}$ is the line element of a ($d-2$)-dimensional
space with constant curvature $(d-2)(d-3)k$ and volume $V_{d-2}$. We should
note that the constant $k$ indicates that the boundary of $t=constant$ and $%
r=constant$ can be a positive (elliptic), zero (flat) or negative
(hyperbolic) constant curvature hypersurface.

Here, we consider the ansatz of Ref. \cite{Cai2015} for the auxiliary
reference metric
\begin{equation}
\Psi _{\mu \nu }=diag(0,0,c^{2}h_{ij}),  \label{f11}
\end{equation}%
where $c$ is a positive constant. Using the ansatz (\ref{f11}), $\mathcal{U}%
_{i}$'s may be written as \cite{Cai2015}
\begin{eqnarray}
\mathcal{U}_{1} &=&\frac{d_{2}c}{r},  \notag \\
\mathcal{U}_{2} &=&\frac{d_{2}d_{3}c^{2}}{r^{2}},  \notag \\
\mathcal{U}_{3} &=&\frac{d_{2}d_{3}d_{4}c^{3}}{r^{3}},  \notag \\
\mathcal{U}_{4} &=&\frac{d_{2}d_{3}d_{4}d_{5}c^{4}}{r^{4}},  \notag
\end{eqnarray}%
in which we used the notation $d_{i}=d-i$.

I order to study the effects of massive gravity, it is necessary to consider
a reference metric, denoted by $\Psi _{\mu \nu }$. In a special class, one
may treat the reference metric in the context of a bimetric theory to credit
it with a dynamical spin-$2$ tensor field (see \cite{Hassan2} for more
details). Another class of considering reference metric comes from the gauge
invariance property with using of Stuckelberg scalar fields \cite%
{Arkani-Hamed,Dubovsky,Chamseddine}. It was shown that the number of degrees
of freedom propagated by any dRGT-type massive gravity in $d$-dimensional
space-time ($d\geq 3$) with $N$ scalar fields (Stuckelberg fields) equals to
$\frac{1}{2}d(d-3)+N$. Using the Stuckelberg language it was shown that the
number of propagators is less than the number of fields \cite{dRGT,Kluson}.
Vanishing of the determinant of the kinetic (Hessian) matrix of the scalar
field Lagrangian helps us to reduce the number of propagators. Regarding the
determinant of the kinetic matrix ($detS$), the so-called singular solutions
are related to $detS=0$ \cite{detS}. In general, we expect that all regular
solutions ($detS\neq 0$) to be tangential to the surface $detS=0$ at some
point of time. It means that the singular solutions ($detS=0$) are related
to the families of regular solutions. In other words, we can set the
conditions in vicinity of the singular surface and apply the infinitesimal
evolution in time to obtain regular solutions which are tangential to the
singular ones \cite{Alberte,Alberte1}.

For example, $4$-dimensional massive gravity theories admit, generally, up
to six propagating modes where five of them are physical and one may be a
ghost. Depending on the reference metric, the number of degrees of freedom
that can be propagated, hence the existence of ghost may vary. The existence
of the ghost determines whether the theory under consideration is actually
stable or not. In this paper, the reference metric has the form of $\Psi
=(0,0,ch_{ij})$ where in a simpler case for $4$-dimensional spacetime it
will have a form of $\Psi =(0,0,1,1)$. This is a degenerate metric which is
equivalent to a theory of massive gravity with Stuckelberg fields in the
unitary gauge \cite{Vegh}. The reason for such consideration is the fact
that under a specific coordinate transformation the spatial mass term breaks
in spatial dimensions while the general covariance is preserved in radial
and temporal coordinates. Since this choice of the reference metric leads to
preservation of the Hamiltonian constraint, one of the degrees of freedoms
is removed. In addition, the diffeomorphism is not broken in this specific
theory of massive gravity. This leads to elimination of another degrees of
freedom. Therefore, $4$-degrees of freedom exist for the propagation with
this reference metric. Now, considering that only two Stuckelberg fields
exist in the diffeomorphism invariant formulation of this theory, two
degrees of freedom are absent which leads to absence of Boulware-Deser ghost
\cite{Vegh,Zhang}. The existence of ghost for this theory was investigated
in Ref. \cite{Hassan}. The mentioned reference metric is a singular one. It
was shown that this theory with any singular metric is a ghost free theory
\cite{Zhang}. In addition, a case of reduced massive gravity with two
Stuckelberg fields was investigated in Ref. \cite{Alberte}. The extension of
the theory to a negative cosmological constant and a Maxwell field was done
in Ref. \cite{Alberte1}. The stability of the black brane solutions and the
degrees of freedoms that are provided by Maxwell field and massive gravity
was investigated in detail and it was pointed out that the stability of the
solutions and degrees of freedoms depends on the choices of the parameters.
Regarding the above statements, we do not concern for the existence of
regular solutions. Therefore in this paper we relax the dynamical stability
criteria and focus on the thermodynamical properties as well as critical
behavior. Use the electrical gauge potential ansatz, $A_{\mu }=H(r)\delta
_{\mu }^{0}$, with Maxwell equation (\ref{Maxwell equation}). we obtain
\begin{equation}
H(r)=\frac{-q}{d_{3}r^{d_{3}}},
\end{equation}%
where $q$ is an integration constant which is related to the electric
charge. Also, the Maxwell equation implies that the electric field in $d$%
-dimensions is given by
\begin{equation}
F_{tr}=\frac{q}{r^{d_{2}}}.
\end{equation}

\begin{figure}[tbp]
$%
\begin{array}{cc}
\epsfxsize=7cm \epsffile{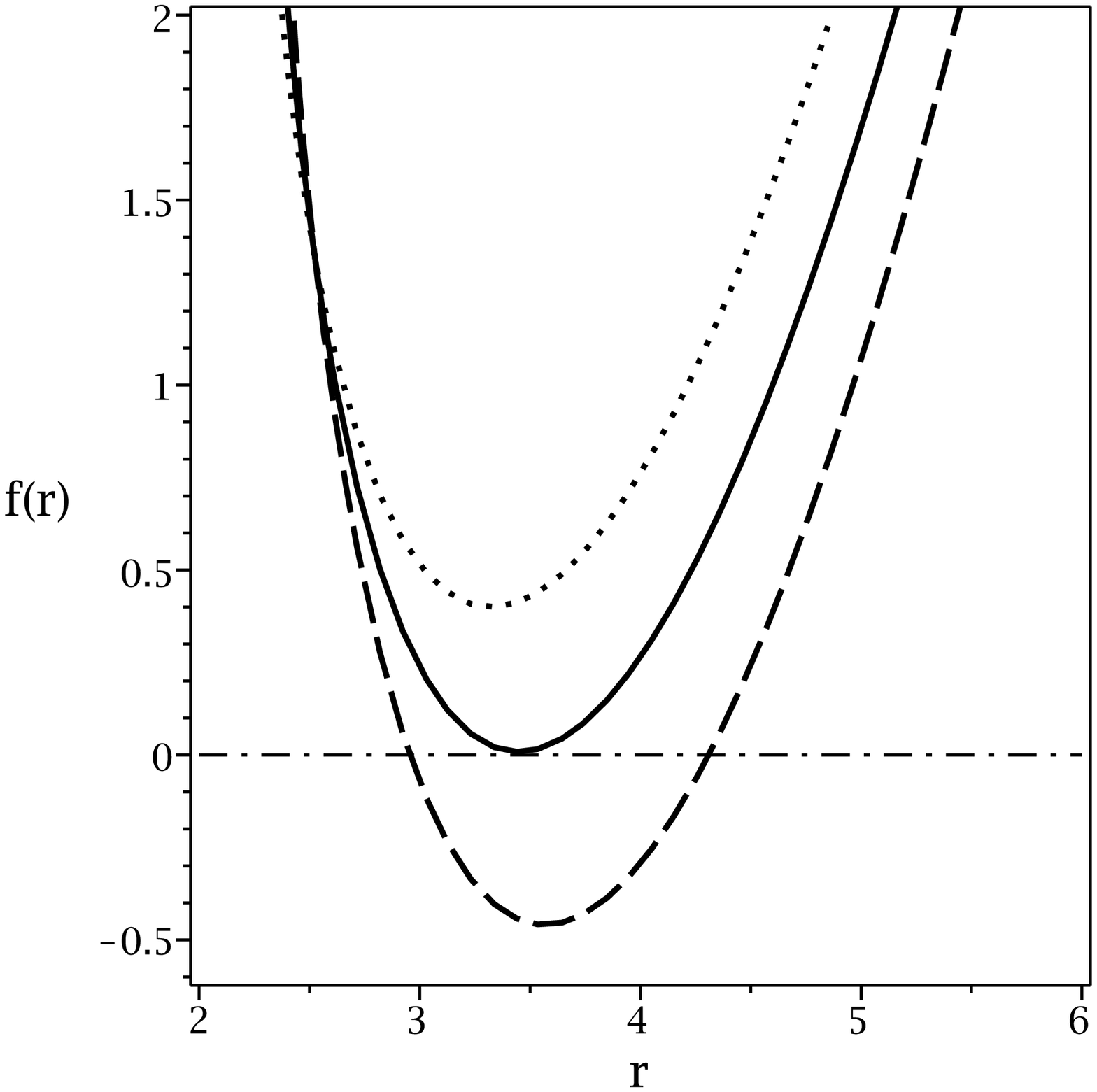} & \epsfxsize=7cm %
\epsffile{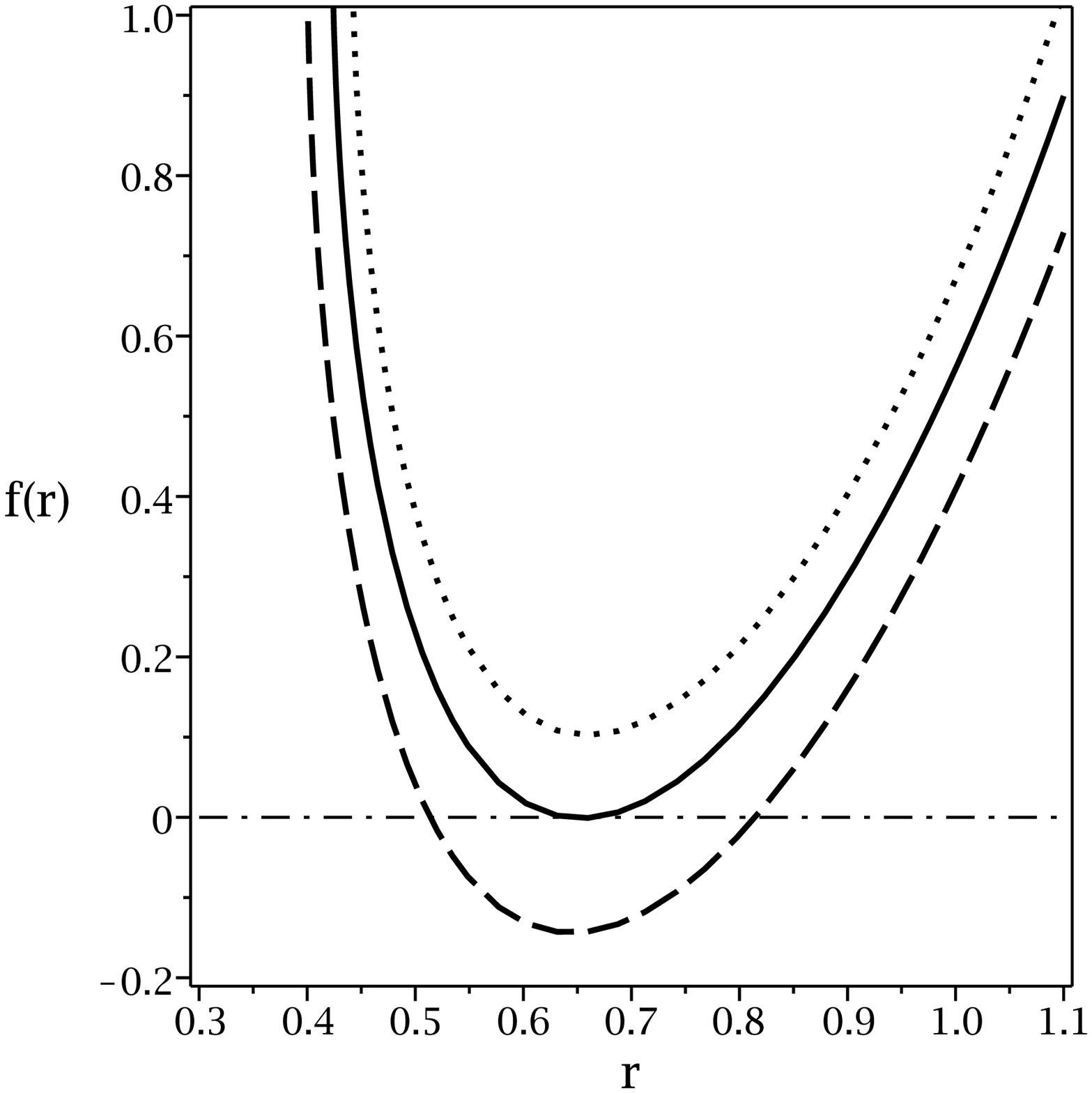}%
\end{array}
$%
\caption{$f(r)$ versus $r$ for $\Lambda =-1$, $q=1$, $\protect\alpha =0.3$, $%
c=1.2$, $c_{1}=0.6$ and $d=5$. \newline
Left panel for $k=0$, $c_{2}=-1.8$, $c_{3}=0.5$, $c_{4}=2$, $m_{0}=2$, $%
m=2.100$ (dashed line), $m=1.895$ (continues line) and $m=1.700$ (dotted
line). \newline
Right panel for $k=1$, $c_{2}=1$, $c_{3}=-0.5$, $m_{0}=1.5$, $m=1.4$, $%
c_{4}=-0.21$ (dashed line), $c_{4}=-0.18$ (continues line) and $c_{4}=-0.16$
(dotted line).}
\label{Figfr2}
\end{figure}

\begin{figure}[tbp]
\epsfxsize=8cm \centerline{\epsffile{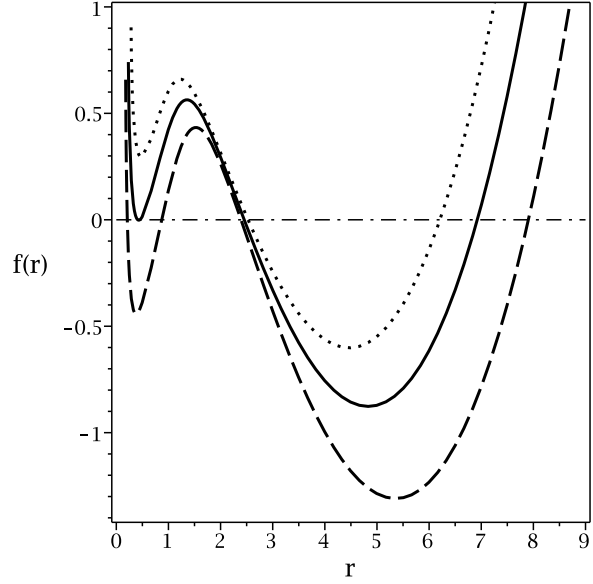}}
\caption{$f(r)$ versus $r$ for $\Lambda =-1$, $q=1$, $\protect\alpha =1.4$, $%
m=2$, $c_{1}=-0.8$, $c_{2}=-0.5$, $c_{3}=1$, $c_{4}=-0.8$, $m_{0}=0.4$, $k=1$
and $d=5$. \newline
diagrams for $c=1.200$ (dashed line), $c=1.082$ (continues line) and $%
c=1.000 $ (dotted line).}
\label{Figfr4}
\end{figure}

Now, we want to obtain the static black hole solutions. For this purpose,
one may use any components of Eq. (\ref{Field equation}) and obtain metric
function $f(r)$. One can use different components ($tt$ and $x_{1}x_{1}$) of
Eq. (\ref{Field equation}) which can be written as
\begin{eqnarray}
tt &=&\frac{d_{2}d_{3}f}{2r^{2d_{2}}}\{d_{4}d_{5}\left[ m^{2}c^{4}c_{4}+%
\alpha f^{2}\right] r^{2d_{4}}+d_{4}\left[ m^{2}c^{3}c_{3}+2\alpha ff{%
^{\prime }}\right] r^{2d_{7/2}}+\left[ m^{2}c^{2}c_{2}-f\right] r^{2d_{3}}
\notag \\
&&+\frac{\left[ m^{2}cc_{1}-f{^{\prime }}\right] r^{2d_{5/2}}}{d_{3}}-\frac{%
2\Lambda r^{2d_{2}}}{d_{2}d_{3}}-\frac{2q^{2}}{d_{2}d_{3}}-r^{2d_{3}}k\left(
\alpha d_{4}d_{5}\left[ 2f-k\right] r^{-2}+2\alpha d_{4}f^{\prime
}r^{-1}-1\right) \},  \label{tteq} \\
&&  \notag \\
x_{1}x_{1} &=&\frac{-d_{3}d_{4}}{2r^{2d_{3}}}\{d_{5}d_{6}\left[
m^{2}c^{4}c_{4}+\alpha f^{2}\right] r^{2d_{4}}+d_{5}\left[
m^{2}c^{3}c_{3}+4\alpha ff{^{\prime }}\right] r^{2d_{7/2}}-\frac{\left(
2\Lambda +f{^{\prime \prime }}\right) r^{2d_{2}}}{d_{3}d_{4}}  \notag \\
&&+\left[ m^{2}c^{2}c_{2}+2\alpha f{^{\prime }+2\alpha f}f{^{\prime \prime }}%
-f\right] r^{2d_{3}}\frac{\left[ m^{2}cc_{1}-2f{^{\prime }}\right]
r^{2d_{5/2}}}{d_{4}}+\frac{2q^{2}}{d_{3}d_{4}}  \notag \\
&&-r^{2d_{3}}k\left( \alpha d_{5}d_{6}\left[ 2f-k\right] r^{-2}+4\alpha
d_{5}f^{\prime }r^{-1}+2\alpha f^{\prime \prime }-1\right) \},
\label{x1x1eq}
\end{eqnarray}
where the prime and double prime are, respectively, the first and second
derivatives with respect to $r$. We can obtain the metric function $f(r)$
using the Eqs. (\ref{tteq}) and (\ref{x1x1eq}), yielding
\begin{eqnarray}
f\left( r\right) &=&k+\frac{r^{2}}{2\alpha d_{3}d_{4}}\left\{ 1-\sqrt{1+%
\frac{8\alpha d_{3}d_{4}}{d_{1}d_{2}}\left[ \Lambda +\frac{d_{1}d_{2}m_{0}}{%
2r^{d_{1}}}-\frac{q^{2}d_{1}}{d_{3}r^{2d_{2}}}+\Upsilon \right] }\right\} ,
\label{f(r)} \\
\Upsilon &=&-m^{2}d_{1}d_{2}\left[ \frac{d_{3}d_{4}c^{4}c_{4}}{2r^{4}}+\frac{%
d_{3}c^{3}c_{3}}{2r^{3}}+\frac{c^{2}c_{2}}{2r^{2}}+\frac{cc_{1}}{2d_{2}r}%
\right] ,  \notag
\end{eqnarray}%
where $m_{0}$ is an integration constant which is related to the total mass
of the black hole. It is notable that, obtained metric function (\ref{f(r)}%
), satisfy the all components of the Eq. (\ref{Field equation}). Also, in
the absence of massive parameter ($m=0$), the solution (\ref{f(r)}) reduces
to
\begin{equation}
f\left( r\right) =k+\frac{r^{2}}{2\alpha d_{3}d_{4}}\left\{ 1-\sqrt{1+\frac{%
8\alpha d_{3}d_{4}}{d_{1}d_{2}}\left[ \Lambda +\frac{d_{1}d_{2}m_{0}}{%
2r^{d_{1}}}-\frac{q^{2}d_{1}}{d_{3}r^{2d_{2}}}\right] }\right\} ,
\end{equation}%
which describes a $d$-dimensional asymptotically (A)dS topological black
hole with a negative, zero or positive constant curvature hypersurface. In
order to study the effects of the massive gravity on metric function, we
have plotted following diagrams (Figs. \ref{Figfr2} and \ref{Figfr4}).

It is evident that for specific values of different parameters, the metric
function will have the usual behavior that one can see in GB-Maxwell
gravity: there may be two horizons, one extreme horizon and no horizon
(naked singularity) (Fig. \ref{Figfr2}). Interestingly, considering specific
set of values for metric function parameters will lead to modification in
metric function behavior and variation from the usual behavior. In other
words, the black holes in this configuration may have following behavior:
two normal and one extreme (outer) horizons, four horizons (Fig. \ref{Figfr4}%
). The variation in number of the horizons emphasizes the contribution of
the massive part. In other words, due to existence of the massive part, the
geometrical structures of the black holes are modified and hence, their
corresponding phenomenology is also altered. It should be pointed out that
having three or four horizons is a function of the massive coefficients.
%%%%%%%%%%%%%%%%%%%%%%%%%%%%%%%%%%%%%%%%%%%%%%%%%%%%%%%%%%%%%%%

%%%%%%%%%%%%%%%%%%%%%%%%%%%%%%%%%%%%%%%%%%%%%%%%%%%%%%%%%%%%%%%
%%%%%%%%%%%%%%%%%%%%%%%%%%%%%%%%%%%%%%%%%%%%%%%%%%%%%%%%%%%%%%%
%%%%%%%%%%%%%%%%%%%%%%%%%%%%%%%%%%%%%%%%%%%%%%%%%%%%%%%%%%%%%%%%

%%%%%%%%%%%%%%%%%%%%%%%%%%%%% PENROSE %%%%%%%%%%%%%%%%%%%%%%%%%%
%%%%%%%%%%%%%%%%%%%%%%%%%%%%%%%%%%%%%%%%%%%%%%%%%%%%%%%%%%%%%%%%
\begin{figure}[tbp]
\epsfxsize=7cm \epsffile{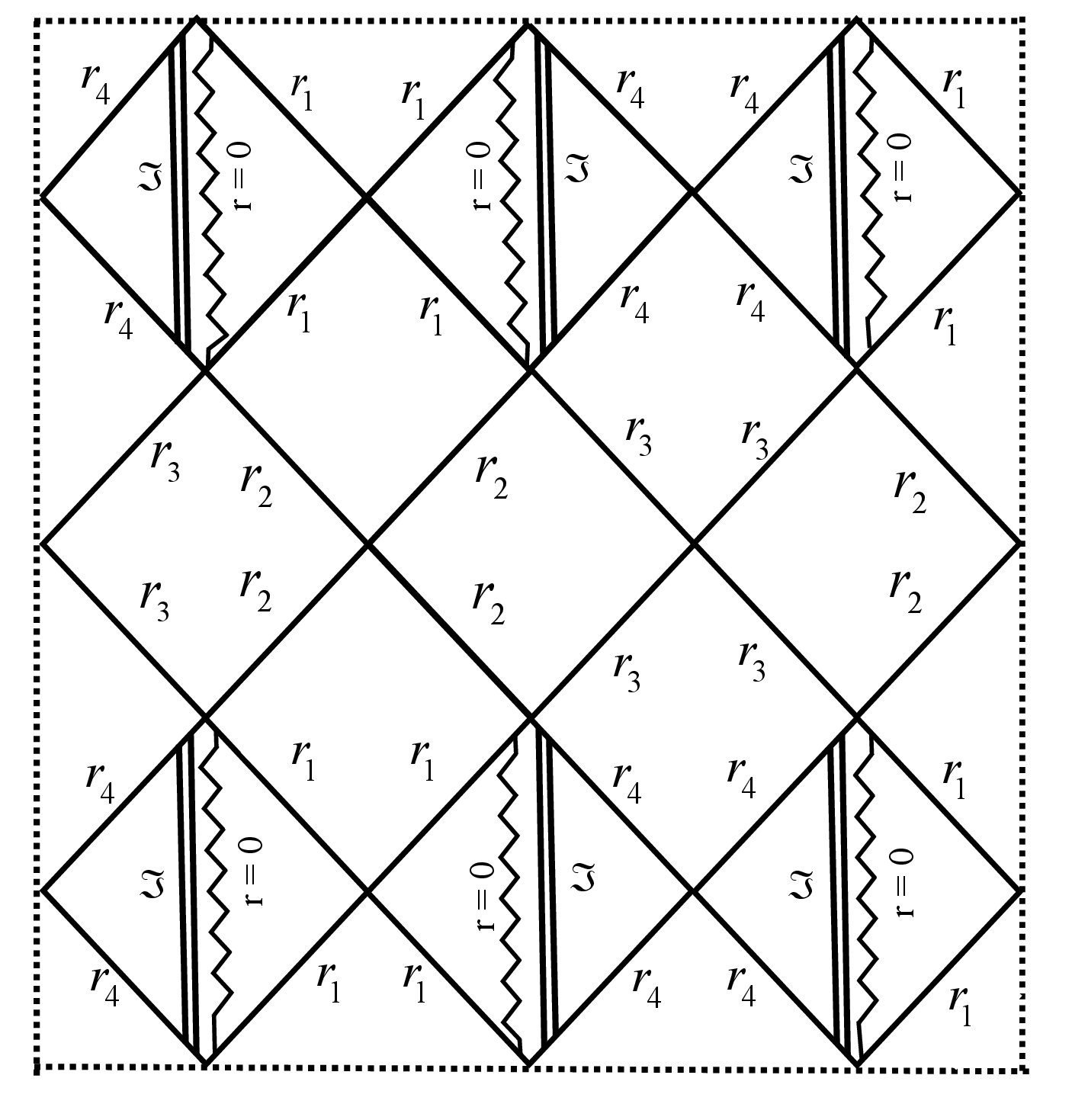}
\caption{Carter--Penrose diagram for the asymptotically adS black holes when
the metric function has four real positive roots ($r_{1}<r_{2}<r_{3}<r_{4}$%
). }
\label{PenI}
\end{figure}
%%%%%%%%%%%%%%%%%%%%%%%%%%%%%%%%%%%%%%%%%%%%%%%%%%%%%%%%%%%%%%%
%%%%%%%%%%%%%%%%%%%%%%%%%%%%%%%%%%%%%%%%%%%%%%%%%%%%%%%%%%%%%%%
\begin{figure}[tbp]
$%
\begin{array}{cc}
\epsfxsize=6cm \epsffile{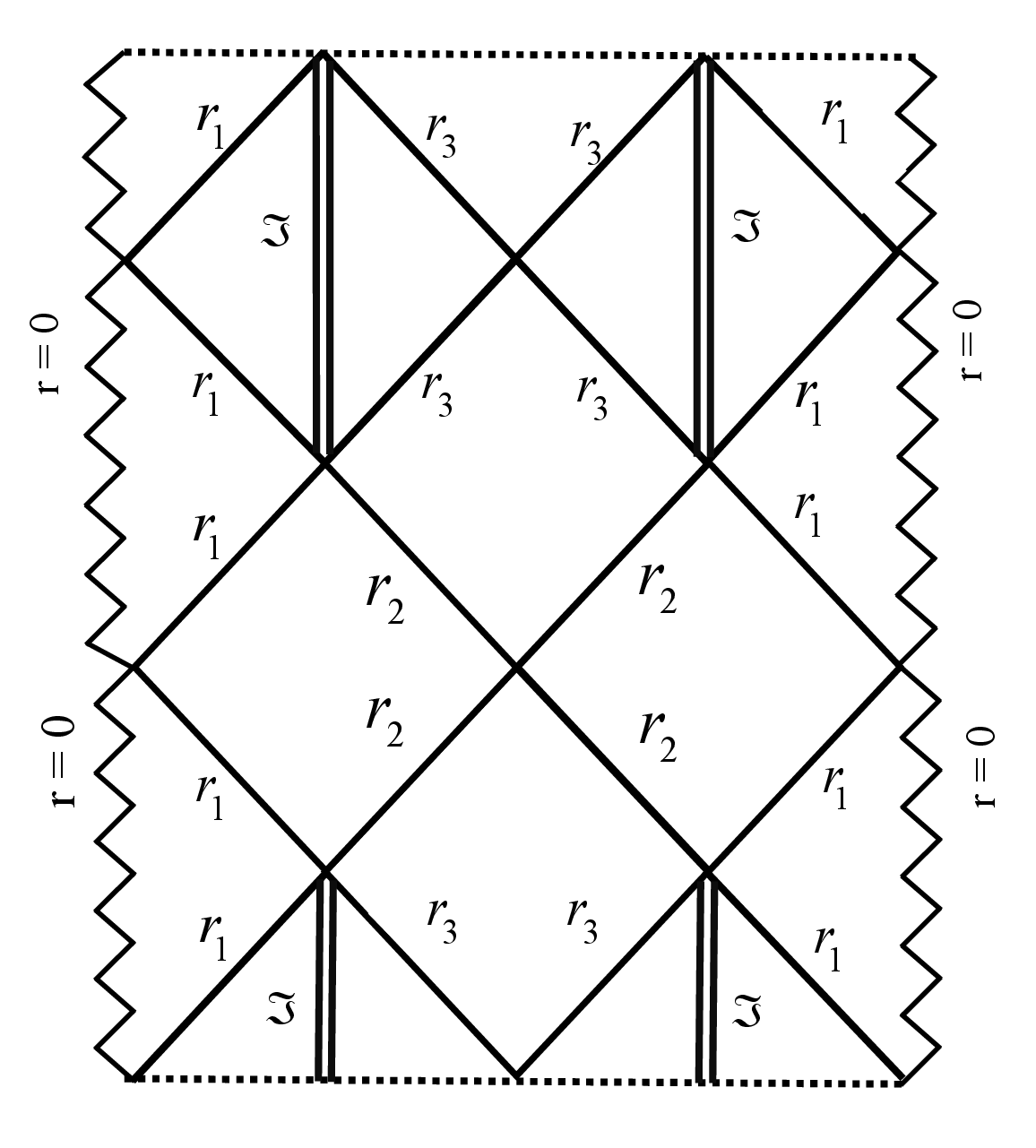} & \epsfxsize=5.5cm %
\epsffile{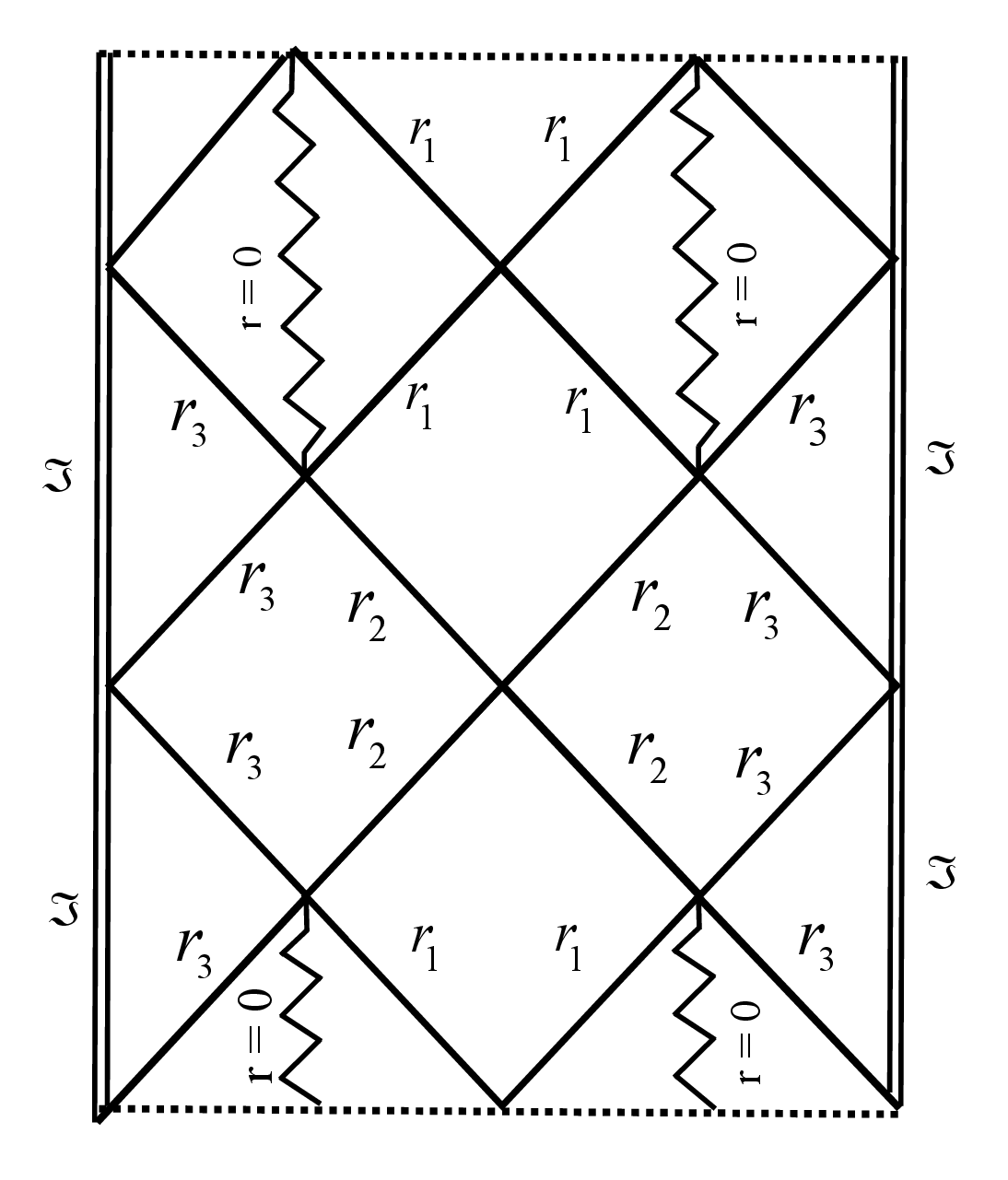}%
\end{array}
$%
\caption{Carter--Penrose diagram for the asymptotically adS black holes when
the metric function has three real positive roots ($r_{1}<r_{2}<r_{3}$). $%
r_{1}$ in left panel and $r_{3}$ in right panel are extreme roots. }
\label{PenII-III}
\end{figure}

%%%%%%%%%%%%%%%%%%%%%%%%%%%%%%%%%%%%%%%%%%%%%%%%%%%%%%%%%%%%%%%
%%%%%%%%%%%%%%%%%%%%%%%%%%%%%%%%%%%%%%%%%%%%%%%%%%%%%%%%%%%%%%%
\begin{figure}[tbp]
$%
\begin{array}{cc}
\epsfxsize=3.5cm \epsffile{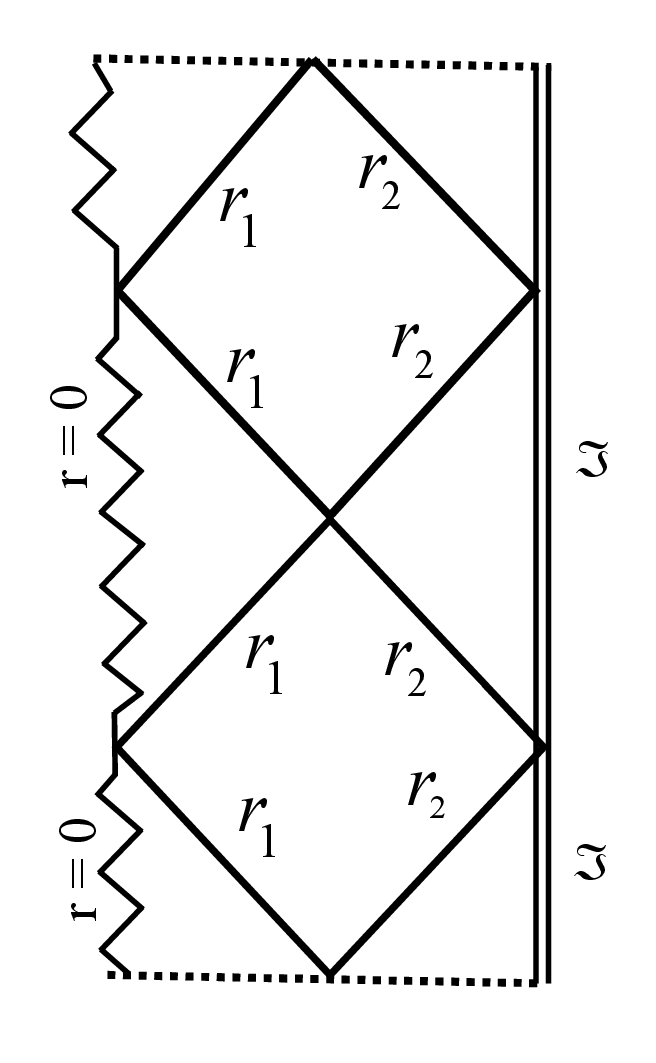} & \epsfxsize=5.5cm %
\epsffile{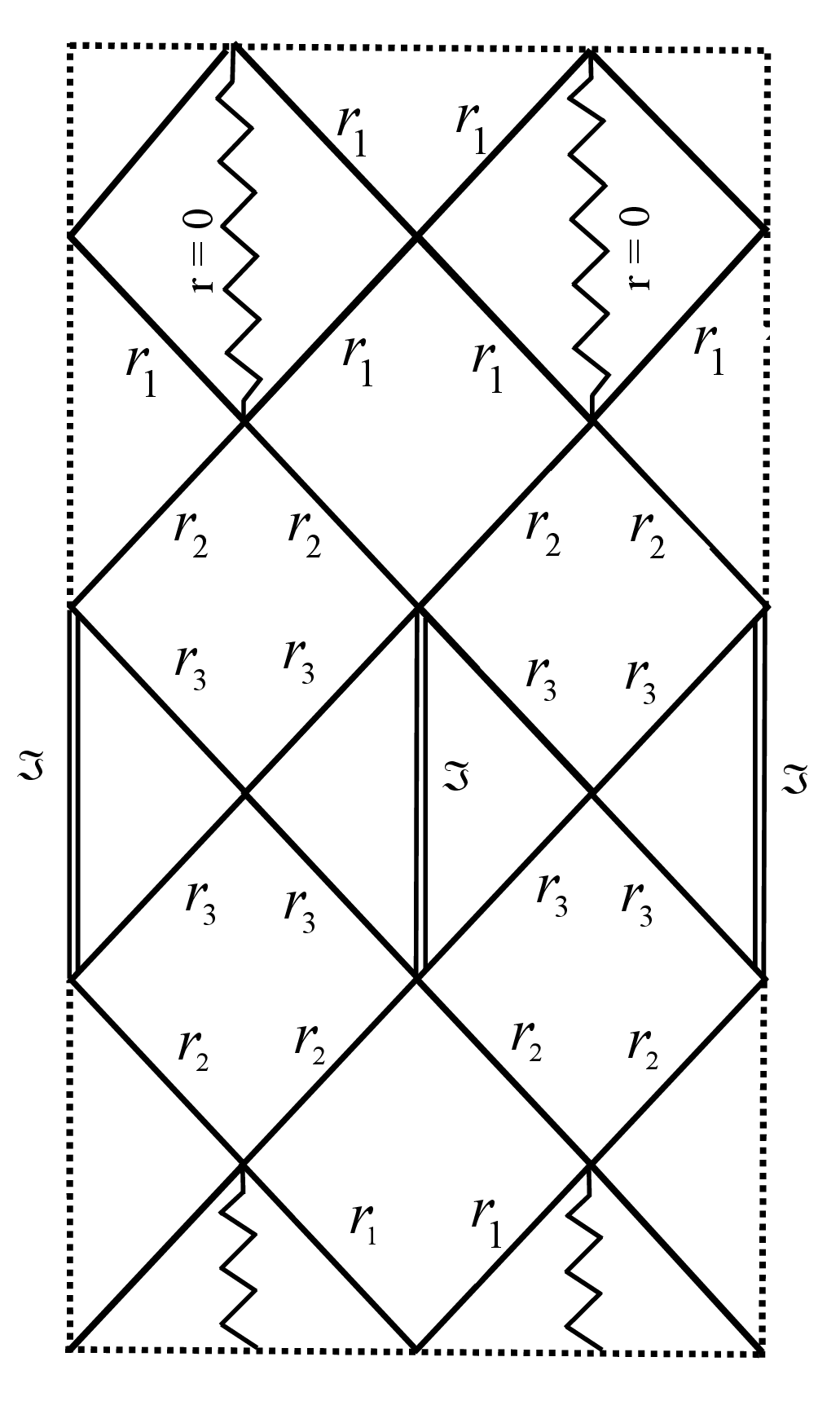}%
\end{array}
$%
\caption{Carter--Penrose diagram for the asymptotically adS black holes when
the metric function has two extreme roots (left panel: $r_{1}<r_{2}$) and
three real positive roots (right panel: $r_{1}<r_{2}<r_{3}$). $r_{2}$ in
right panel is an extreme root. }
\label{PenIV-V}
\end{figure}

%%%%%%%%%%%%%%%%%%%%%%%%%%%%%%%%%%%%%%%%%%%%%%%%%%%%%%%%%%%%%%%
Now, we are in a position to study the existence of the singularity. It is
straightforward to show that all curvature invariants of the spacetime such
as the Weyl square, Ricci square, Ricci and Kretschmann scalars are only the
functions of $f^{\prime \prime }(r)$\textbf{, }$\frac{f^{\prime }(r)}{r}$
and $\frac{f(r)}{r^{2}}$, thus it is sufficient to study one of them. It is
easy to show that
\begin{equation}
R_{\alpha \beta \gamma \delta }R^{\alpha \beta \gamma \delta }={{f^{\prime
\prime 2}(r)}}+2d_{2}{\left( {\frac{{f^{\prime }(r)}}{{r}}}\right) ^{2}}%
+2d_{2}d_{3}{\left( {\frac{{f(r)-k}}{{{r^{2}}}}}\right) ^{2}},  \label{RR}
\end{equation}%
where
\begin{equation}
f^{\prime }(r)=\frac{r\left( 1-\Omega \right) }{\alpha d_{3}d_{4}}-\frac{%
2r^{2}\Xi }{d_{1}d_{2}\Omega },
\end{equation}
\begin{equation}
f^{\prime \prime }(r)=\frac{1-\Omega }{\alpha d_{3}d_{4}}-\frac{8r\Xi }{%
d_{1}d_{2}\Omega }+\frac{8\alpha d_{3}d_{4}r^{2}\Xi ^{2}}{%
d_{1}^{2}d_{2}^{2}\Omega ^{3}}-\frac{2r^{2}\Gamma }{d_{1}d_{2}\Omega },
\end{equation}
in which
\begin{equation*}
\Xi =m^{2}d_{1}d_{2}\left[ \frac{2d_{3}d_{4}c^{4}c_{4}}{r^{5}}+\frac{%
3d_{3}c^{3}c_{3}}{2r^{4}}+\frac{c^{2}c_{2}}{r^{3}}+\frac{cc_{1}}{2d_{2}r^{2}}%
\right] -\frac{d_{1}^{2}d_{2}m_{0}}{2r^{d}}+\frac{2q^{2}d_{1}d_{2}}{%
d_{3}r^{2d-3}},
\end{equation*}
\begin{equation*}
\Gamma =\frac{dd_{1}^{2}d_{2}m_{0}}{2r^{d_{-1}}}-m^{2}d_{1}d_{2}\left[ \frac{%
10d_{3}d_{4}c^{4}c_{4}}{r^{6}}+\frac{6d_{3}c^{3}c_{3}}{r^{5}}+\frac{%
3c^{2}c_{2}}{r^{4}}+\frac{cc_{1}}{d_{2}r^{3}}\right] -\frac{%
2q^{2}d_{1}d_{2}\left( 2d-3\right) }{d_{3}r^{2d_{1}}},
\end{equation*}
\begin{equation*}
\Omega =\sqrt{1+\frac{8\alpha d_{3}d_{4}}{d_{1}d_{2}}\left[ \Lambda +\frac{%
d_{1}d_{2}m_{0}}{2r^{d_{1}}}-\frac{q^{2}d_{1}}{d_{3}r^{2d_{2}}}+\Upsilon %
\right] }.
\end{equation*}

Since the metric function diverges at $r=0$ and is a smooth regular function
for $r>0$, its first and second derivatives are regular functions for $r>0$.
Therefore, we expect to have only one physical curvature singularity located
at $r=0$. To confirm it, one can investigate the Kretschmann scalar for
various ranges of $r$. Straightforward calculations show that $R_{\alpha
\beta \gamma \delta }R^{\alpha \beta \gamma \delta }\propto r^{-2\left(
n-1\right) }$ for small values of radial coordinate. In other words, one
concludes
\begin{equation}
{\lim_{r\longrightarrow 0}}R_{\alpha \beta \gamma \delta }R^{\alpha \beta
\gamma \delta }=\infty ,  \label{RRzero}
\end{equation}%
which shows that there is an essential singularity located at the origin.
For asymptotical behavior of the metric solutions, one can series expand the
Kretschmann scalar for large values of radial coordinate which leads to
\begin{equation}
{\lim_{r\longrightarrow \infty }}R_{\alpha \beta \gamma \delta }R^{\alpha
\beta \gamma \delta }=\frac{\left[ d_{-1}d_{1}d_{2}+\left( 4\Lambda
d_{-1}+d_{1}\right) \alpha d_{3}d_{4}\right] d_{2}-\left(
d_{-1}d_{2}+d_{3}d_{4}\alpha \right) \sqrt{d_{1}d_{2}\left[
8d_{3}d_{4}\alpha \Lambda +d_{1}d_{2}\right] }}{d_{1}d_{2}d_{3}^{2}d_{4}^{2}%
\alpha ^{2}},  \label{RRlarge}
\end{equation}%
in which for small values of GB parameter it will yield
\begin{equation}
{\lim_{r\longrightarrow \infty }}R_{\alpha \beta \gamma \delta }R^{\alpha
\beta \gamma \delta }=\frac{8d_{-1}}{d_{1}^{2}d_{2}}\Lambda ^{2}-\frac{4}{%
d_{1}d_{2}}\Lambda +O\left( \alpha \right) .  \label{RRexpand}
\end{equation}

Eqs. (\ref{RRlarge}) and (\ref{RRexpand}) confirm that asymptotical behavior
of the solutions is (A)dS with an effective cosmological constant $%
\Lambda_{eff}=\Lambda_{eff}(\Lambda,\alpha)$.

Here we give a brief discussion regarding Carter--Penrose diagram. It is
believed that for investigating the conformal structure of the solutions,
one may use the conformal compactification method to plot the
Carter--Penrose or conformal diagram (see Figs. \ref{PenI}, \ref{PenII-III}
and \ref{PenIV-V}). According to the Carter--Penrose diagrams one finds
that, the singularity is timelike such as that of Reissner--Nordstr\"{o}m
black holes. In other words, although massive part of metric function may
change the horizon structure of black holes, it does not affect the type of
singularity and asymptotical behavior of the solutions. Drawing the
Carter--Penrose diagrams, one can find that the causal structure of the
solutions are asymptotically well behaved. For completeness, we should note
that the curvature singularity comes from the nature of metric function and
the geometrical nature of spacetime. In order to investigate physical
singularity, one should study the Riemann tensor. It is a matter of
calculation to show that the nonzero components of Riemann tensor for
topological black holes are
\begin{equation*}
\begin{array}{c}
R_{trtr}=\frac{1}{2}f^{\prime \prime }(r) \\
\\
R_{tx_{i}tx_{i}}=\frac{1}{2}rf^{\prime }(r)\Theta  \\
\\
R_{rx_{i}rx_{i}}=-\frac{r}{2}\frac{f^{\prime }(r)}{f(r)}\Theta =-\frac{1}{2}%
\left( \frac{\left[ rf(r)\right] ^{^{\prime }}}{f(r)}-1\right) \Theta  \\
\\
R_{x_{1}x_{i}x_{1}xi}=r^{2}\left[ k-f(r)\right] \Theta
\end{array}%
,
\end{equation*}%
where
\begin{equation*}
\Theta =\left\{
\begin{array}{cc}
\prod\limits_{i=1}^{n-1}\sin ^{2}\left( x_{i}\right)  & k=1 \\
1 & k=0 \\
\prod\limits_{i=1}^{n-1}\sinh ^{2}\left( x_{i}\right)  & k=-1%
\end{array}%
\right. ,
\end{equation*}%
and the first and second derivatives of the metric function are given
before. Although we cannot analytically check the finiteness properties of
the Riemann tensor, numerical calculations show that all singularities of
the metric function ($r=0$ and the roots of $f(r)$) are coordinate
singularity except $r=0$. The Kretschmann scalar and also the nature of
Carter-Penrose diagrams confirm that the only physical singularity is
located at $r=0$.

\section{Thermodynamics \label{Thermo}}

In this section, we calculate the conserved and thermodynamics quantities of
the static black hole solutions in $d$-dimensional GB-massive context and
then examine the first law of thermodynamics.

The Hawking temperature of the black hole at the outer horizon $r_{+}$, may
be obtained through the use of the definition of surface gravity.
Straightforward calculations lead to
\begin{eqnarray}
T &=&\frac{1}{4\pi d_{2}r_{+}^{2d_{3/2}}\left[ r_{+}^{2}+2kd_{3}d_{4}\right]
}\left\{ \Xi _{2}-2r_{+}^{4}\left( \Lambda r_{+}^{2d_{2}}+q^{2}\right)
+d_{2}m^{2}cc_{1}r_{+}^{2d_{1/2}}\right\} ,  \label{TotalT} \\
&&  \notag \\
\Xi _{2} &=&d_{2}d_{3}d_{4}d_{5}\left[ \alpha k+m^{2}c^{4}c_{4}\right]
r_{+}^{2d_{2}}+d_{2}d_{3}\left[ k+m^{2}c^{2}c_{2}\right]
r_{+}^{2d_{1}}+d_{2}d_{3}d_{4}m^{2}c^{3}c_{3}r_{+}^{2d_{3/2}}.  \notag
\end{eqnarray}

In order to calculate the electric charge of the black hole, one can use the
flux of the electromagnetic field at infinity, yielding
\begin{equation}
Q=\frac{V_{d_{2}} \ q}{4\pi }.  \label{TotalQ}
\end{equation}

Next, for the electric potential, $U$, we use the following definition \cite%
{blackholesGB}
\begin{equation}
U=A_{\mu }\chi ^{\mu }\left\vert _{r\rightarrow \infty }\right. -A_{\mu
}\chi ^{\mu }\left\vert _{r\rightarrow r_{+}}\right. =\frac{V_{d_{2}}\ q}{%
d_{3}r_{+}^{d_{3}}}.  \label{TotalU}
\end{equation}

In order to obtain the entropy of the black holes, due to generalization to
Gauss-Bonnet gravity, one can use Wald's formula to calculate it \cite%
{blackholesGB}. This leads to
\begin{equation}
S=\frac{V_{d_{2}}}{4}r_{+}^{d_{2}}\left( 1+\frac{2d_{2}d_{3}}{r_{+}^{2}}%
k\alpha \right) ,  \label{TotalS}
\end{equation}%
which shows that area law is violated for GB black holes with non-flat
horizons ($k \neq 0$).

In order to obtain total mass of the black holes, one can use Hamiltonian
approach which results into
\begin{equation}
M=\frac{V_{d_{2}}\ d_{2}\ m_{0}}{16\pi }.  \label{TotalM}
\end{equation}

Having conserved and thermodynamic quantities at hand, we are in a position
to check the first law of thermodynamics for our solutions. We obtain the
mass as a function of the extensive quantities $S$ and $Q$. One may then
regard the parameters $S$ and $Q$ as a complete set of extensive parameters
for the mass $M(S,Q)$
\begin{equation}
M(S,Q)=\frac{2d_{2}\left( 4S\right) ^{\frac{d_{1}}{d_{2}}}\left[ \frac{%
d_{3}S^{2}}{2}\mathcal{A}+d_{1}\pi ^{2}Q^{2}\left( 4S\right) ^{\frac{4}{d_{2}%
}}\right] }{\pi d_{1}d_{2}d_{3}\left( 4S\right) ^{\frac{2d}{d_{2}}}},
\label{MM}
\end{equation}%
where $\mathcal{A}$ is
\begin{equation}
\mathcal{A}=-2\Lambda \left( 4S\right) ^{\frac{4}{d_{2}}}+d_{1}c m^{2}\left(
4S\right) ^{\frac{3}{d_{2}}}\left( c_{1}+\frac{d_{2}cc_{2}}{\left( 4S\right)
^{\frac{1}{d_{2}}}}\right) +d_{1}d_{2}d_{3} m^{2}c^{3}\left( 4S\right) ^{%
\frac{1}{d_{2}}}\left( c_{3}+\frac{d_{4}cc_{4}}{\left( 4S\right) ^{\frac{1}{%
d_{2}}}}\right) .  \notag
\end{equation}

We define the intensive parameters conjugate to $S$ and $Q$. These
quantities are the temperature and the electric potential
\begin{equation}
T=\left( \frac{\partial M}{\partial S}\right) _{Q}\ \ \ \ \ \ \ \& \ \ \ \ \
\ \ \ \ U=\left( \frac{\partial M}{\partial Q}\right) _{S}.  \label{TU}
\end{equation}

The results of Eq. (\ref{TU}) coincide with Eqs. (\ref{TotalT}) and (\ref%
{TotalU}) and, therefore, we find that these conserved and thermodynamic
quantities satisfy the first law of black hole thermodynamics with the
following form
\begin{equation}
dM=TdS+UdQ.
\end{equation}

\section{Heat capacity and stability in canonical ensemble \label{Stability}}

In this section, we study the stability of the solutions. In the context of
black hole physics, there are two types of stability examination that one
can employ, the so-called dynamical and thermodynamical stabilities. In this
paper, we are only interested in thermodynamical stability of the solutions.
We investigate thermal stability in context of canonical ensemble by
calculating the heat capacity. Stability condition states that in order to a
black hole being thermally stable, its heat capacity must be positive. There
are two cases that may happen for an unstable black hole: it may have a
phase transition and acquire stable state or the obtained solution is not a
physical one (no phase transition takes place and the system is always
unstable). Another important property of the heat capacity that motivates
one to study it, is phase transition. It is stated that roots and divergence
points of the heat capacity represent type one and type two phase
transition, respectively. In other words, one can employ the numerator and
denominator of the heat capacity for finding different phase transition
points.

One can use following relation for calculating the heat capacity
\begin{equation}
C_{Q}=\frac{T}{\left( \frac{\partial ^{2}M}{\partial S^{2}}\right) _{Q}}%
=T\left(\frac{\partial S}{\partial T}\right)_{Q}=T\frac{\left( \frac{%
\partial S}{\partial r_{+}}\right)_{Q} }{\left( \frac{\partial T}{\partial
r_{+}}\right)_{Q} }.  \label{CQ}
\end{equation}

Considering Eqs. (\ref{TotalT}) and (\ref{TotalS}), it is a matter of
calculation to show that heat capacity will be%
\begin{equation}
C_{Q}=\frac{B}{C},  \label{Heat}
\end{equation}%
where
\begin{eqnarray}
B &=&-\left[ \left( 2kd_{3}d_{4}\alpha +r_{+}^{2}\right)
^{2}d_{2}^{2}c\left(
r_{+}^{3d_{2}}d_{3}d_{4}d_{5}c_{4}c^{3}+r_{+}^{3d_{5/3}}d_{3}d_{4}c_{3}c^{2}+r_{+}^{3d_{4/3}}d_{3}c_{2}c+r_{+}^{3d_{1}}c_{1}\right) %
\right] m^{2}  \notag \\
&&+2d_{2}q^{2}\left( 2kd_{3}d_{4}\alpha +r_{+}^{2}\right)
^{2}r_{+}^{d_{-2}}+2\Lambda d_{2}\left( 2kd_{3}d_{4}\alpha +r_{+}^{2}\right)
^{2}r_{+}^{3d_{2/3}}  \notag \\
&&-kd_{2}^{2}d_{3}\left( 2kd_{3}d_{4}\alpha +r_{+}^{2}\right) ^{2}\left[
r_{+}^{3d_{2}}kd_{4}d_{5}\alpha +r_{+}^{3d_{4/3}}\right] ,  \notag \\
&&  \notag \\
C &=&\left[ c_{4}d_{4}d_{5}\left( 2kd_{3}d_{4}\alpha +3r_{+}^{2}\right)
c^{3}+2r_{+}^{3}d_{4}c_{3}c^{2}-(2r_{+}^{2}kd_{3}d_{4}\alpha
c_{2}-c_{2}r_{+}^{4})c-4r_{+}^{3}kd_{4}\alpha c_{1}\right]
4d_{2}d_{3}cm^{2}r_{+}^{2d_{1}}  \notag \\
&&-8q^{2}\left[ 4kd_{7/2}d_{3}d_{4}\alpha +2d_{5/2}r_{+}^{2}\right]
r_{+}^{6}+8\Lambda \left[ 6kd_{3}d_{4}\alpha +r_{+}^{2}\right]
r_{+}^{2d_{-1}}  \notag \\
&&+4kd_{2}d_{3}\left[ 2k^{2}d_{3}d_{4}^{2}d_{5}\alpha
^{2}+kr_{+}^{2}d_{4}d_{9}\alpha +r_{+}^{4}\right] r_{+}^{2d_{1}}.  \notag
\end{eqnarray}

\begin{figure}[tbp]
$%
\begin{array}{cc}
\epsfxsize=6cm \epsffile{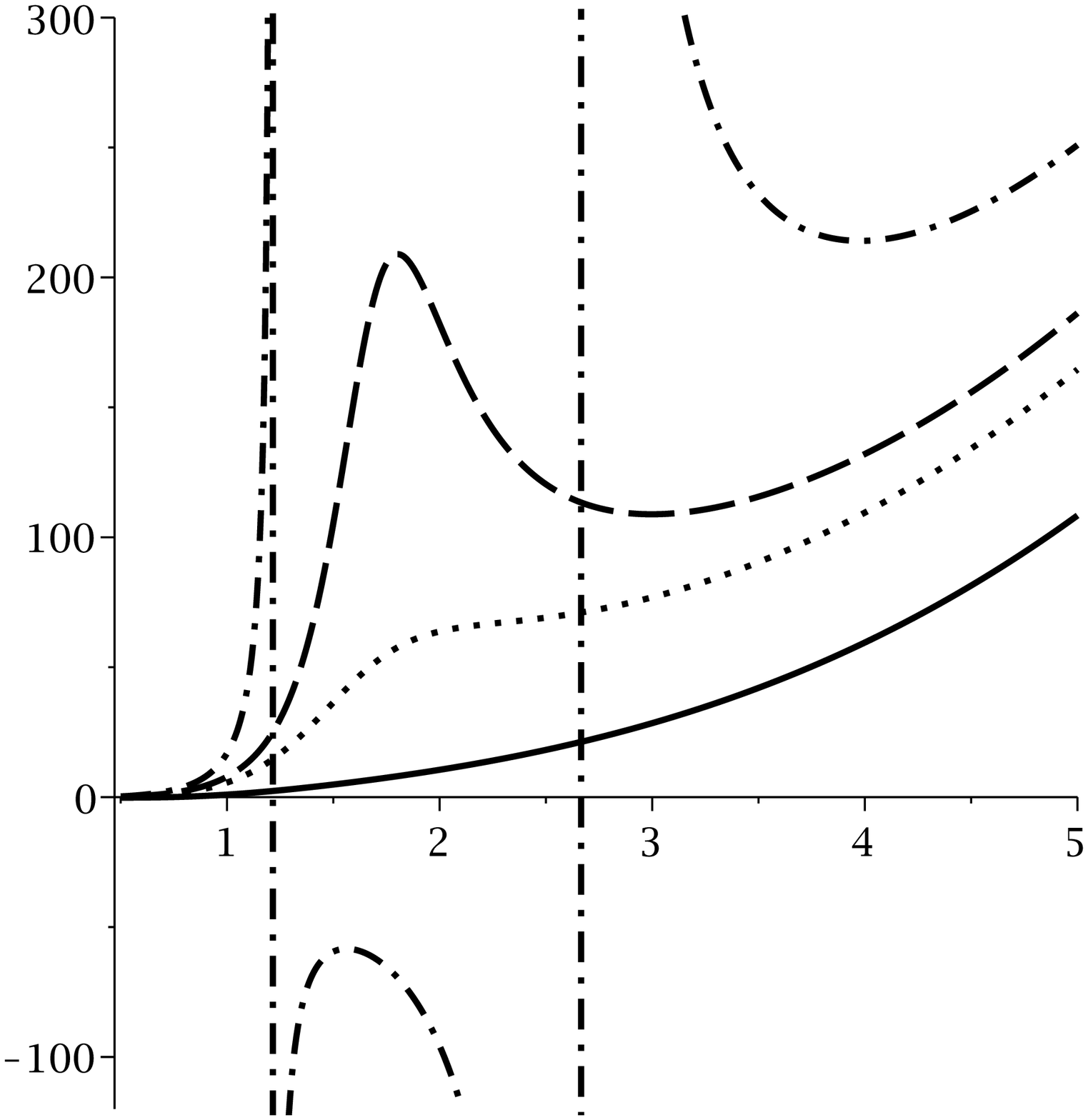} & \epsfxsize=6cm %
\epsffile{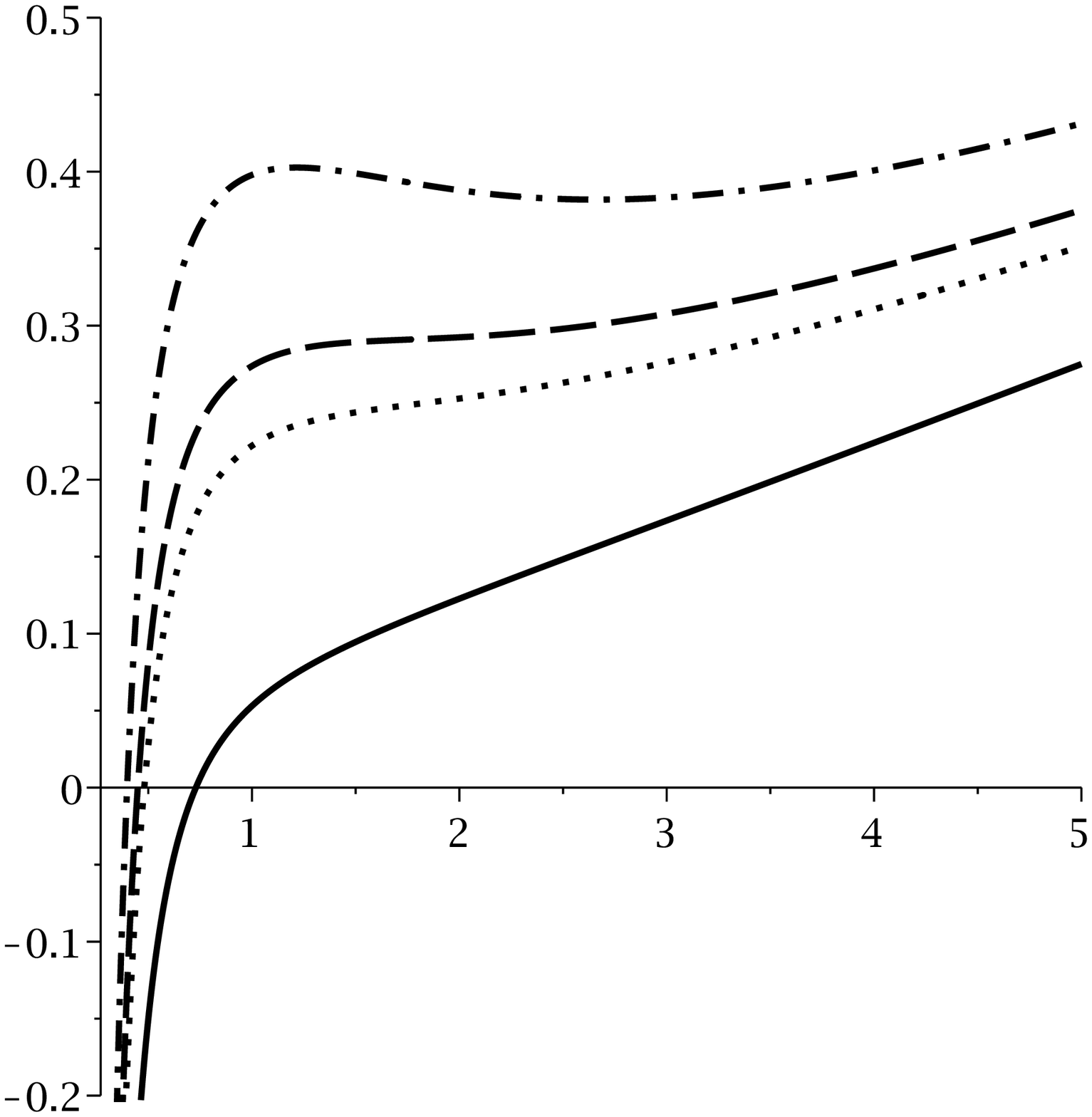}%
\end{array}
$%
\caption{$C_{Q}$ (left panel) and $T$ (right panel) versus $r_{+}$ for $q=1$%
, $\Lambda=-1$ $c=c_{1}=c_{2}=c_{3}=2$, $c_{4}=0$, $\protect\alpha=0.5$, $d=5
$ and $k=1$; $m=0$ (continues line), $m=0.35$ (dotted line), $m=0.40$
(dashed line) and $m=0.50$ (dashed-dotted line). \emph{"different scales"}}
\label{Fig1}
\end{figure}

\begin{figure}[tbp]
$%
\begin{array}{cc}
\epsfxsize=6cm \epsffile{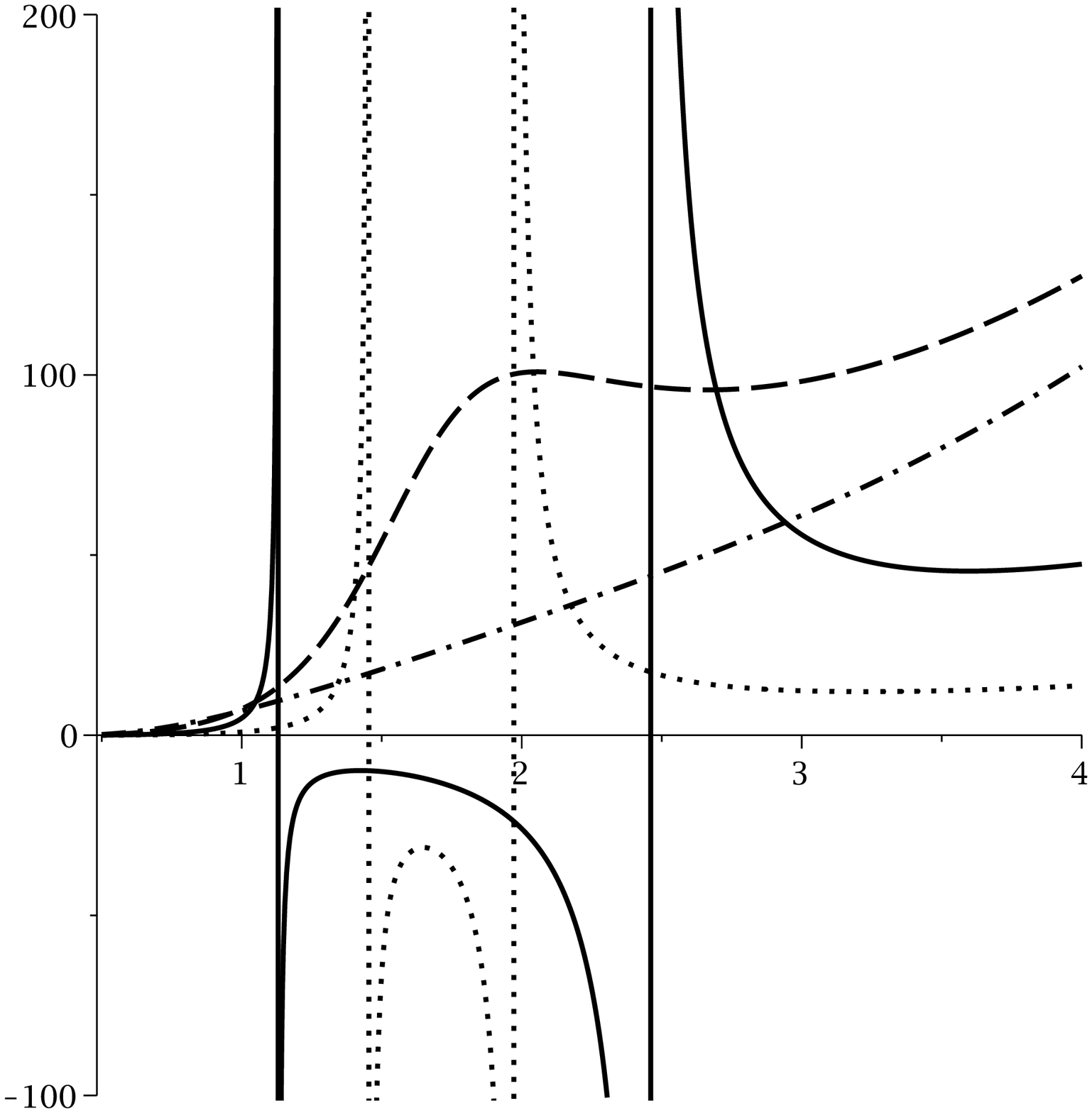} & \epsfxsize=6cm %
\epsffile{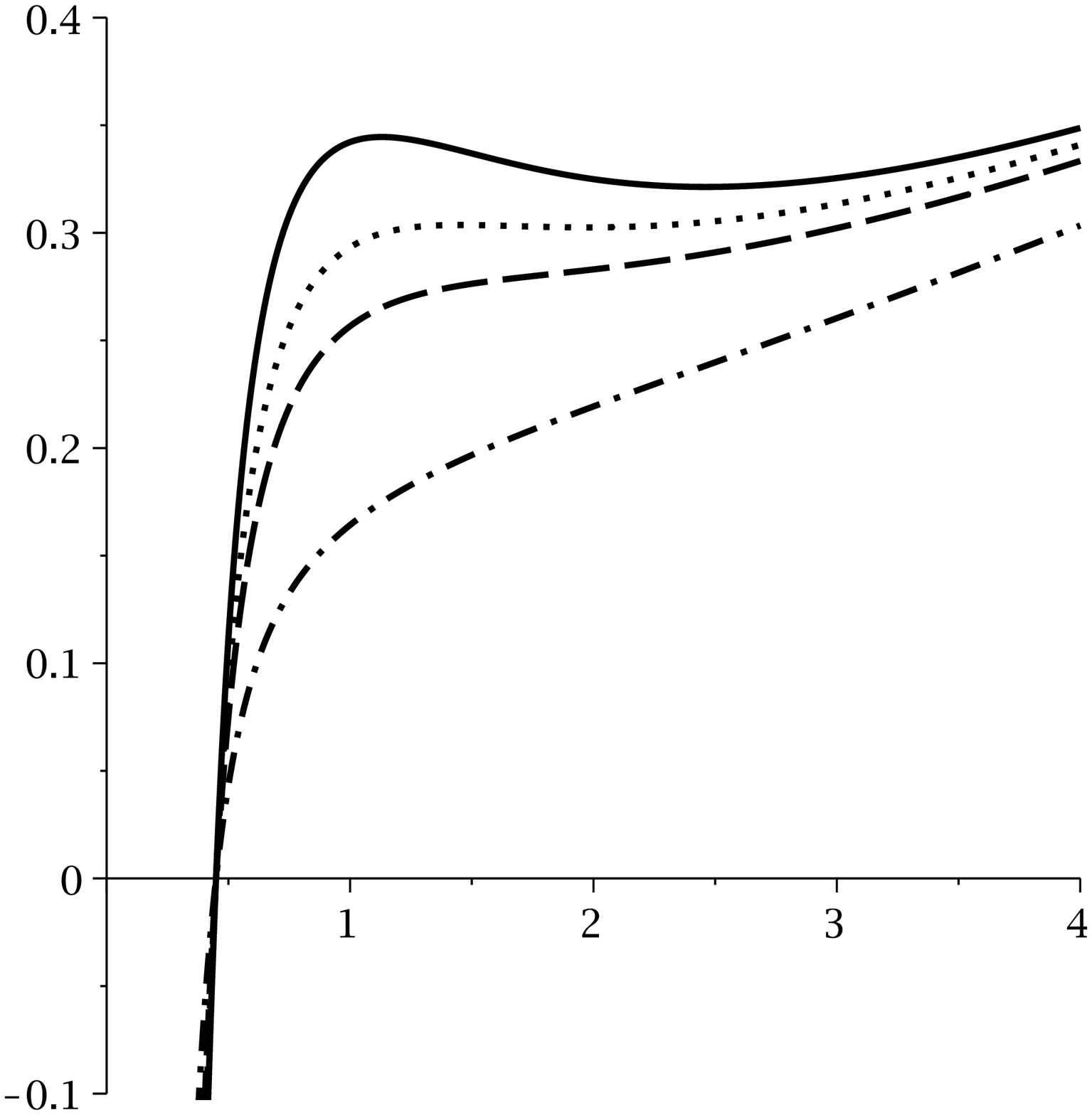}%
\end{array}
$%
\caption{$C_{Q}$ (left panel) and $T$ (right panel) versus $r_{+}$ for $q=1$%
, $\Lambda=-1$, $c=c_{1}=c_{2}=c_{3}=2$, $c_{4}=0$, $m=0.40$, $d=5$ and $k=1$%
; $\protect\alpha=0.35$ (continues line), $\protect\alpha=0.45$ (dotted
line), $\protect\alpha=0.55$ (dashed line) and $\protect\alpha=1$
(dashed-dotted line). \emph{"different scales"}}
\label{Fig2}
\end{figure}

\begin{figure}[tbp]
$%
\begin{array}{cccc}
\epsfxsize=5cm \epsffile{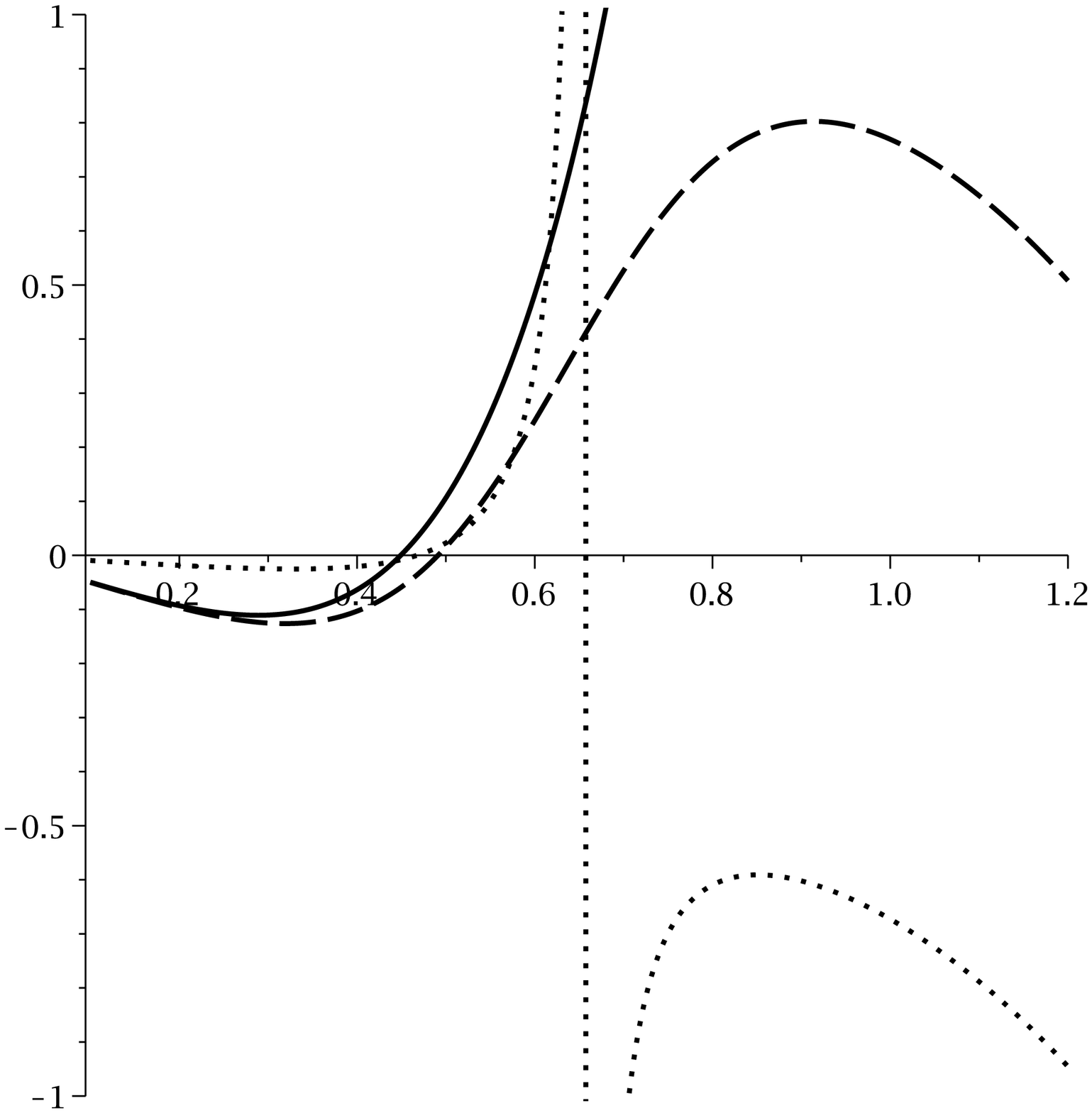} & \epsfxsize=5cm %
\epsffile{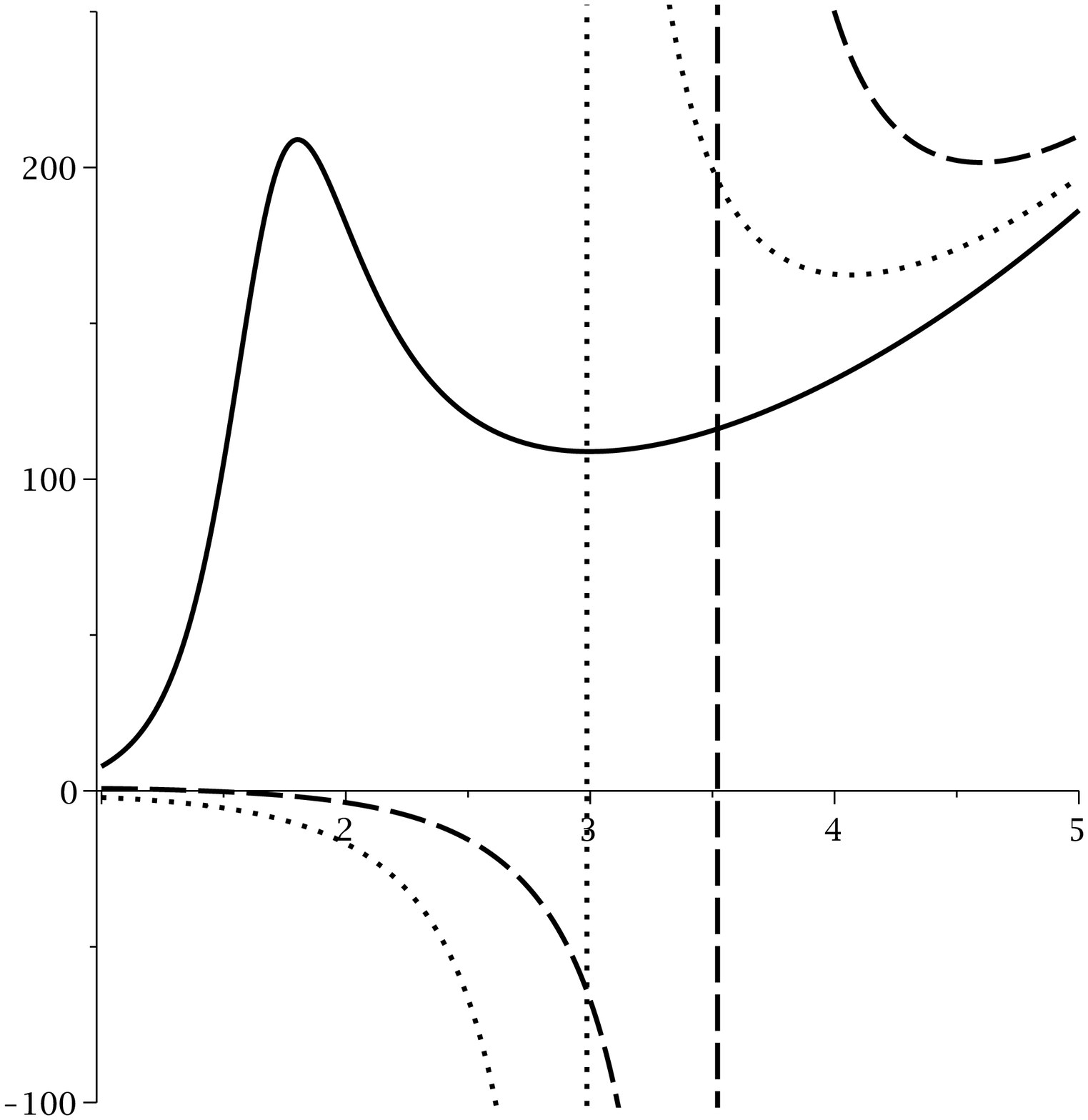} & \epsfxsize=5cm \epsffile{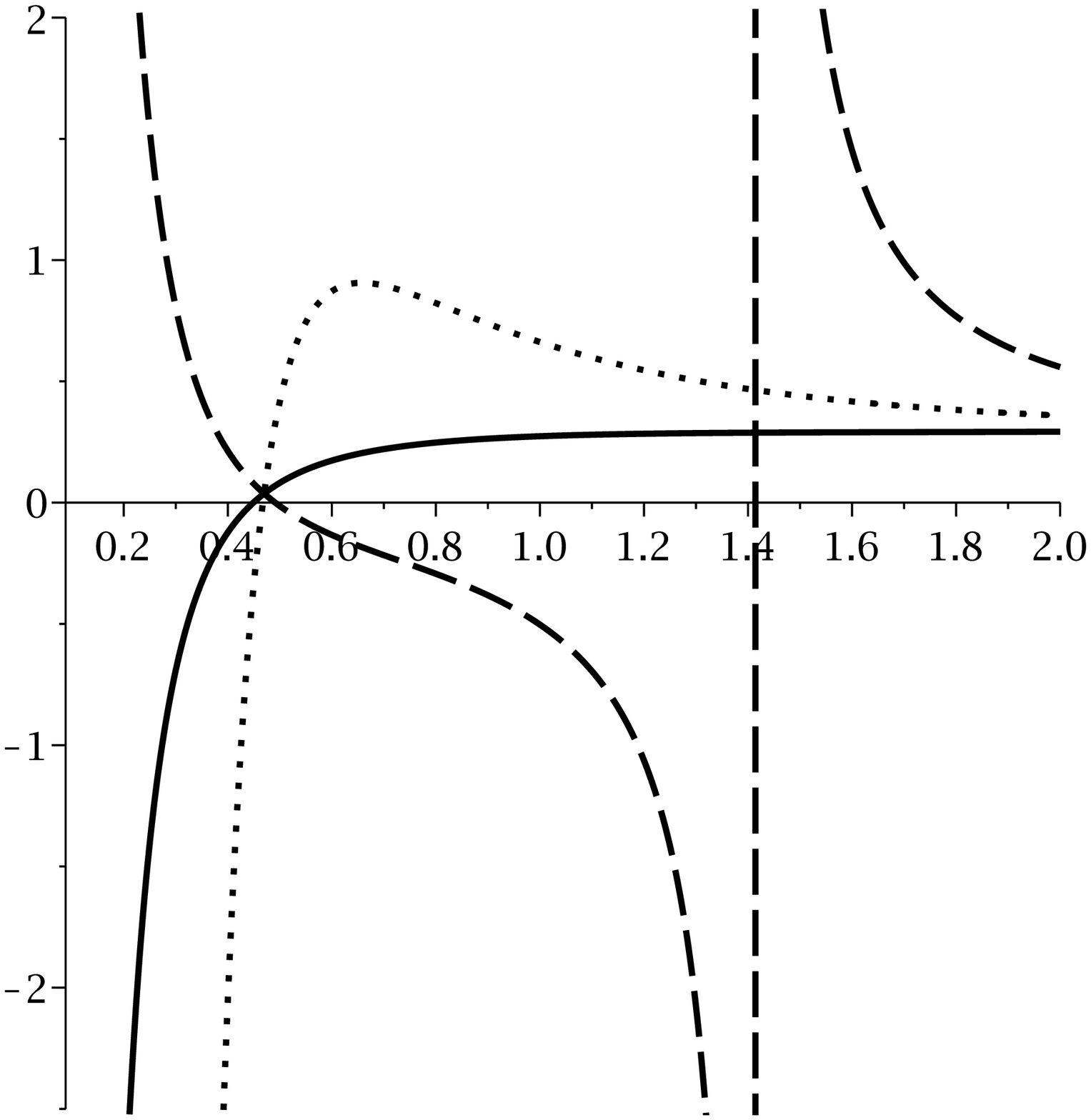} &
\end{array}
$%
\caption{$C_{Q}$ (left and middle panels) and $T$ (right panel) versus $%
r_{+} $ for $q=1$, $\Lambda=-1$, $c=c_{1}=c_{2}=c_{3}=2$, $c_{4}=0$, $m=0.40$%
, $d=5$ and $\protect\alpha=0.5$; $k=1$ (continues line), $k=0$ (dotted
line) and $k=-1$ (dashed line). \emph{"different scales"}}
\label{Fig3}
\end{figure}

\begin{figure}[tbp]
$%
\begin{array}{cc}
\epsfxsize=6cm \epsffile{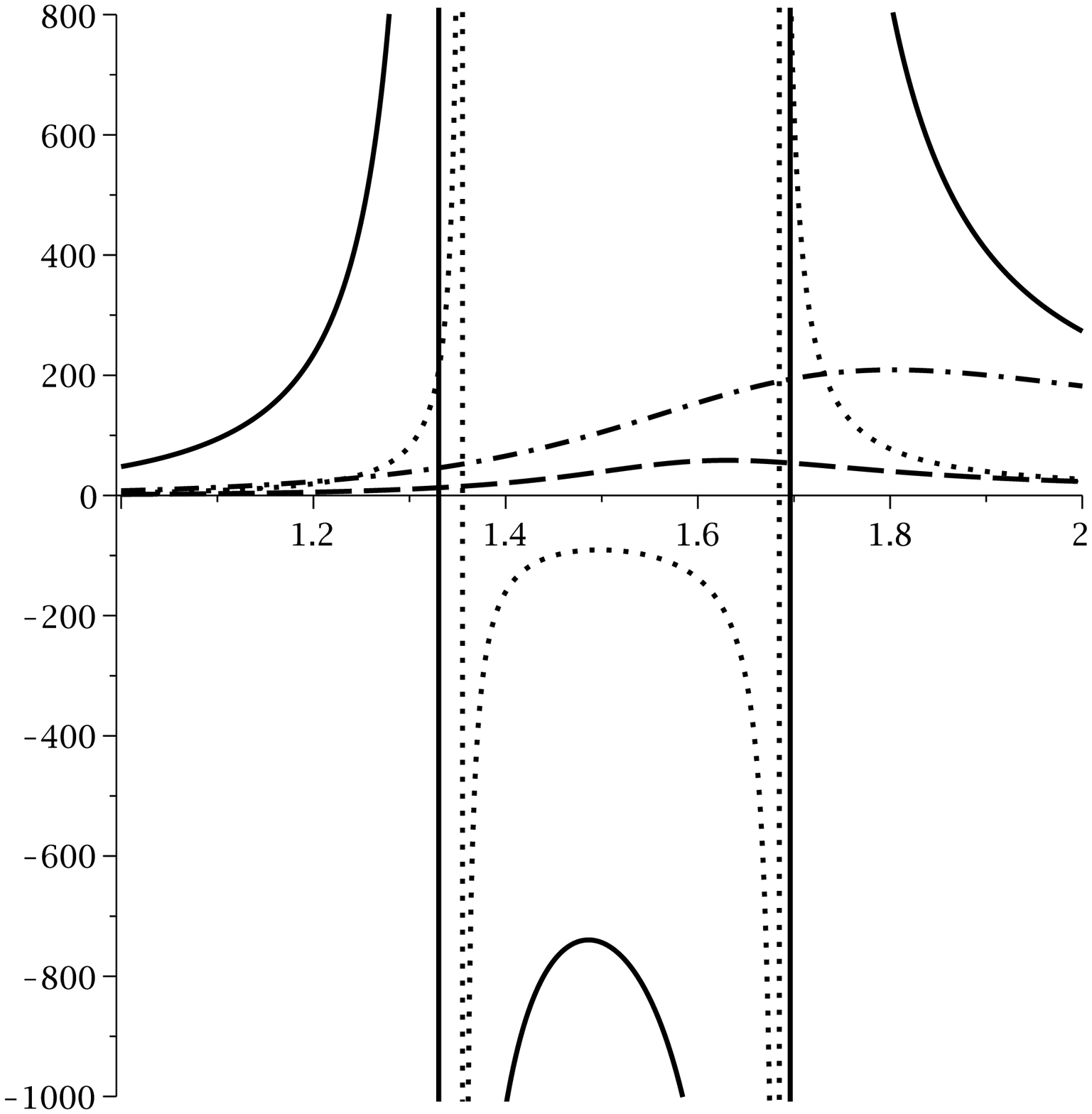} & \epsfxsize=6cm %
\epsffile{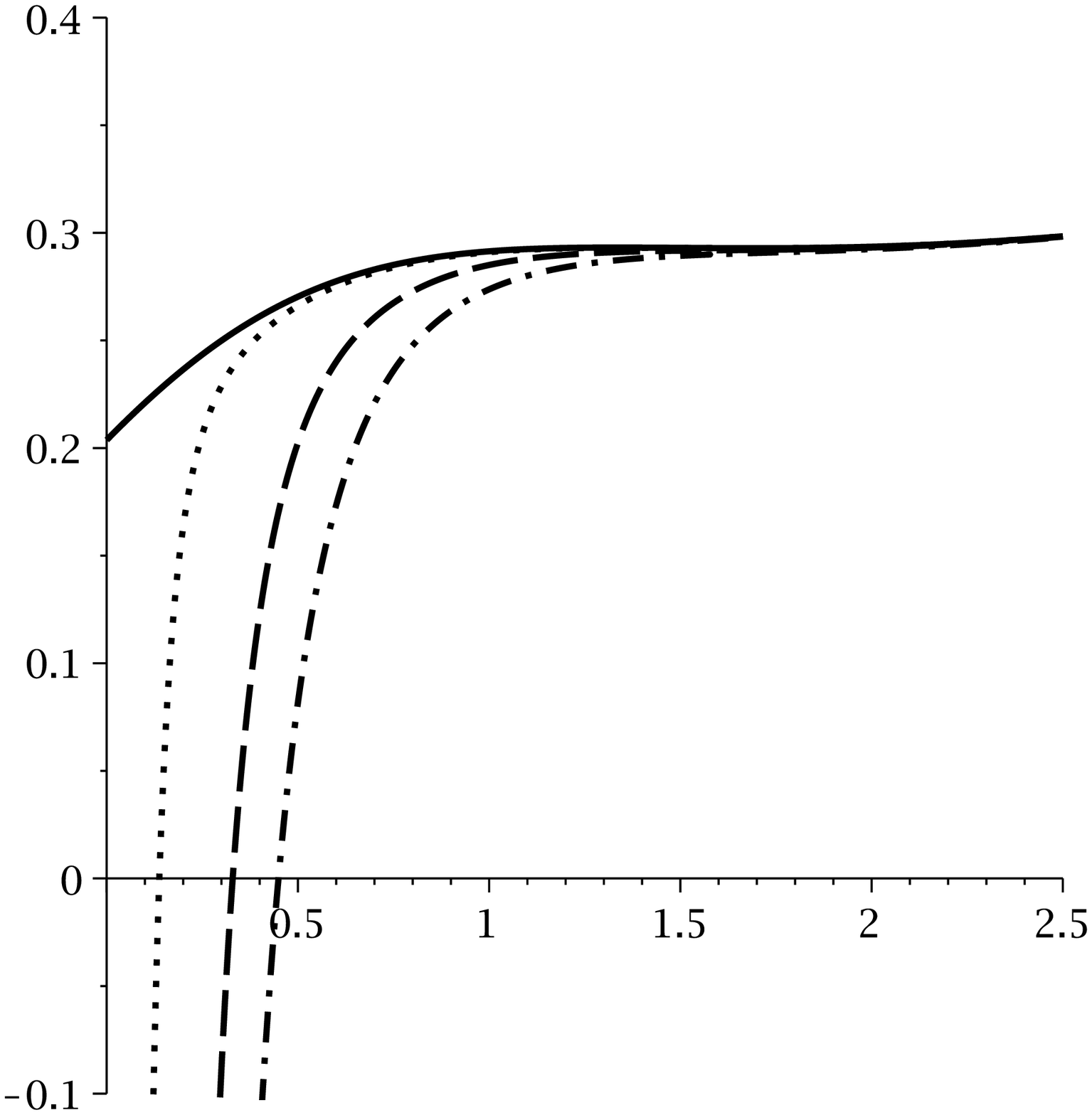}%
\end{array}
$%
\caption{$C_{Q}$ (left panel) and $T$ (right panel) versus $r_{+}$ for $%
\Lambda=-1$, $c=c_{1}=c_{2}=c_{3}=2$, $c_{4}=0$, $\protect\alpha=0.5$, $%
m=0.40 $, $d=5 $ and $k=1$; $q=0$ (continues line), $q=0.15$ (dotted line), $%
q=0.60$ (dashed line) and $q=1$ (dashed-dotted line). \emph{"different
scales"}}
\label{Fig4}
\end{figure}

\begin{figure}[tbp]
$%
\begin{array}{cc}
\epsfxsize=6cm \epsffile{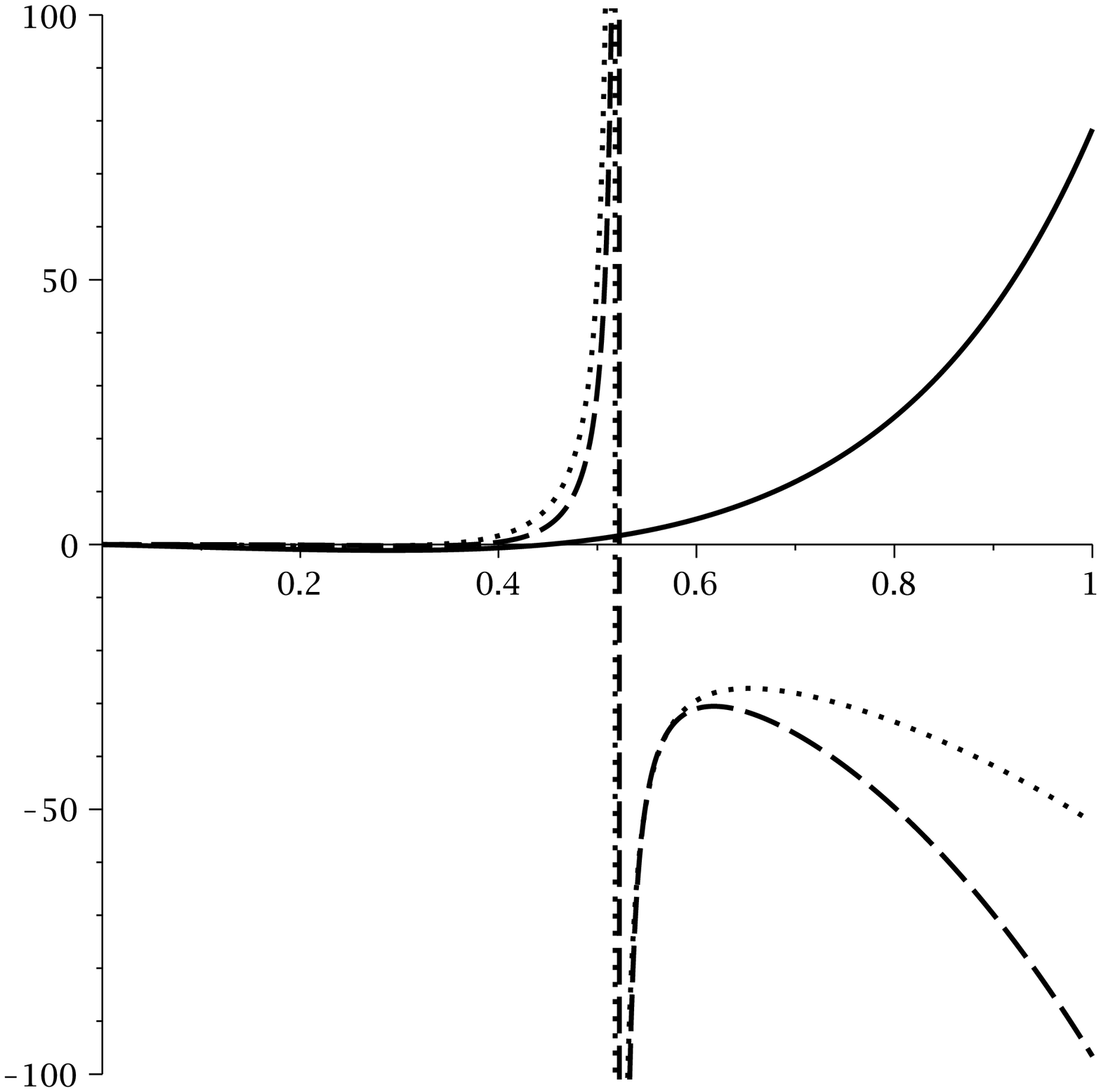} & \epsfxsize=6cm %
\epsffile{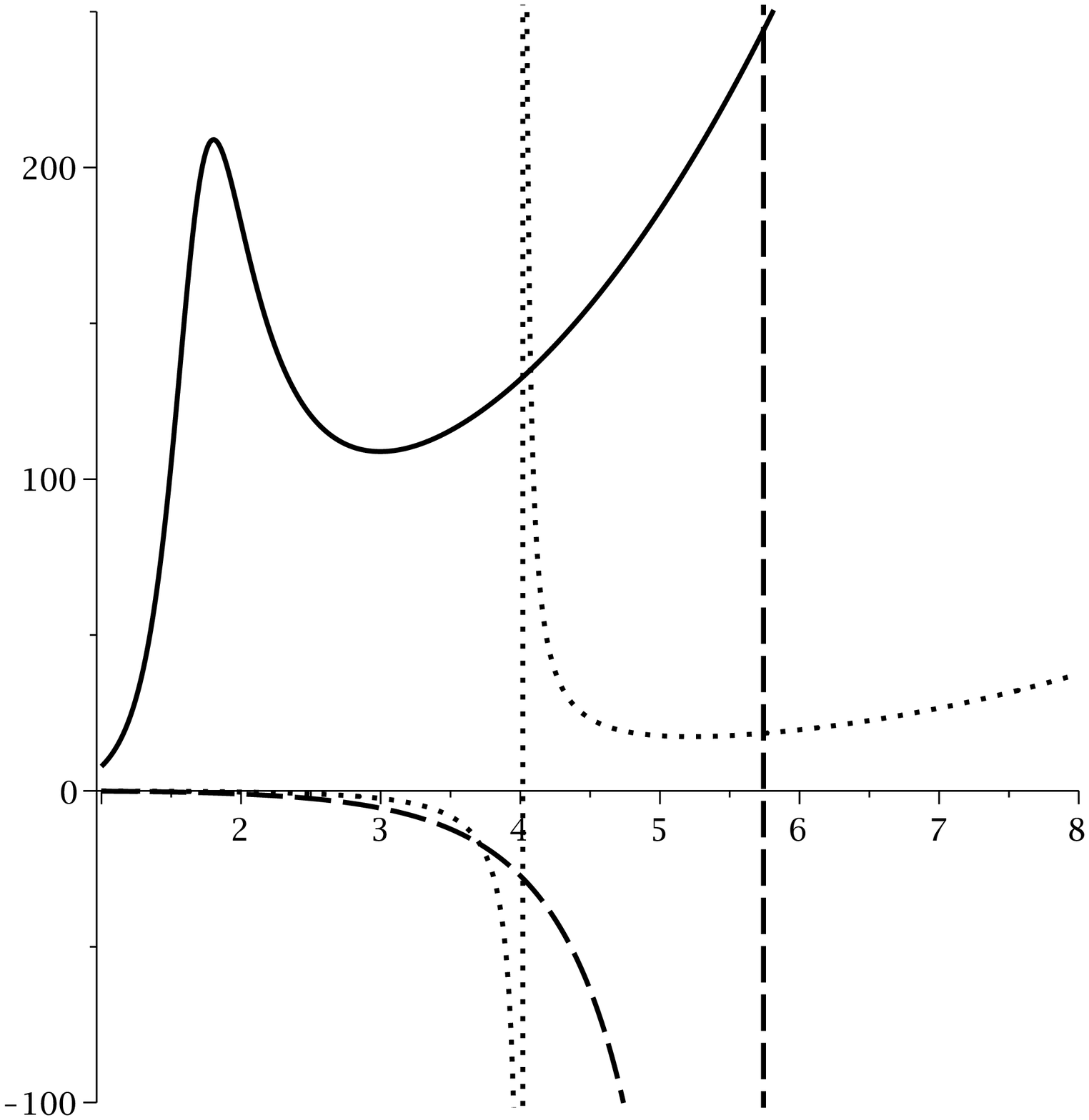}%
\end{array}
$%
\caption{$C_{Q}$ versus $r_{+}$ for $q=1$, $\Lambda =-1$, $%
c=c_{1}=c_{2}=c_{3}=c_{4}=2$, $\protect\alpha =0.5$, $m=0.40$ and $k=1$; $%
d=5 $ (continues line), $d=6$ (dotted line) and $d=7$ (dashed line). \emph{%
"different scales"}}
\label{Fig5}
\end{figure}

As one can see, GB parameter and the topology factor ($k$) are coupled with
each other. It shows that in case of flat horizon ($k=0$), the effect of GB
gravity is vanished. In other words, in horizon flat solutions, the heat
capacity is independent of the GB gravity effects. In order to study the
thermodynamical behavior of the system, stability and phase transition
points, we have plotted Figs. \ref{Fig1} - \ref{Fig5}.

It is evident that in case of spherical horizon ($k=1$), there is a root for
temperature, $r_{R}$, in which for $r_{+}<r_{R}$, temperature is negative.
Therefore, in this region, obtained solutions are non-physical. $r_{R}$ is a
decreasing function of massive parameter (Fig. \ref{Fig1} right panel), an
increasing function of electric charge (right panels of Figs. \ref{Fig4})
and finally independent of GB parameter (Fig. \ref{Fig2} right panel). It
should be pointed out that interestingly, increasing and decreasing
mentioned parameters, will lead to formation of a maximum and minimum.
Considering Eq. (\ref{CQ}), these extrema are places in which heat capacity
will have divergencies. In case of the topology, interestingly, for $k=-1$,
temperature have a root and one divergence point (Fig. \ref{Fig3} right
panel) which is not observed in case of horizon flat (Fig. \ref{Fig3} middle
panel) and spherical symmetric (Fig. \ref{Fig3} right panel).

For small values of massive parameter, the heat capacity is an increasing
function of horizon radius (Fig. \ref{Fig1} left panel). For specific values
of the massive parameter, a maximum will be formed which is highly sensitive
to variation of this parameter. For sufficiently large enough of $m$, there
will be two divergencies for the heat capacity and between these two
divergencies heat capacity is negative, hence, the system is unstable (Fig. %
\ref{Fig1} left panel).

Interestingly, the variation of the GB parameter has opposite effect on heat
capacity comparing to variation of massive parameter. In other words, for
small values of the GB parameter, black holes enjoy two divergencies with
being unstable between these divergencies (Fig. \ref{Fig2} left panel). The
interval between these divergencies is a decreasing function of GB
parameter. For sufficiently small values of $\alpha $, the heat capacity is
only an increasing function of the horizon radius without divergency (Fig. %
\ref{Fig2} left panel).

As for topological effects, in case of spherical symmetric, no divergency is
observed and for the $r_{+}>r_{R}$, system is in physical stable state (Fig. %
\ref{Fig3} left panel). On the other hand, in case of horizon flat, two
phase transitions are observed for heat capacity with black holes being
unstable between two divergencies (Fig. \ref{Fig3} left and middle panels).
Finally, for $k=-1$, two roots and one divergence point are found. Between
these roots and after divergence point, system is physical and stable
whereas between larger root and divergence point, black holes are thermally
unstable (Fig. \ref{Fig3} left and middle panels).

Similar behavior to variation of GB parameter is observed for variation of
the charge except for one difference. For large values of GB parameter, heat
capacity is only an increasing function of the horizon radius whereas by
increasing the electric charge, the heat capacity will have maximum again
(Fig. \ref{Fig4} left panel).

Finally, in case of dimensions, it is observed that higher dimensionality
has contribution to number and place of the divergencies for heat capacity.
For $5$-dimensions, only one root is observed, whereas in case of $6$ and $7$%
-dimensions, heat capacity has one root and two divergencies which are
increasing functions of dimensions (Fig. \ref{Fig5}).

Now we are in a position to discuss phase transitions of the solutions. As
it was stated before, roots and divergencies of the heat capacity are
representing phase transition points of the system. In addition, it should
be pointed out that thermodynamical principles state that systems in
unstable states go under phase transition to stabilize. Therefore, for
systems with one root and two divergence points, the following phase
transitions happen: for the root there is a phase transition from unstable
non-physical system to stable and physical one, for smaller divergence point
a phase transition from larger black holes to smaller ones takes place and
finally for larger divergency, systems go under phase transition from
smaller black holes to larger ones.

%%%%%%%%%%%%%%%%%%%%%%%%%%%%%%%%%%%%%%%%%%%%%%%%%%%%%%%%%%%%%%%

%%%%%%%%%%%%%%%%%%%%%%%%%%%%%%%%%%%%%%%%%%%%%%%%%%%%%%%%%%%%%%%
%%%%%%%%%%%%%%%%%%%%%%%%%%%%%%%%%%%%%%%%%%%%%%%%%%%%%%%%%%%%%%%

%%%%%%%%%%%%%%%%%%%%%%%%%%%%%%%%%%%%%%%%%%%%%%%%%%%%%%%%%%%%%%%
%%%%%%%%%%%%%%%%%%%%%%%%%%%%%%%%%%%%%%%%%%%%%%%%%%%%%%%%%%%%%%%

%%%%%%%%%%%%%%%%%%%%%%%%%%%%%%%%%%%%%%%%%%%%%%%%%%%%%%%%%%%%%%%
%%%%%%%%%%%%%%%%%%%%%%%%%%%%%%%%%%%%%%%%%%%%%%%%%%%%%%%%%%%%%%%

%%%%%%%%%%%%%%%%%%%%%%%%%%%%%%%%%%%%%%%%%%%%%%%%%%%%%%%%%%%%%%%
%%%%%%%%%%%%%%%%%%%%%%%%%%%%%%%%%%%%%%%%%%%%%%%%%%%%%%%%%%%%%%%

%%%%%%%%%%%%%%%%%%%%%%%%%%%%%%%%%%%%%%%%%%%%%%%%%%%%%%%%%%%%%%%

\section{$P-V$ criticality of charged black holes in GB-Massive gravity
\label{PV}}

In this section, we study the phase transition points of the charged black
holes in GB-Massive gravity through use of $P-V$ criticality and related
phase diagrams in spherical symmetric spacetime ($k=1$). To do so, we
consider following relationship between thermodynamical pressure and
cosmological constant
\begin{equation}
P=-\frac{\Lambda }{8\pi }.  \label{P}
\end{equation}

From thermodynamical point of view, one can indicate that conjugating
thermodynamical variable corresponding to pressure would be thermodynamical
volume. Therefore, in order to calculate the thermodynamical volume of the
solutions, one should use
\begin{equation}
V=\left( \frac{\partial H}{\partial P}\right) _{S,Q}.  \label{V}
\end{equation}

Considering cosmological constant as thermodynamical pressure leads to
interpretation of mass not only as internal energy but also as Enthalpy of
thermodynamical system. This interpretation leads to following relation for
the Gibbs free energy of the system
\begin{equation}
G=H-TS=M-TS,  \label{G}
\end{equation}%
where by considering Eqs. (\ref{TotalT}), (\ref{TotalS}) and (\ref{TotalM})
\begin{equation}
G=-\frac{\Theta _{0}}{16r^{d_{-1}}d_{1}d_{2}d_{3}d_{4}\left(
r_{+}^{2}+2\alpha ^{\prime }\right) },  \label{Gibbs}
\end{equation}%
in which
\begin{eqnarray}
\Theta _{0} &=&16\pi d_{3}r_{+}^{2d}\left( 6d_{2}\alpha ^{\prime
}+d_{4}r_{+}^{2}\right) P-4d_{1}q^{2}\left[ 2d_{2}d_{7/2}+d_{4}d_{5/2}\right]
r_{+}^{4}-d_{1}d_{2}d_{3}\left[ 2d_{2}\alpha ^{\prime 2}+d_{8}\alpha
^{\prime }r_{+}^{2}+d_{4}r_{+}^{2}\right] r_{+}^{2d_{2}}  \notag \\
&&-d_{1}d_{2}d_{3}m^{2}c\left[ \left(
d_{2}d_{3}d_{4}c_{4}c^{3}-d_{2}c_{2}cr_{+}^{2}-2c_{1}r_{+}^{3}\right)
2\alpha ^{\prime }+\left(
3d_{3}d_{4}c_{4}c^{2}+2d_{3}c_{3}cr_{+}+c_{2}r_{+}^{2}\right) d_{4}cr_{+}^{2}%
\right] r_{+}^{2d_{2}},  \notag
\end{eqnarray}%
and $\alpha ^{\prime }=d_{3}d_{4}\alpha $.

By regarding this interpretation and using Eqs. (\ref{TotalM}), (\ref{P})
and (\ref{V}), one can obtain volume as
\begin{equation}
V=\frac{\omega _{d_{2}}}{d_{1}}r_{+}^{d_{1}},
\end{equation}%
which is in agreement with the volume of the spacetime with spherical
boundary and radius $r_{+}$. Because of the relation between volume and
radius of the black hole, we use horizon radius (specific volume) for
obtaining critical values. To calculate critical values, one can employ the
inflection point properties
\begin{equation}
\left( \frac{\partial P}{\partial r_{+}}\right)_{T} =\left( \frac{\partial
^{2}P}{\partial r_{+}^{2}}\right)_{T} =0.  \label{infel}
\end{equation}

As for thermodynamical pressure, by using Eqs. (\ref{TotalT}) and (\ref{P}),
one can obtain the following equation of state
\begin{equation}
P=\frac{d_{2}\left( 2\alpha ^{\prime }+r_{+}^{2}\right) T}{4r_{+}^{3}}+\frac{%
q^{2}}{8\pi r_{+}^{2d_{2}}}-\frac{d_{2}\left( d_{5}\alpha ^{\prime
}+d_{3}r_{+}^{2}\right) }{16\pi r_{+}^{4}}-\frac{m^{2}\left[
d_{4}d_{5}c_{4}c^{3}+d_{4}c_{3}c^{2}r_{+}c_{2}cr_{+}^{2}+c_{1}r_{+}^{3}%
\right] d_{2}d_{3}c}{16\pi r_{+}^{4}}.  \label{PP}
\end{equation}

Now, we are in a position to calculate critical horizon radius. To do so,
one should employ Eqs. (\ref{infel}) and (\ref{PP}) which lead to following
relation%
\begin{eqnarray}
&&4d_{2}q^{2} \left[ 6d_{7/2}\alpha ^{\prime }+d_{5/2}r_{+}^{2}\right]
r_{+}^{4}=d_{2}\left[ 12d_{5}\alpha ^{\prime 2}-12\alpha ^{\prime
}r_{+}^{2}+d_{3}r_{+}^{4}\right]
r_{+}^{2d_{2}}+m^{2}[6d_{2}d_{3}d_{4}d_{5}c_{4}c^{4}\left( 2\alpha ^{\prime
}+r_{+}^{2}\right) r_{+}^{2d_{2}}  \notag \\
&&+3d_{2}d_{3}d_{4}c_{3}c^{3}r_{+}^{2d_{1/2}}-d_{2}d_{3}\left( 6\alpha
^{\prime }-r_{+}^{2}\right) _{4}c_{2}c^{2}r_{+}^{2d_{1}}-6d_{2}c_{1}c\alpha
^{\prime }r_{+}^{2d_{1/2}}].
\end{eqnarray}

It is evident that obtaining critical horizon is not possible, analytically.
Therefore, we use numerical method for obtaining critical horizon radius and
corresponding critical temperature and pressure. The results of this
numerical evaluation are presented in tables 1 and 2.

\begin{center}
\begin{tabular}{ccccc}
\hline\hline
$m$ & $r_{c}$ & $T_{c}$ & $P_{c}$ & $\frac{P_{c}r_{c}}{T_{c}}$ \\
\hline\hline
$0.00000$ & $2.13039$ & $0.08558$ & $0.01089$ & $0.27118$ \\ \hline
$0.10000$ & $1.98493$ & $0.10527$ & $0.01415$ & $0.26697$ \\ \hline
$1.00000$ & $0.91380$ & $2.63516$ & $0.77052$ & $0.26719$ \\ \hline
$2.00000$ & $0.72963$ & $10.78040$ & $3.78491$ & $0.25616$ \\ \hline
$3.00000$ & $0.65579$ & $24.51754$ & $9.25479$ & $0.24754$ \\ \hline
\end{tabular}
\\[0pt]
\vspace{0.1cm} Table ($1$): $q=1$, $\alpha ^{\prime }=0.5$, $%
c_{1}=c_{2}=c_{3}=2$ and $d=5$. \vspace{0.5cm}
\end{center}

%%%%%%%%%%%%%%%%%%%%%%%%%%%%%%%%%%%%%%%%%%%%%%%%%%%%%%%%%%%%%%%%%%%%%%%%%%%%%%
%%%%%%%%%%%%%%%%%%%%%%%%%%%%%%%%%%%%%%%%%%%%%%%%%%%%%%%%%%%%%%%%%%%%%%%%%%%%%%

\begin{center}
\begin{tabular}{ccccc}
\hline\hline
$\alpha ^{\prime }$ & $r_{c}$ & $T_{c}$ & $P_{c}$ & $\frac{P_{c}r_{c}}{T_{c}}
$ \\ \hline\hline
$0.00000$ & $0.70743$ & $3.44914$ & $1.82560$ & $0.37444$ \\ \hline
$0.10000$ & $0.85114$ & $1.70944$ & $0.68803$ & $0.34257$ \\ \hline
$0.50000$ & $1.19717$ & $0.67256$ & $0.15085$ & $0.26853$ \\ \hline
$1.00000$ & $1.62237$ & $0.41635$ & $0.05625$ & $0.21922$ \\ \hline
$2.00000$ & $2.67012$ & $0.26191$ & $0.01729$ & $0.17628$ \\ \hline
\end{tabular}
\\[0pt]
\vspace{0.1cm} Table ($2$): $q=1$, $m=0.5$, $c_{1}=c_{2}=c_{3}=2$ and $d=5$.
\vspace{0.5cm}
\end{center}

\begin{figure}[tbp]
$%
\begin{array}{ccc}
\epsfxsize=5cm \epsffile{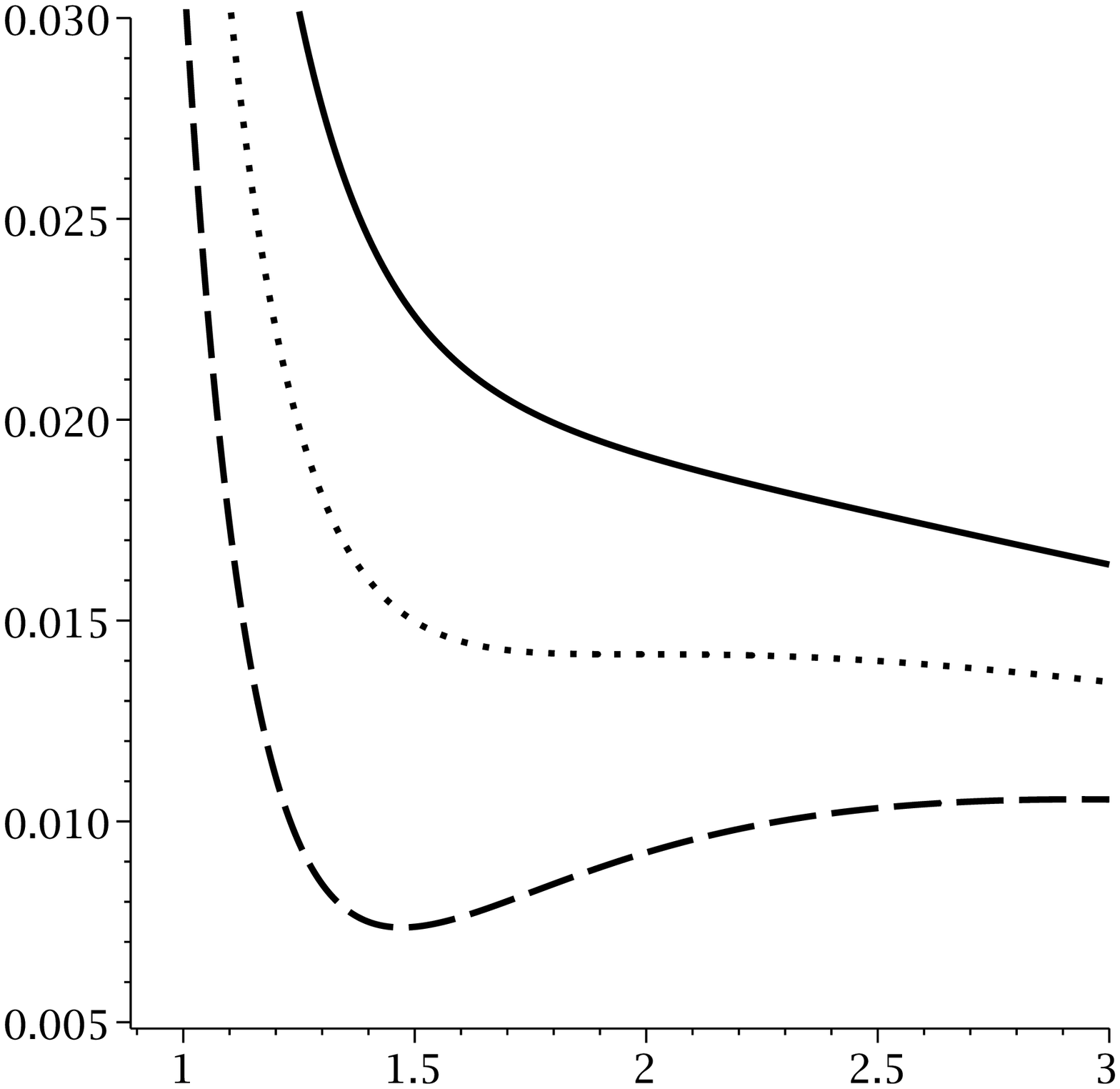} & \epsfxsize=5cm %
\epsffile{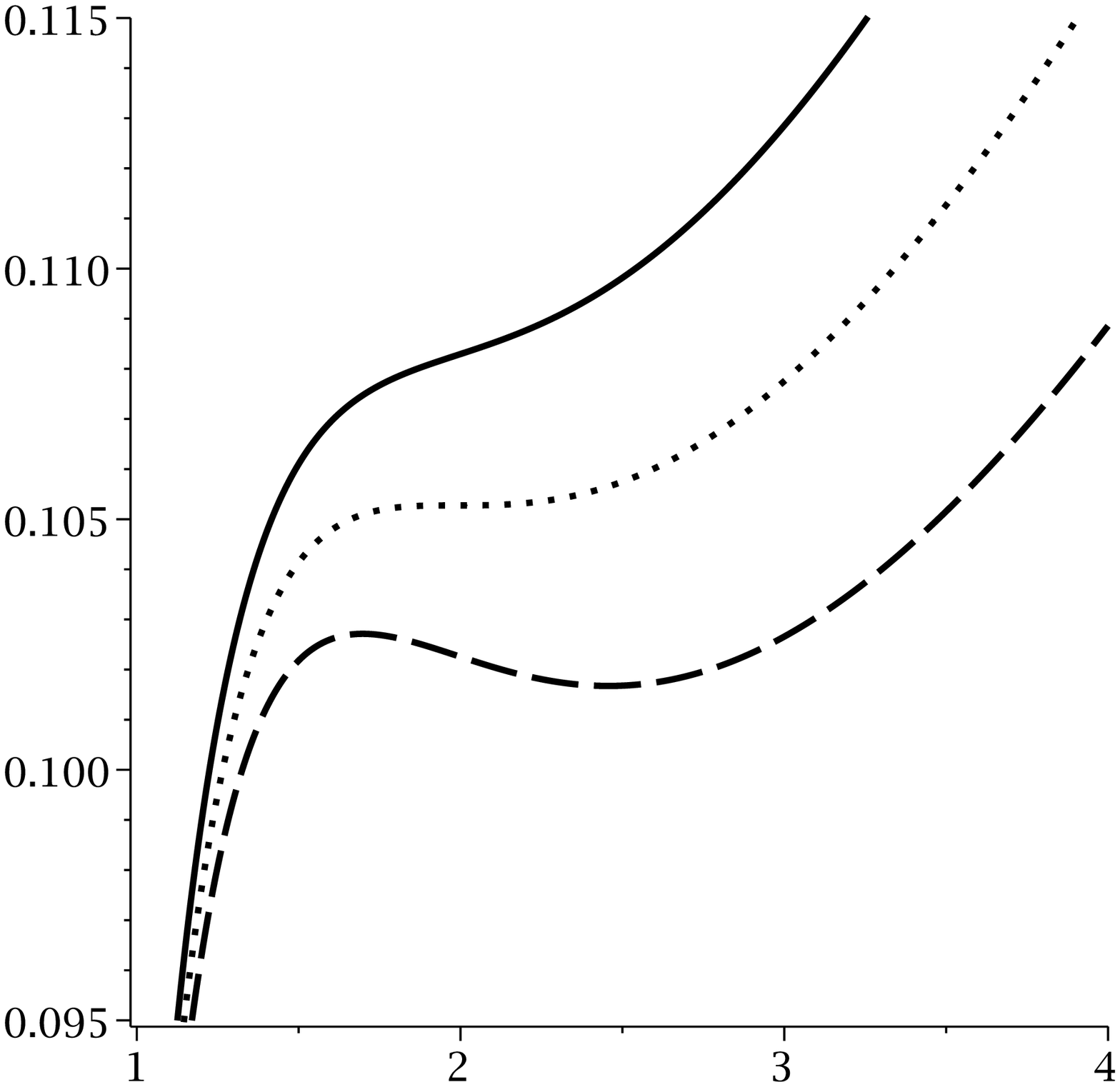} & \epsfxsize=5cm \epsffile{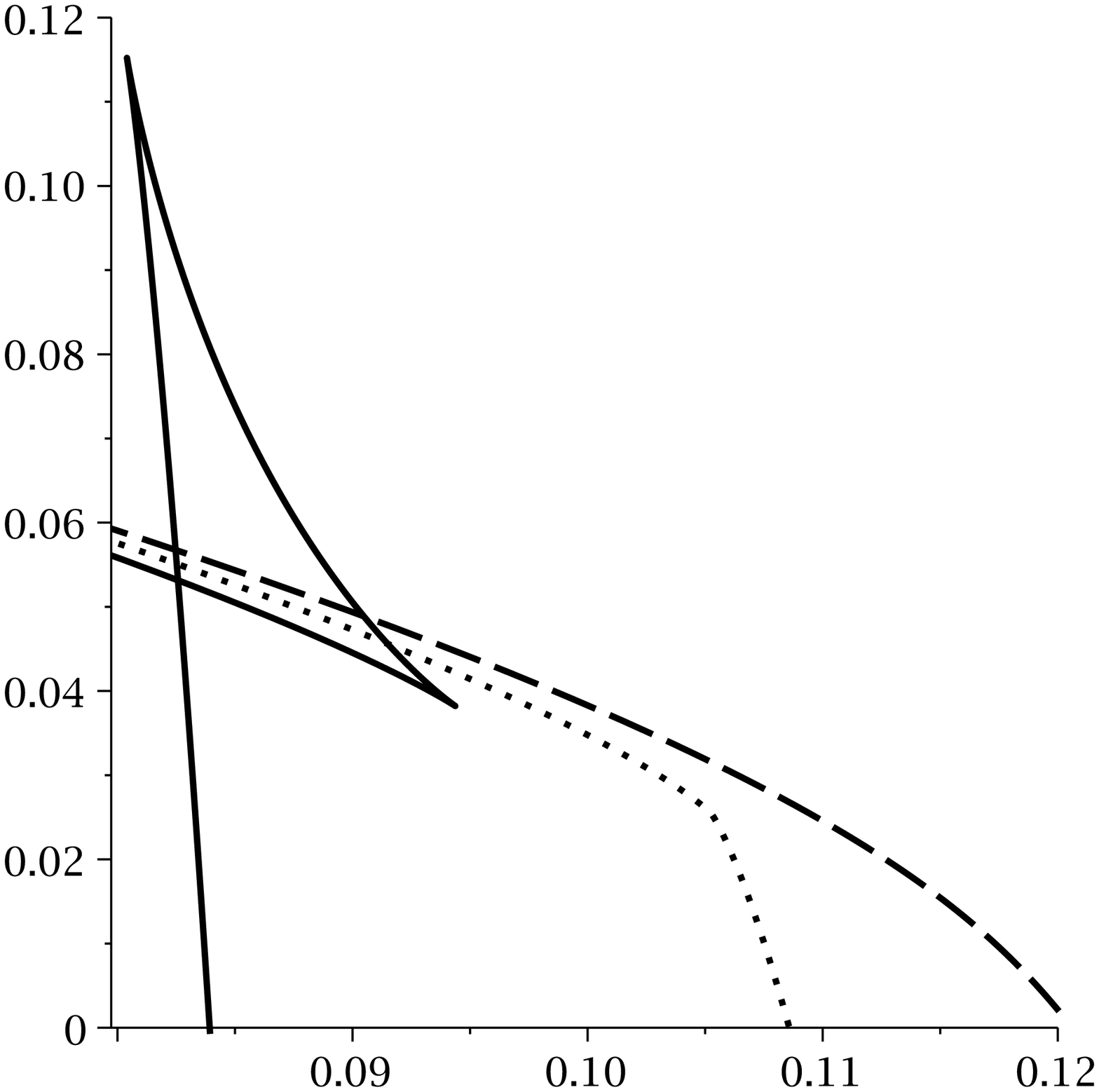}%
\end{array}
$%
\caption{ $P-r_{+}$ (left), $T-r_{+}$ (middle) and $G-T$ (right) diagrams
for $\protect\alpha ^{\prime }=0.5$, $q=1$, $m=0.1$, $c=c_{1}=c_{2}=c_{3}=2$%
, $c_{4}=0$ and $d=5$. \newline
$P-r_{+}$ diagram, from up to bottom $T=1.1T_{c}$, $T=T_{c}$ and $T=0.9T_{c}$%
, respectively. \newline
$T-r_{+}$ diagram, from up to bottom $P=1.1P_{c}$, $P=P_{c}$ and $P=0.9P_{c}$%
, respectively. \newline
$G-T$ diagram for $P=0.5P_{c}$ (continuous line), $P=P_{c}$ (dotted line)
and $P=1.5P_{c}$ (dashed line). }
\label{Fig6}
\end{figure}

\begin{figure}[tbp]
$%
\begin{array}{ccc}
\epsfxsize=5cm \epsffile{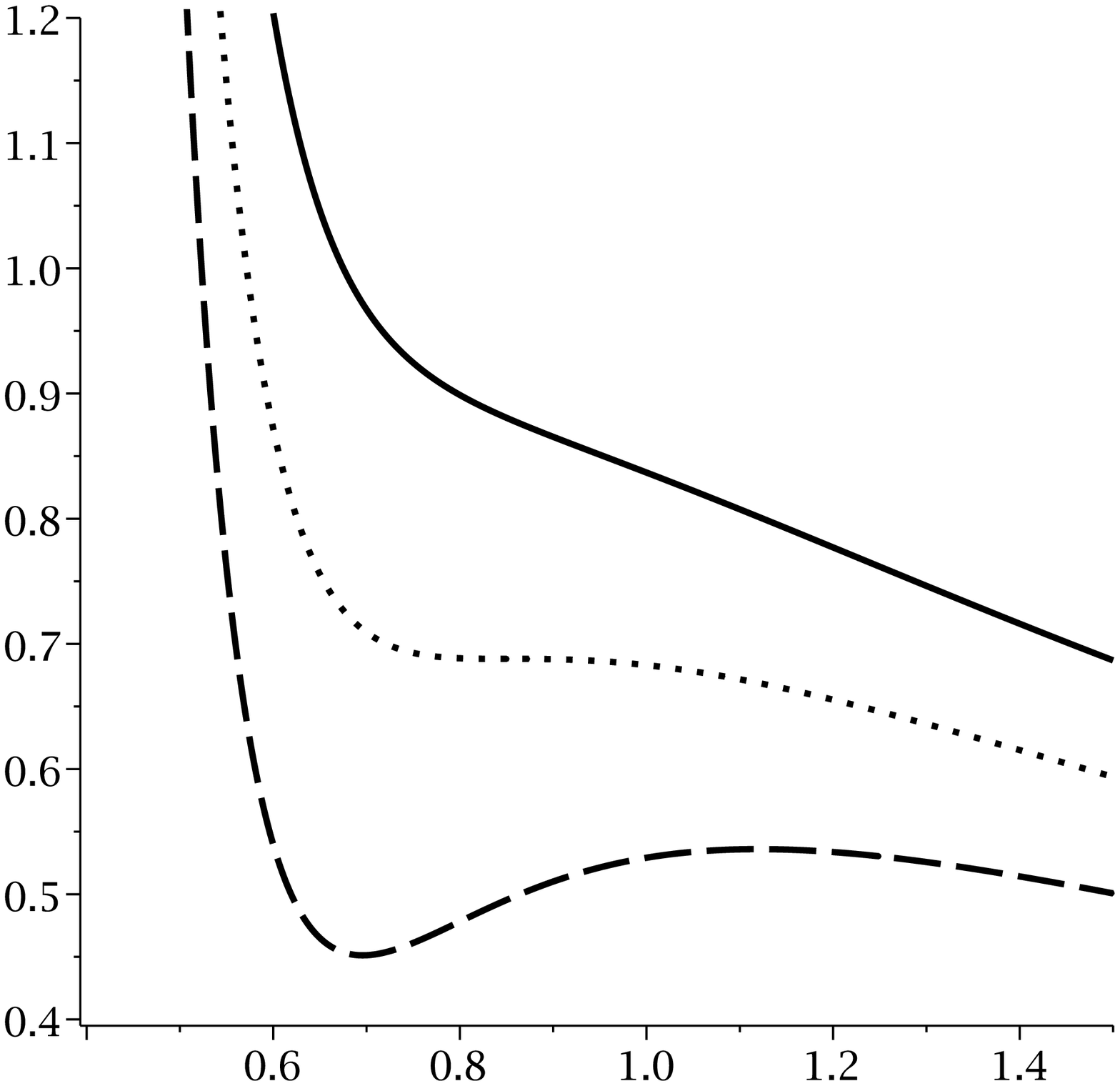} & \epsfxsize=5cm %
\epsffile{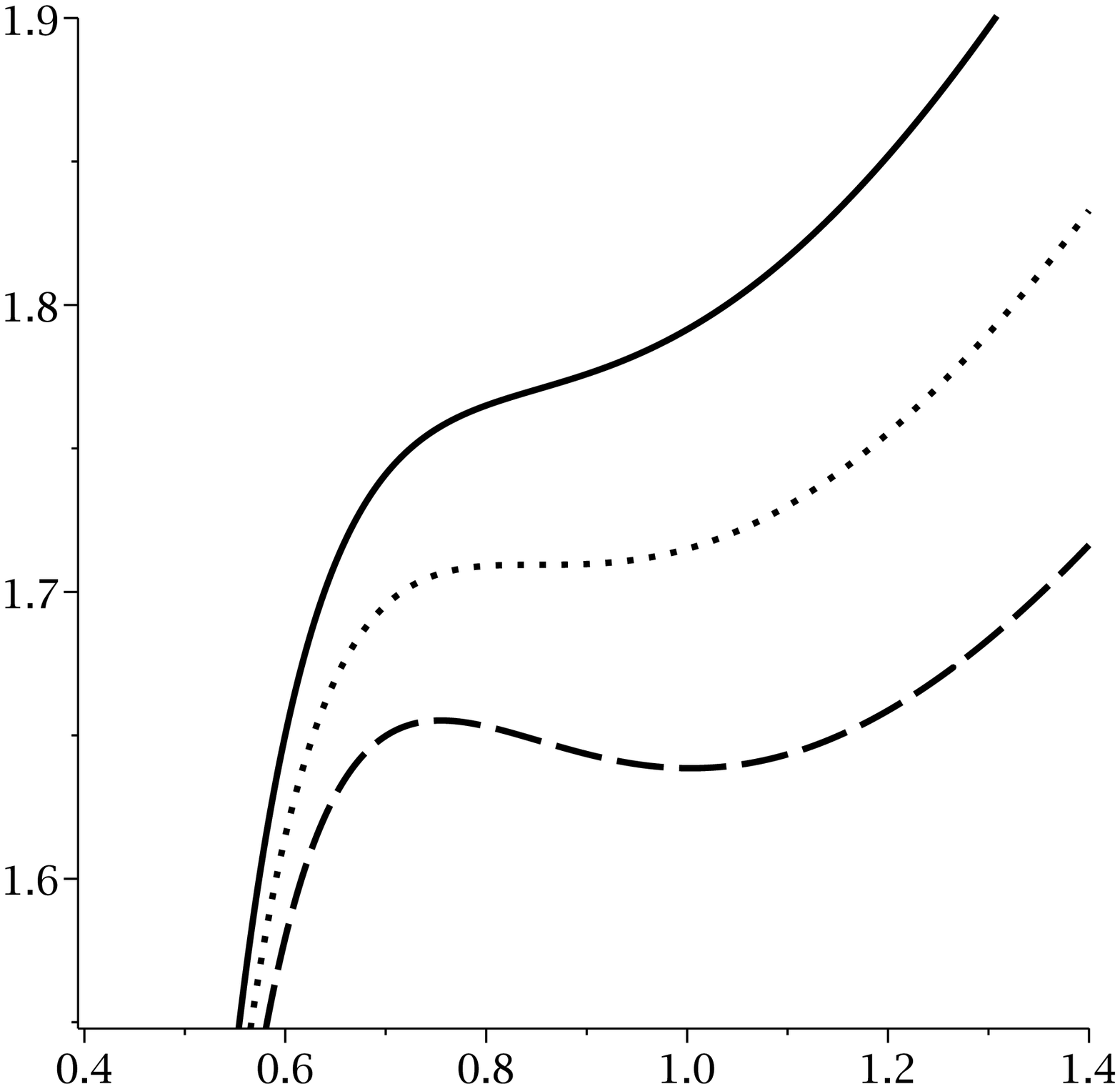} & \epsfxsize=5cm \epsffile{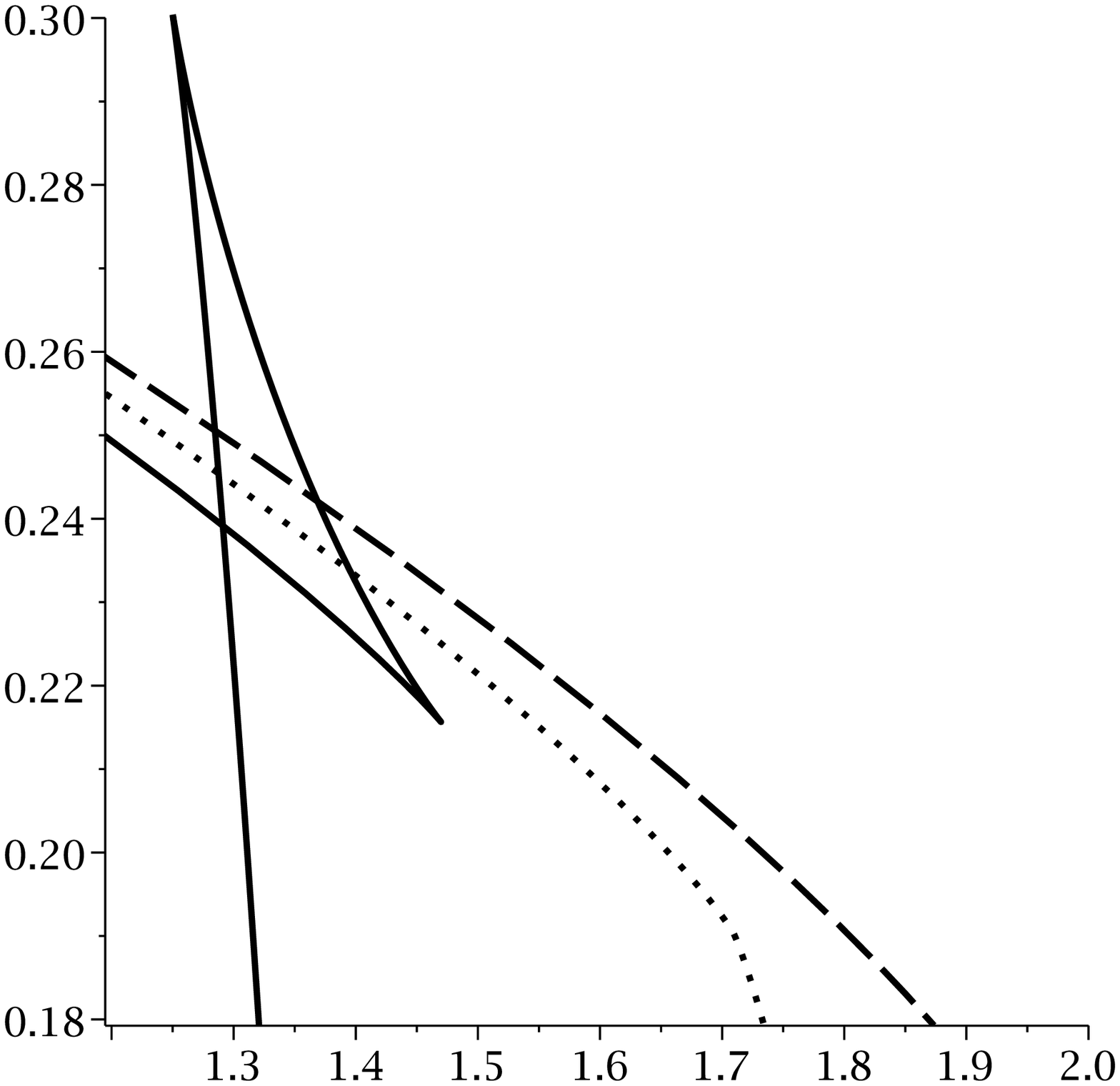}%
\end{array}
$%
\caption{ $P-r_{+}$ (left), $T-r_{+}$ (middle) and $G-T$ (right) diagrams
for $\protect\alpha ^{\prime }=0.1$, $q=1$, $m=0.5$, $c=c_{1}=c_{2}=c_{3}=2$%
, $c_{4}=0$ and $d=5$. \newline
$P-r_{+}$ diagram, from up to bottom $T=1.1T_{c}$, $T=T_{c}$ and $T=0.9T_{c}$%
, respectively. \newline
$T-r_{+}$ diagram, from up to bottom $P=1.1P_{c}$, $P=P_{c}$ and $P=0.9P_{c}$%
, respectively. \newline
$G-T$ diagram for $P=0.5P_{c}$ (continuous line), $P=P_{c}$ (dotted line)
and $P=1.5P_{c}$ (dashed line). }
\label{Fig7}
\end{figure}

\begin{figure}[tbp]
$%
\begin{array}{ccc}
\epsfxsize=5cm \epsffile{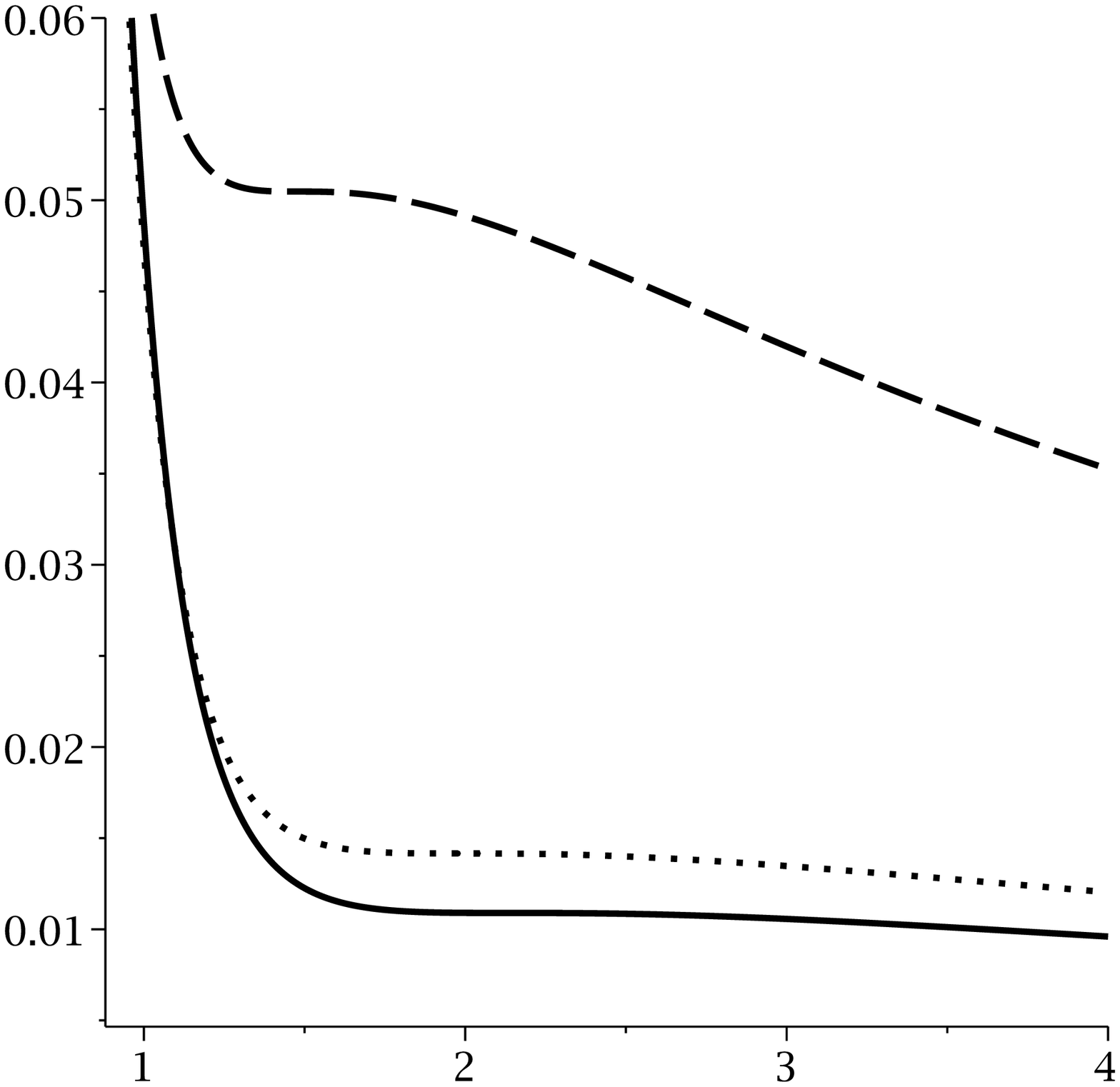} & \epsfxsize=5cm %
\epsffile{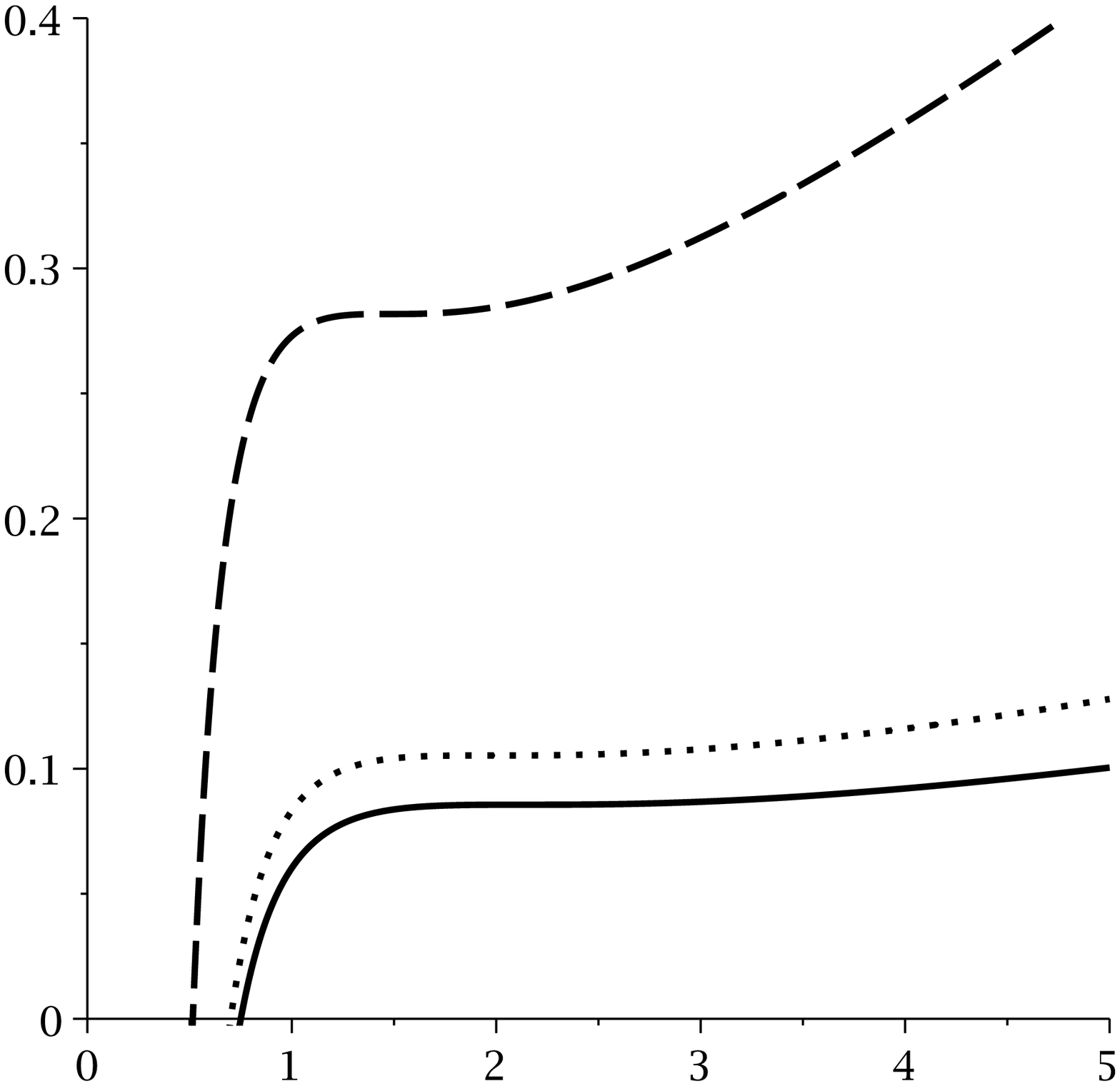} & \epsfxsize=5cm \epsffile{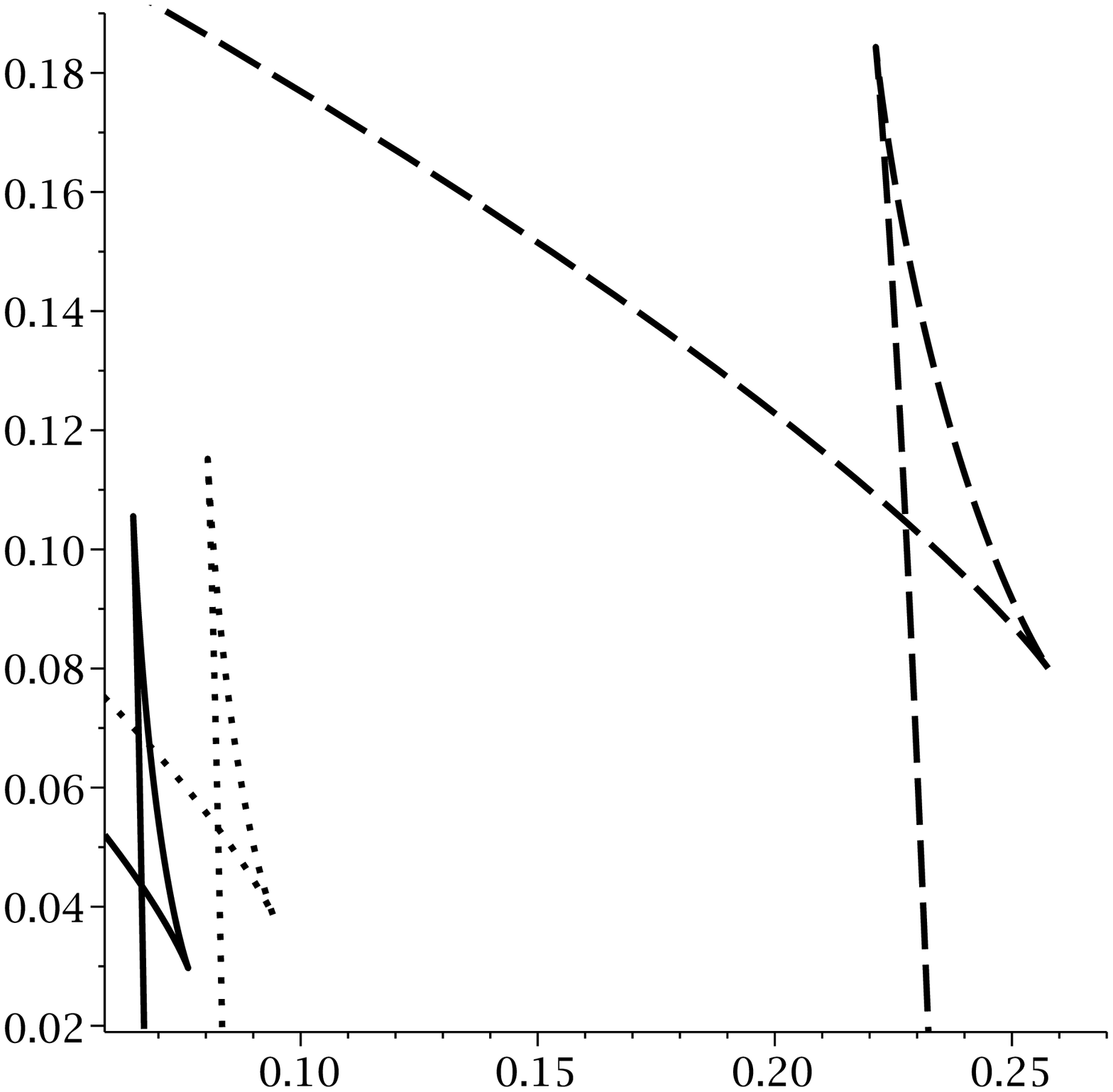}%
\end{array}
$%
\caption{ $P-r_{+}$ (left), $T-r_{+}$ (middle) and $G-T$ (right) diagrams
for $\protect\alpha ^{\prime }=0.5$, $q=1$, $c=c_{1}=c_{2}=c_{3}=2$, $%
c_{4}=0 $, $d=5$, $m=0$ (continuous line), $m=0.1$ (dotted line) and $m=0.3$
(dashed line). \newline
$P-r_{+}$ diagram for $T=T_{c}$, $T-r_{+}$ diagram for $P=P_{c}$ and $G-T$
diagram for $P=0.5P_{c}$. }
\label{Fig8}
\end{figure}

\begin{figure}[tbp]
$%
\begin{array}{ccc}
\epsfxsize=5cm \epsffile{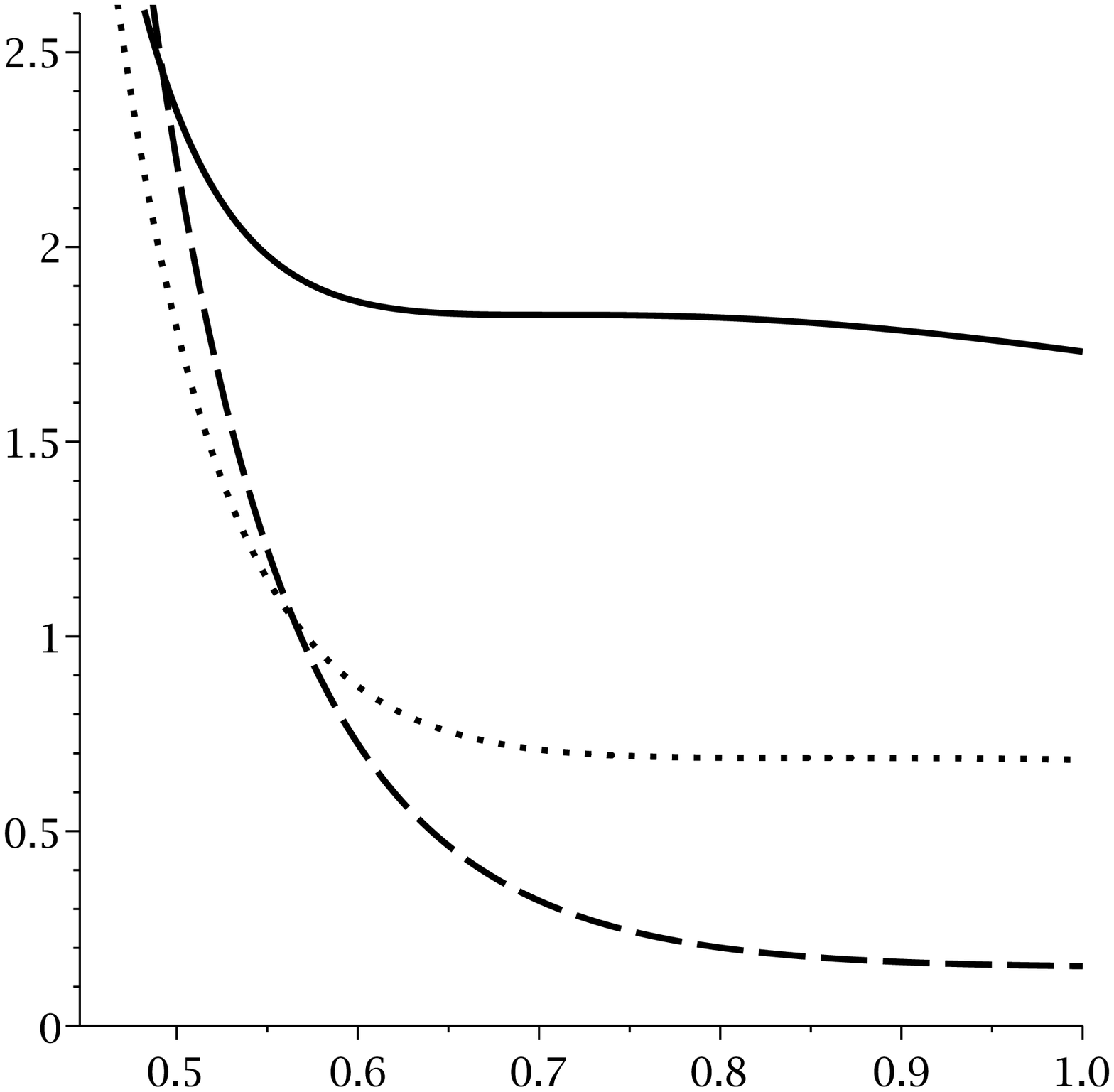} & \epsfxsize=5cm %
\epsffile{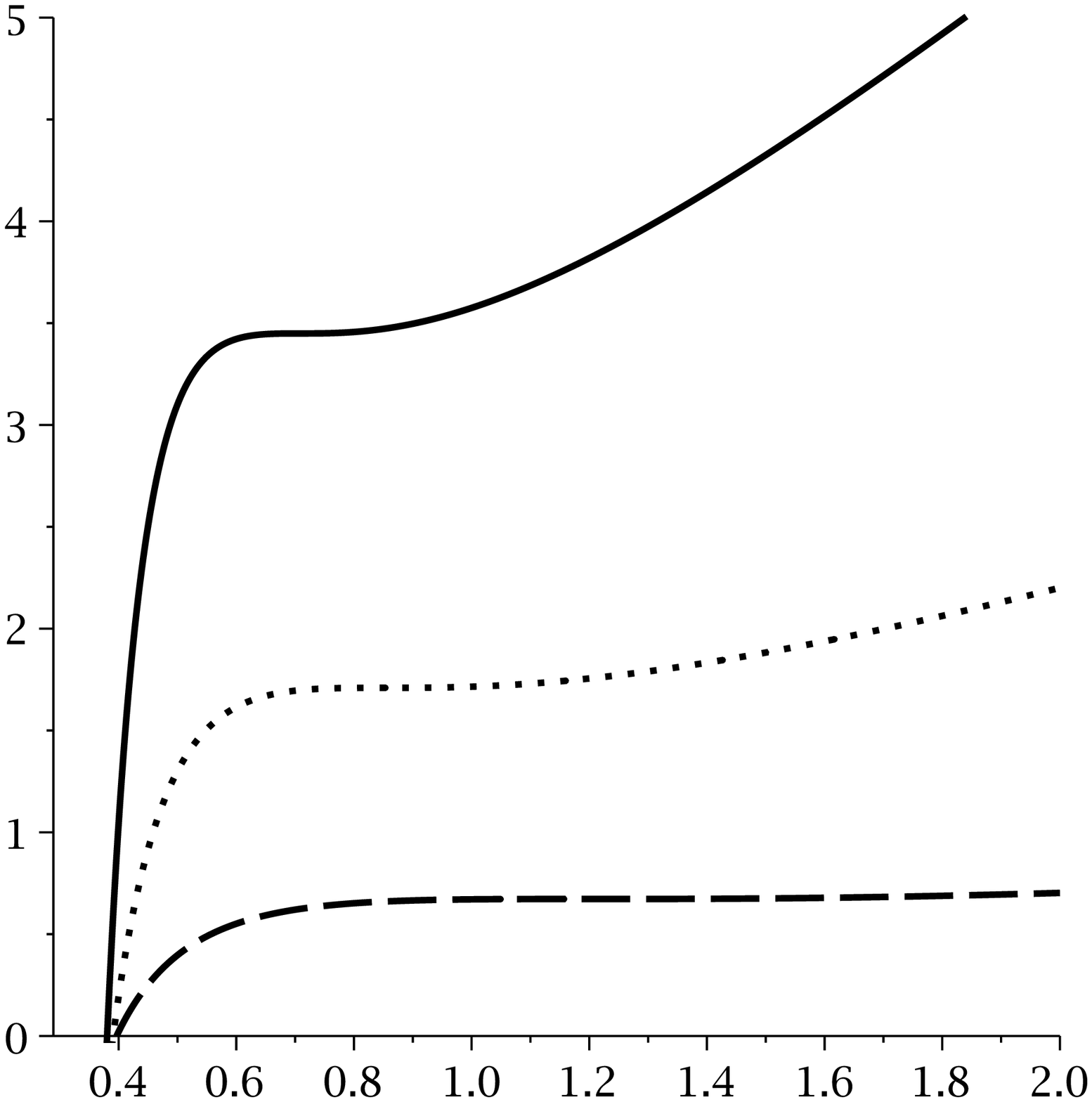} & \epsfxsize=5cm %
\epsffile{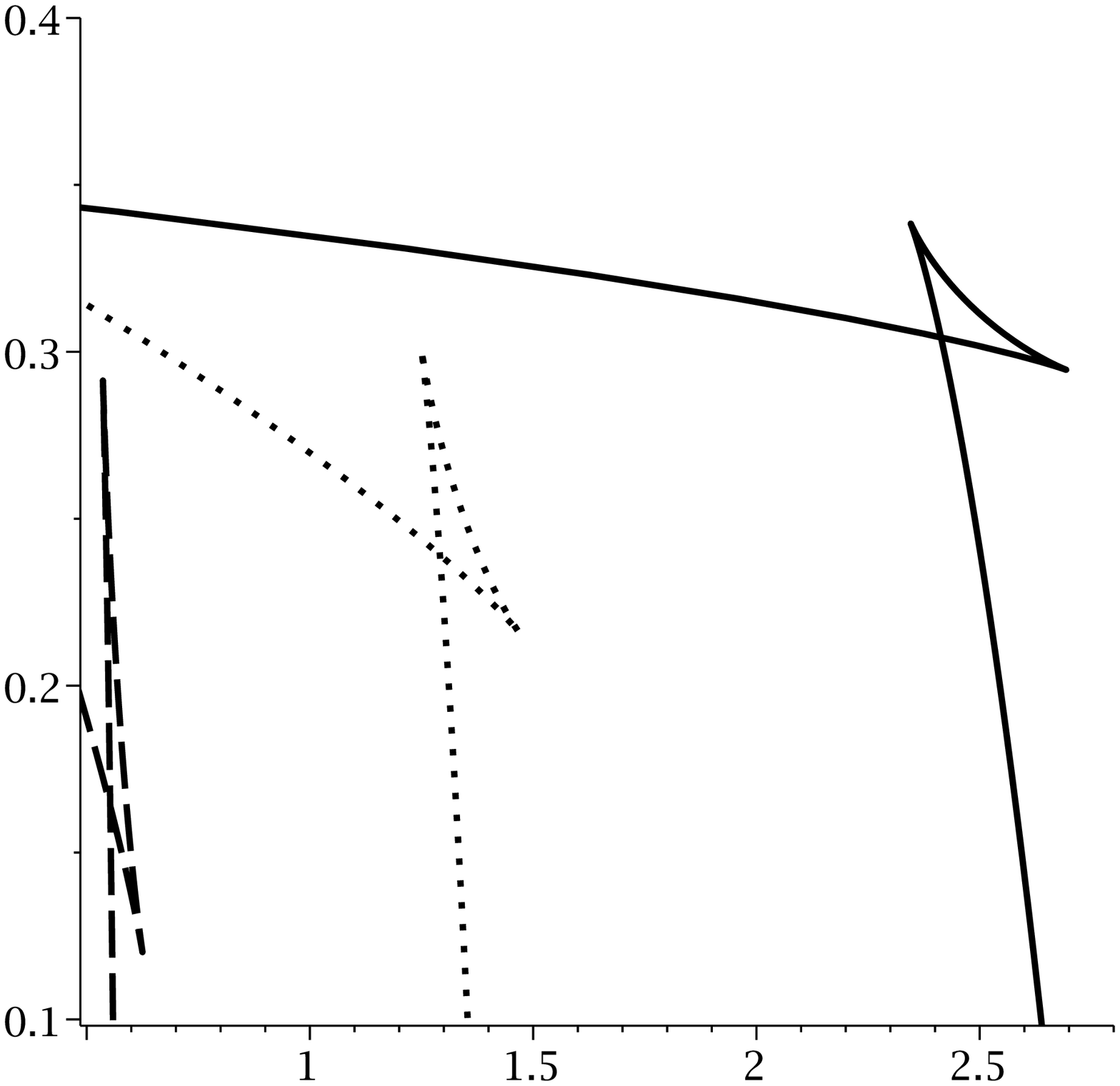}%
\end{array}
$%
\caption{ $P-r_{+}$ (left), $T-r_{+}$ (middle) and $G-T$ (right) diagrams
for $m=0.5$, $q=1$, $c=c_{1}=c_{2}=c_{3}=2$, $c_{4}=0$, $d=5$, $\protect%
\alpha ^{\prime }=0$ (continuous line), $\protect\alpha ^{\prime }=0.1$
(dotted line) and $\protect\alpha ^{\prime }=0.5$ (dashed line). \newline
$P-r_{+}$ diagram for $T=T_{c}$, $T-r_{+}$ diagram for $P=P_{c}$ and $G-T$
diagram for $P=0.5P_{c}$. }
\label{Fig9}
\end{figure}

Considering obtained critical values for horizon radius, temperature and
pressure, one can plot their corresponding phase diagrams ($P-r_{+}$, $%
T-r_{+}$ and $G-T$ diagrams). For the economical reasons, we plot phase
diagrams for two set of critical values (Figs. \ref{Fig6} and \ref{Fig7}).
In order to elaborate the effect of variation of GB and massive parameters
on critical behavior of the system, we plot two sets of comparing diagrams
(Figs. \ref{Fig8} and \ref{Fig9}).

%%%%%%%%%%%%%%%%%%%%%%%%%%%%%%%%%%%%%%%%%%%%%%%%%%%%%%%%%%%%%%%
%%%%%%%%%%%%%%%%%%%%%%%%%%%%%%%%%%%%%%%%%%%%%%%%%%%%%%%%%%%%%%%
%%%%%%%%%%%%%%%%%%%%%%%%%%%%%%%%%%%%%%%%%%%%%%%%%%%%%%%%%%%%%%%
%%%%%%%%%%%%%%%%%%%%%%%%%%%%%%%%%%%%%%%%%%%%%%%%%%%%%%%%%%%%%%%
%%%%%%%%%%%%%%%%%%%%%%%%%%%%%%%%%%%%%%%%%%%%%%%%%%%%%%%%%%%%%%%
%%%%%%%%%%%%%%%%%%%%%%%%%%%%%%%%%%%%%%%%%%%%%%%%%%%%%%%%%%%%%%%
%%%%%%%%%%%%%%%%%%%%%%%%%%%%%%%%%%%%%%%%%%%%%%%%%%%%%%%%%%%%%%%
%%%%%%%%%%%%%%%%%%%%%%%%%%%%%%%%%%%%%%%%%%%%%%%%%%%%%%%%%%%%%%%

The appearance of characteristic swallow tail in $G-T$ diagrams shows that
obtained values are critical ones in which phase transition takes place
(Figs. \ref{Fig6} - \ref{Fig9} right panels). Also, in $P-r_{+}$ diagrams
the existence of region of phase transition and critical behavior of the
system for different factors of the critical temperature are representing
phase transition taking place (Figs. \ref{Fig6} - \ref{Fig9} left panels).
As for $T-r_{+}$, the formation of subcritical isobars which divide the
phase states to three different cases (before, in and after phase
transition), indicates that calculated critical pressure is indeed the one
that system goes under phase transition in (Figs. \ref{Fig6} - \ref{Fig9}
middle panels).

It is evident that, critical horizon radius and universal ratio of $\frac{%
P_{c}r_{c}}{T_{c}}$ are decreasing functions of massive parameter whereas
critical temperature and pressure are increasing functions of it (see table $%
1$ for more details). Interestingly, critical temperature and pressure are
highly sensitive to variation of massive parameter. Increasing massive
parameter for even small value leads to a significant change in the critical
temperature. This shows that for black holes to have phase transition, it
must be heated more significantly. In other words, for acquiring stable
state, significant amount of energy is needed in order to have phase
transition. It is worthwhile to mention that critical horizon also decreases
highly but not with similar rate as temperature changes.

Interestingly, except for universal ratio of $\frac{P_{c}r_{c}}{T_{c}}$, the
effects of variation of GB parameter on critical values, are completely
opposite of massive parameter. In other words, critical temperature and
pressure are decreasing functions of $\alpha $ whereas critical horizon is
an increasing function of it (see table $2$ for more details). It is
worthwhile to mention that the effect of GB parameter is not as significant
as the effect of massive parameter. Although both massive and GB gravities
are extensions for Einstein gravity, their contributions are opposite. The
value of GB parameter is representing the strength of the gravitational
field. Taking this fact into account, one can conclude that strength of
gravity puts restriction on (weakens) the contribution of the massive
gravity.

Finally, it should be pointed out that subcritical isobars are increasing
and decreasing functions of $\alpha $ and $m$, respectively. Subcritical
isobars are representing the region in which phase transition takes place
(Figs. \ref{Fig8} and \ref{Fig9} middle panels). Therefore, their
increments/reductions will decrease/increase single state region of one
phase which in our case it is whether small or large black holes.
Interestingly, the gap between energy of two different phases is an
increasing function of both GB and massive parameters. The only differences
is that Gibbs energy level of two phases are increasing functions of massive
parameter (Fig. \ref{Fig8} right panel) and decreasing functions of $\alpha $
(Fig. \ref{Fig9} right panel).

\section{Geometrical thermodynamics for heat capacity and $P-V$ criticality
\label{GTD}}

\begin{figure}[tbp]
$%
\begin{array}{cc}
\epsfxsize=6cm \epsffile{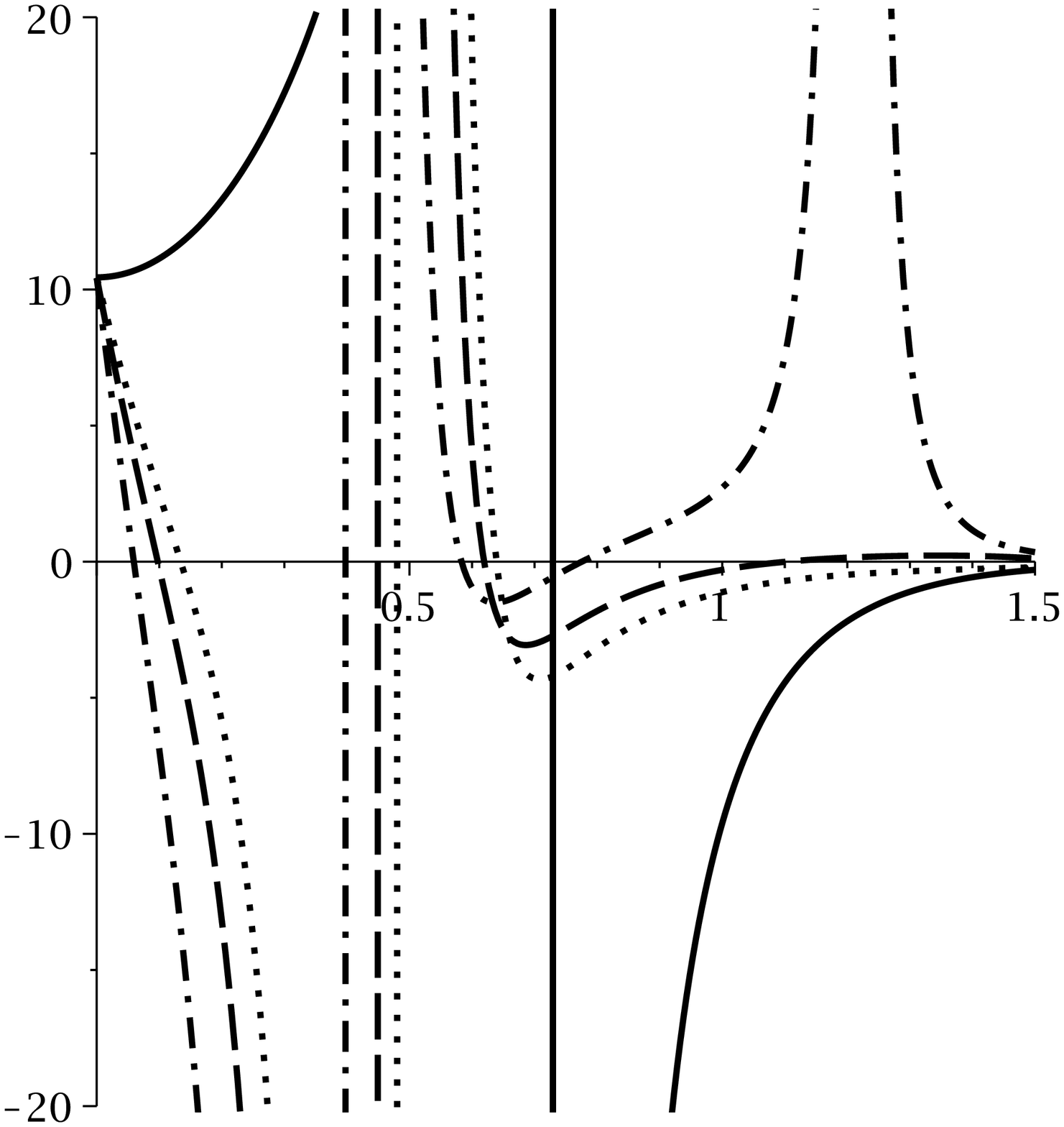} & \epsfxsize=6cm %
\epsffile{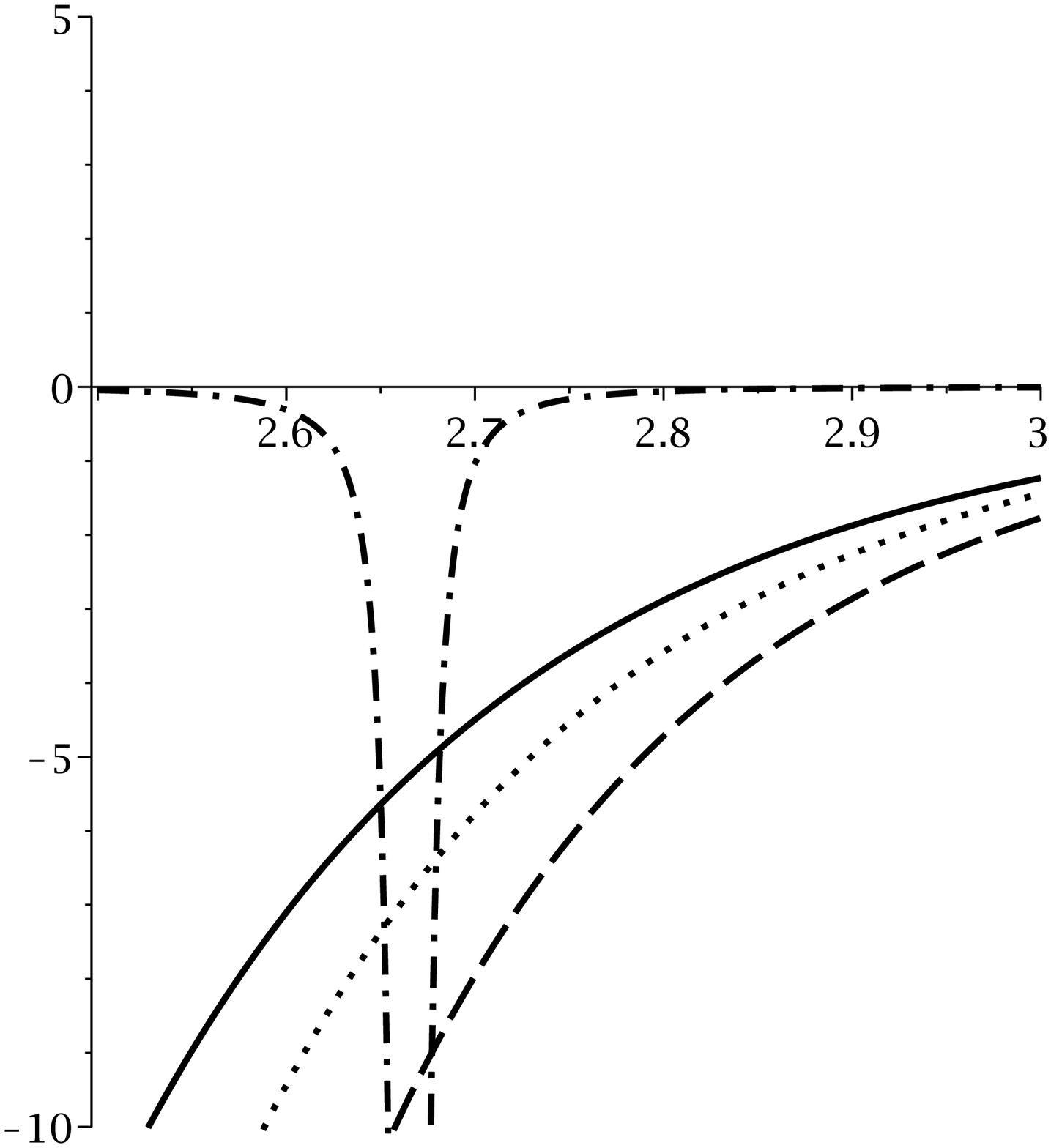}%
\end{array}
$%
\caption{$\mathcal{R}$ versus $r_{+}$ for $q=1$, $\Lambda=-1$ $%
c=c_{1}=c_{2}=c_{3}=2$, $c_{4}=0$, $\protect\alpha=0.5$, $d=5$ and $k=1$; $%
m=0$ (continues line), $m=0.35$ (dotted line), $m=0.40$ (dashed line) and $%
m=0.50$ (dashed-dotted line). \emph{"different scales"}}
\label{Fig10}
\end{figure}

\begin{figure}[tbp]
$%
\begin{array}{cc}
\epsfxsize=6cm \epsffile{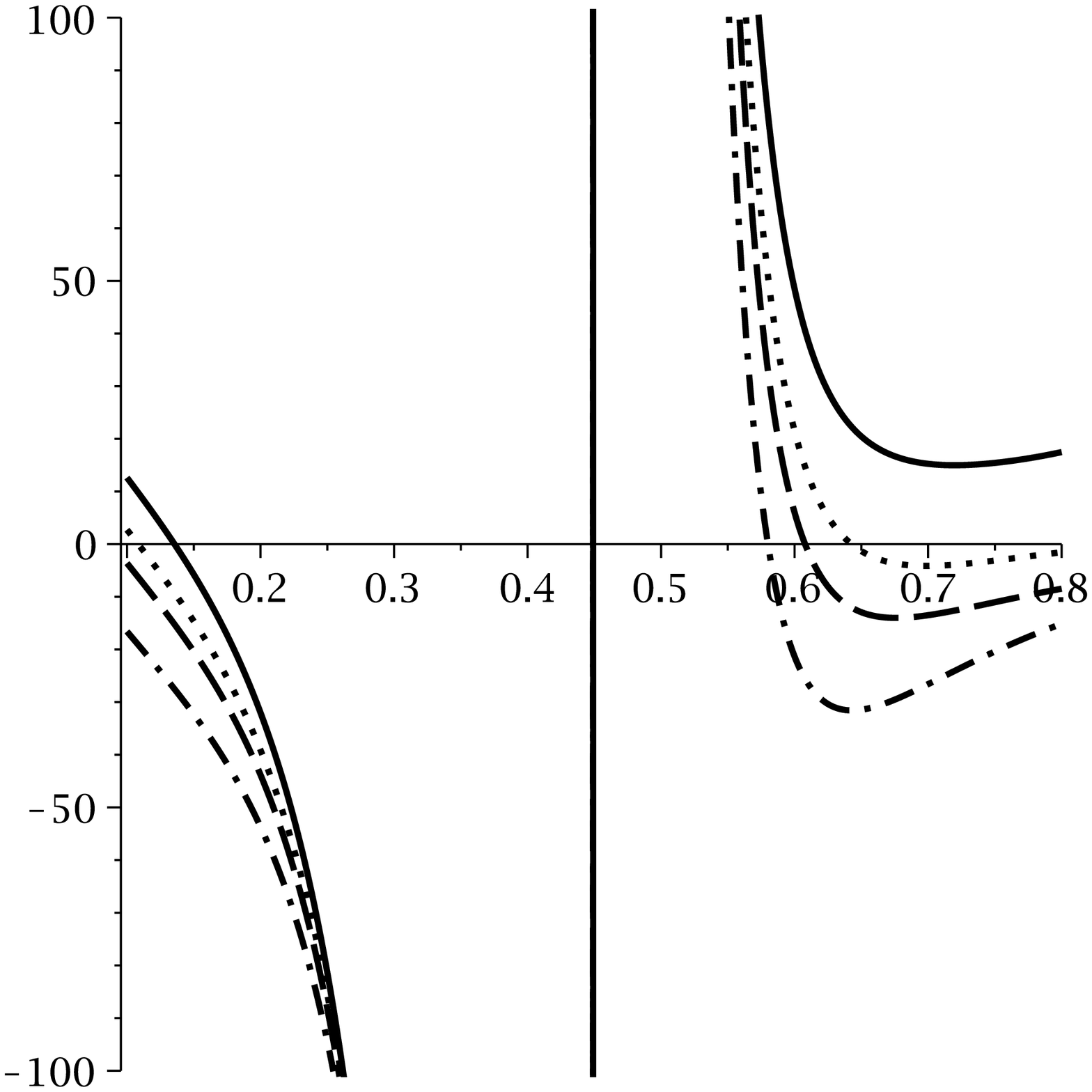} & \epsfxsize=6cm %
\epsffile{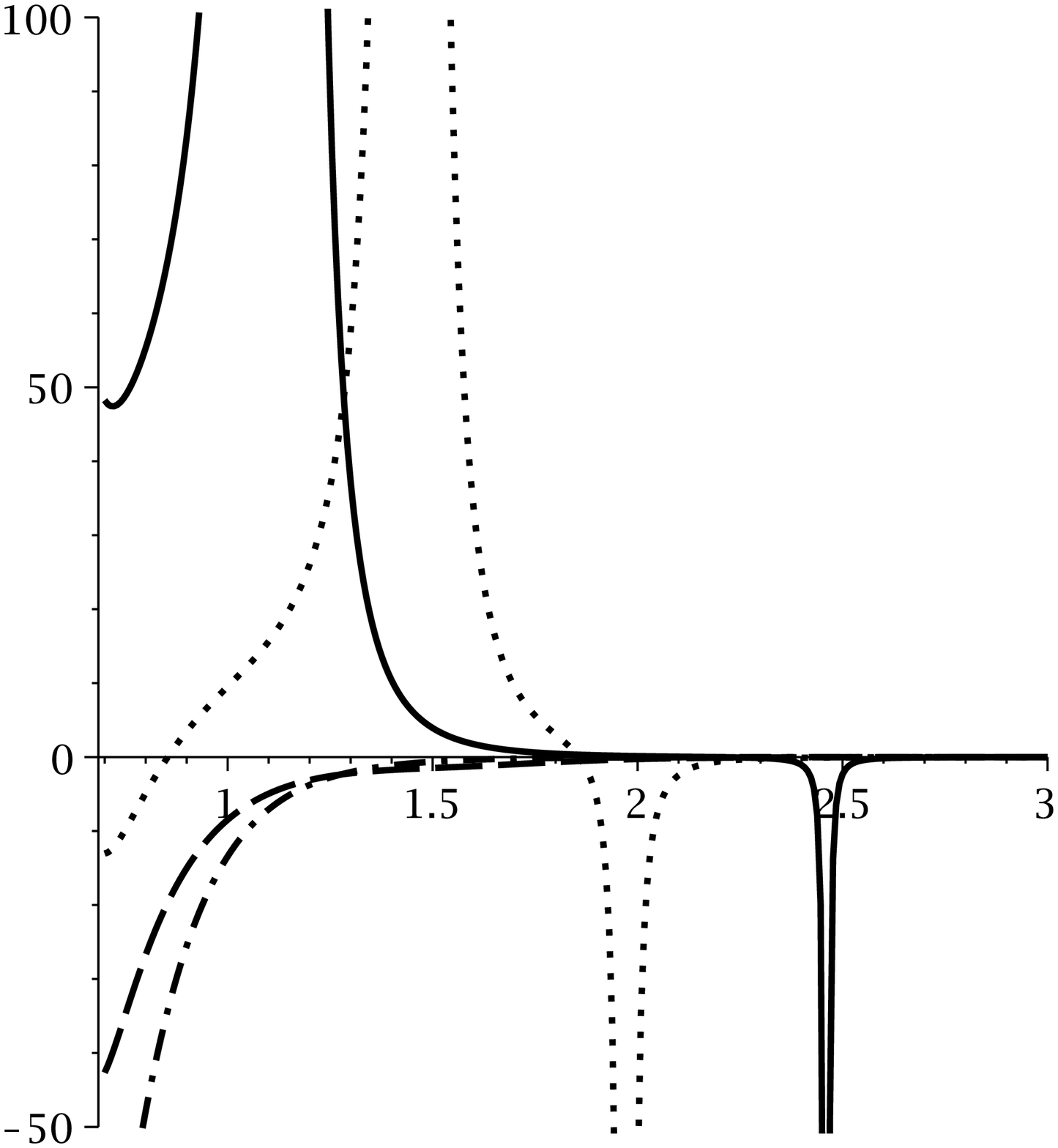}%
\end{array}
$%
\caption{$\mathcal{R}$ versus $r_{+}$ for $q=1$, $\Lambda=-1$, $%
c=c_{1}=c_{2}=c_{3}=2$, $c_{4}=0$, $m=0.40$, $d=5$ and $k=1$; $\protect\alpha%
=0.35$ (continues line), $\protect\alpha=0.45$ (dotted line), $\protect\alpha%
=0.55$ (dashed line) and $\protect\alpha=1$ (dashed-dotted line). \emph{%
"different scales"}}
\label{Fig11}
\end{figure}

In this section we will study thermodynamical behavior of these black holes
through geometrical thermodynamics (GTs). As it was pointed out earlier,
there are several approaches for constructing thermodynamical spacetime. In
this paper, we follow HPEM method for studying phase transition points of
the heat capacity and later another metric for phase transition points of
the extended phase space. The HPEM metric has the following form \cite%
{HendiPEM,HPEM}
\begin{equation}
ds_{New}^{2}=\frac{SM_{S}}{\left( \Pi _{i=2}^{n}\frac{\partial ^{2}M}{%
\partial \chi _{i}^{2}}\right) ^{3}}\left(
-M_{SS}dS^{2}+\sum_{i=2}^{n}\left( \frac{\partial ^{2}M}{\partial \chi
_{i}^{2}}\right) d\chi _{i}^{2}\right) ,  \label{HPEM}
\end{equation}%
where $M_{S}=\partial M/\partial S$, $M_{SS}=\partial ^{2}M/\partial S^{2}$
and $\chi _{i}$ ($\chi _{i}\neq S$) are extensive parameters. For economical
reasons, we only plot corresponding figures for variation of $\alpha$ and $m$
for studying the heat capacity (Figs. \ref{Fig1} and \ref{Fig2}). By
employing Eqs. (\ref{TotalQ}), (\ref{TotalS}) and (\ref{MM}) and HPEM metric
(Eq. (\ref{HPEM})), one can plot following diagrams (Figs. \ref{Fig10} and %
\ref{Fig11}).

%%%%%%%%%%%%%%%%%%%%%%%%%%%%%%%%%%%%%%%%%%%%%%%%%%%%%%%%%%%%%%%

%%%%%%%%%%%%%%%%%%%%%%%%%%%%%%%%%%%%%%%%%%%%%%%%%%%%%%%%%%%%%%%
%%%%%%%%%%%%%%%%%%%%%%%%%%%%%%%%%%%%%%%%%%%%%%%%%%%%%%%%%%%%%%%

%%%%%%%%%%%%%%%%%%%%%%%%%%%%%%%%%%%%%%%%%%%%%%%%%%%%%%%%%%%%%%%

Studying GTs diagrams shows that Ricci scalar of considered metric diverges
in places of phase transition points. In other words, divergencies of the
Ricci scalar coincide with divergencies and roots of the heat capacity
(compare Figs. \ref{Fig10} and \ref{Fig11} with \ref{Fig1} and \ref{Fig2}).
Taking a closer look at the divergencies of the Ricci scalar, one can see
that the behavior of the Ricci scalar around divergence points can be
categorized into three groups: there is a change of sign around divergence
point in which correspondingly there is a change of sign in heat capacity,
divergency toward $+\infty $ where there is a phase transition from larger
black holes to smaller ones and finally divergency toward $-\infty $ in
which system goes under phase transition of smaller to larger black holes.
This property of the Ricci scalar enables one to recognize the type of phase
transition without studying heat capacity.

In Ref. \cite{PVEinstein} a new thermodynamical metric for studying critical
behavior in case of $P-V$ criticality was introduced. It was also shown that
GTs, $P-V$ criticality and heat capacity will lead to consistent results. In
other words, number and places of phase transitions in these three pictures
are uniform and they yield same results. There are two forms of GTs metric
for studying critical behavior in extended phase space that are given as
\cite{PVEinstein}
\begin{equation}
ds^{2}=\left\{
\begin{array}{cc}
S\frac{M_{S}}{M_{QQ}^{3}}\left( -M_{SS}dS^{2}+M_{QQ}dQ^{2}+dP^{2}\right) &
Case\text{ }I \\
S\frac{M_{S}}{M_{QQ}^{3}}\left( -M_{SS}dS^{2}+M_{QQ}dQ^{2}+M_{P}dP^{2}\right)
& Case\text{ }II%
\end{array}%
\right. .  \label{GTDPV}
\end{equation}

\begin{figure}[tbp]
$%
\begin{array}{ccc}
\epsfxsize=5.5cm \epsffile{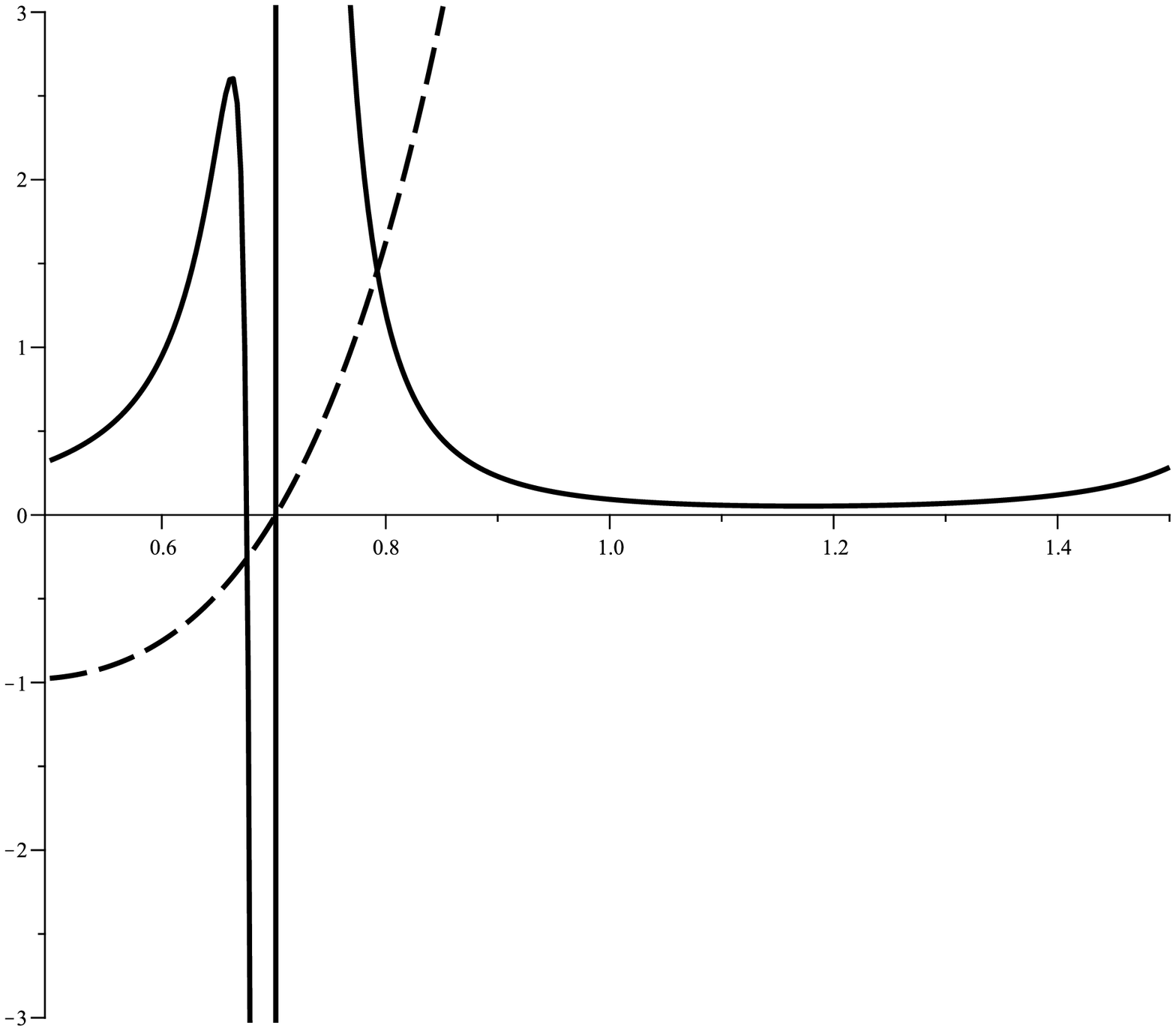} & \epsfxsize=5.5cm %
\epsffile{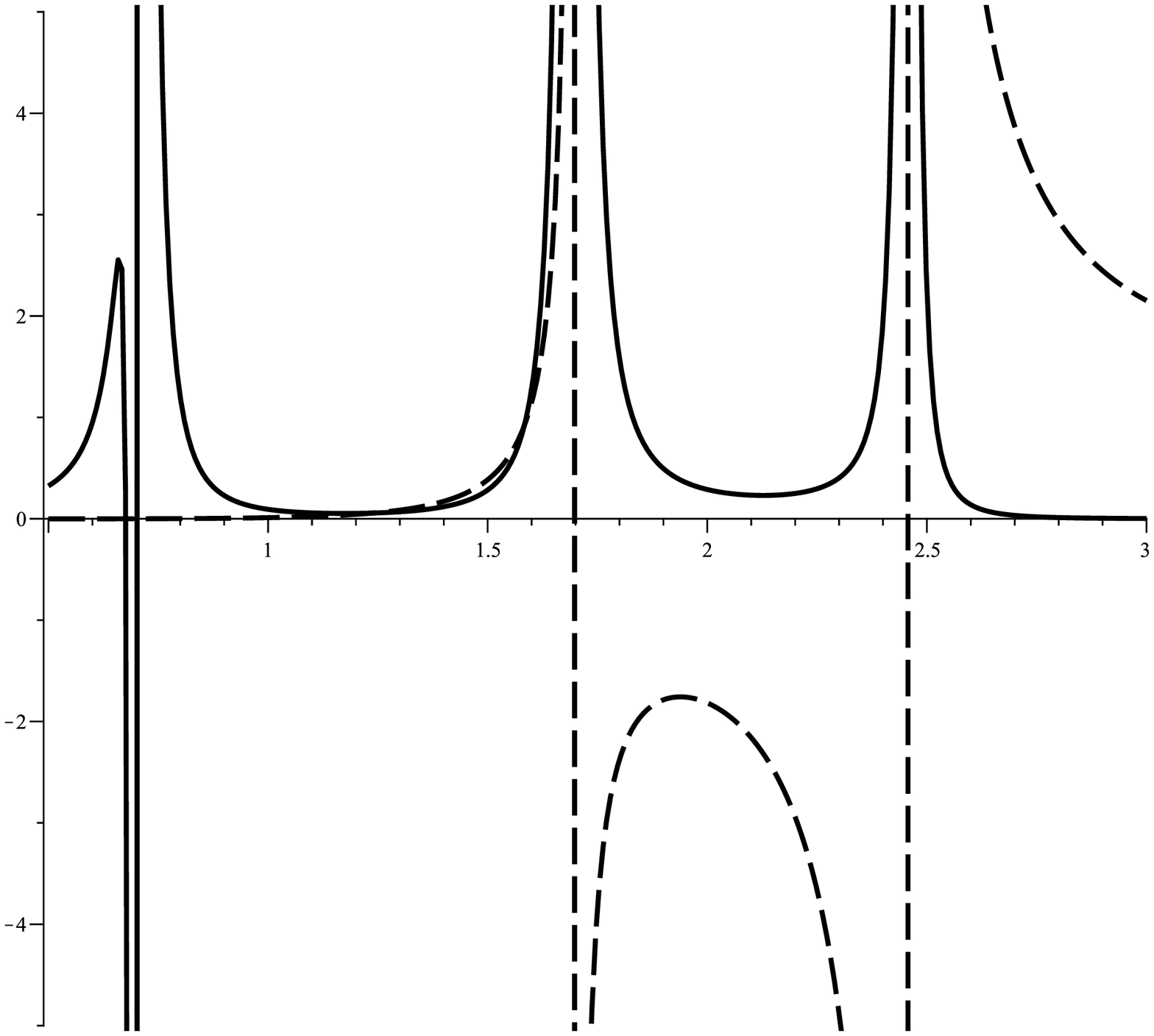} &  \\
\epsfxsize=5.5cm \epsffile{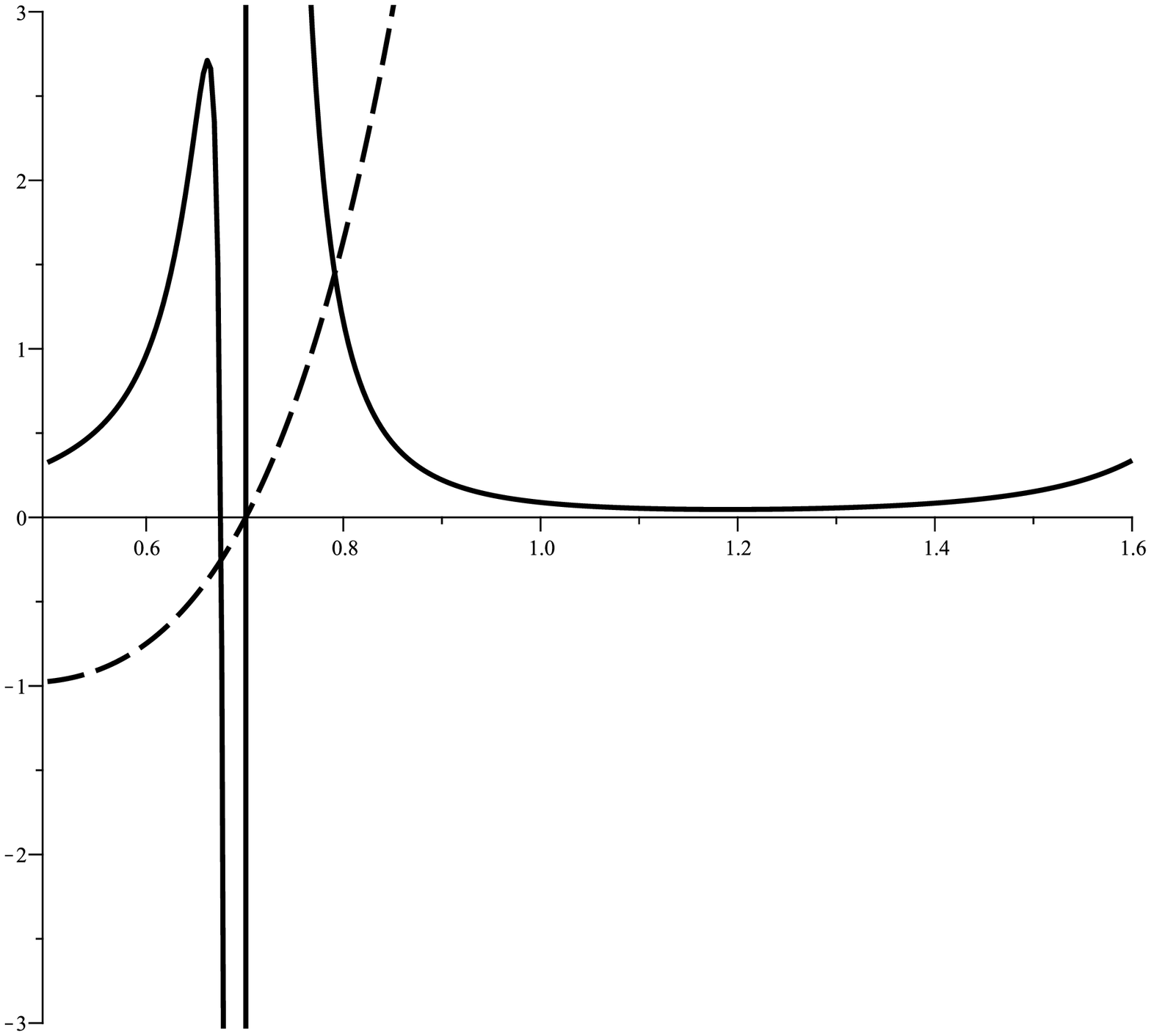} & \epsfxsize=5.5cm %
\epsffile{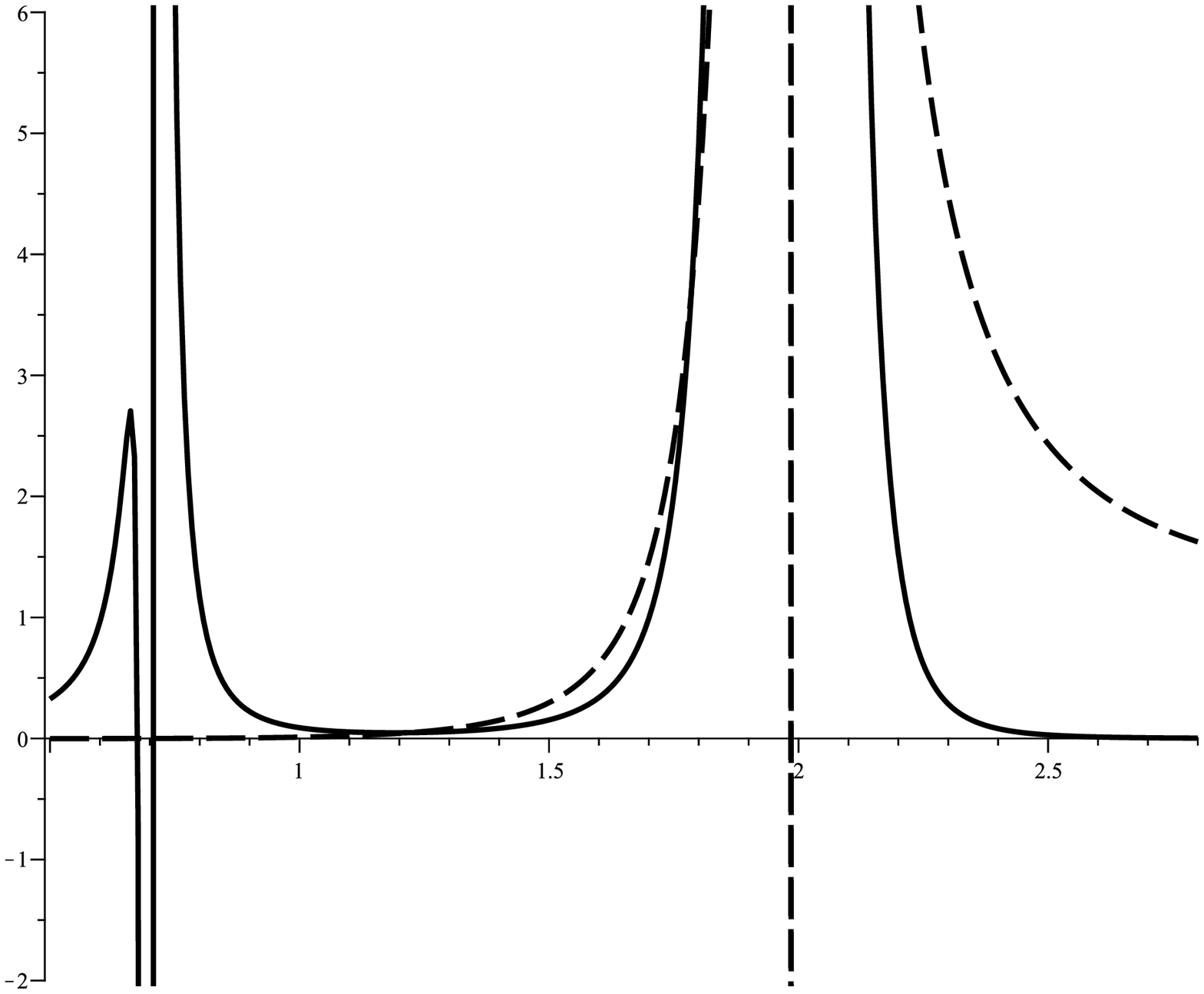} &  \\
\epsfxsize=5.5cm \epsffile{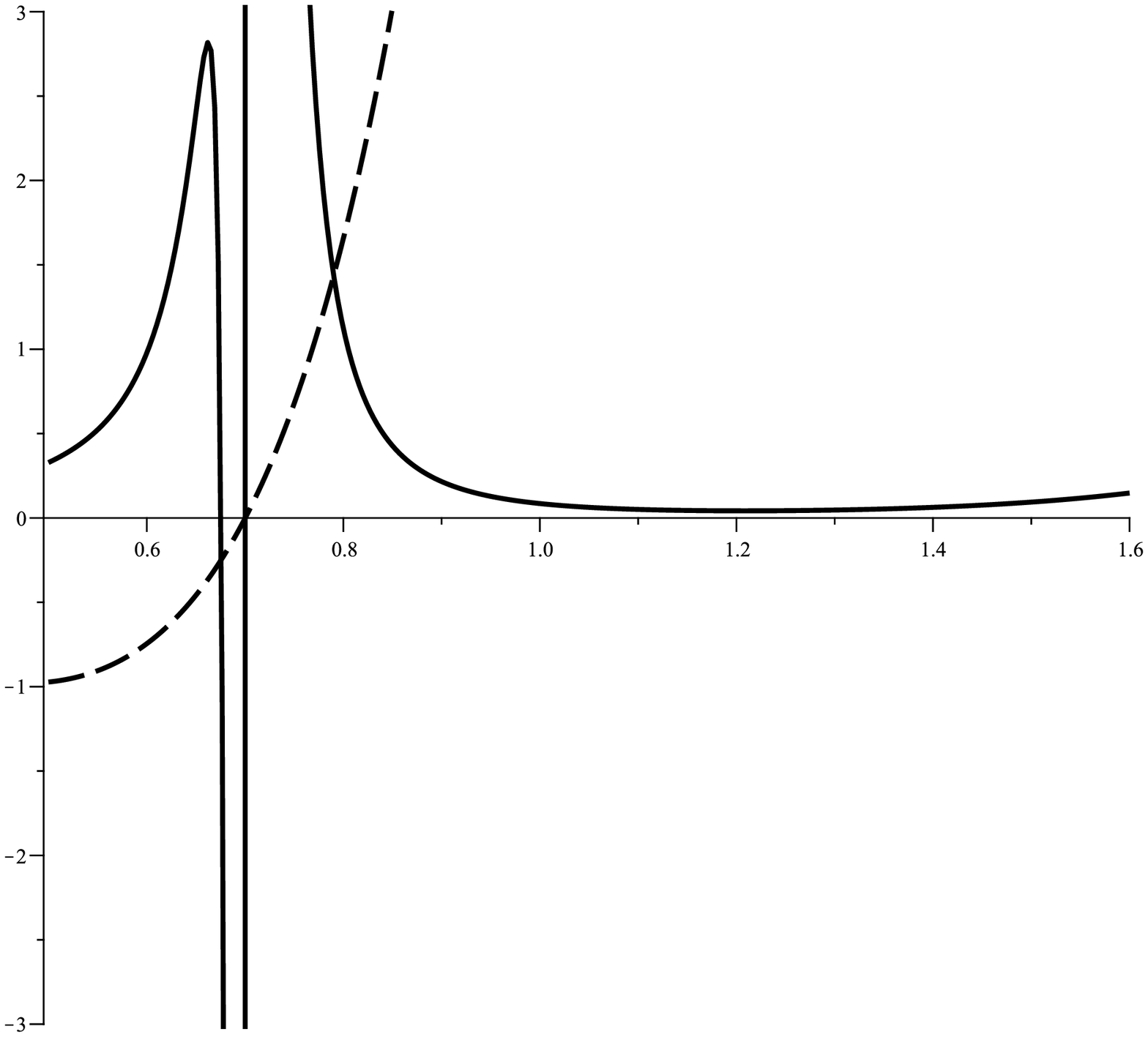} & \epsfxsize=5.5cm %
\epsffile{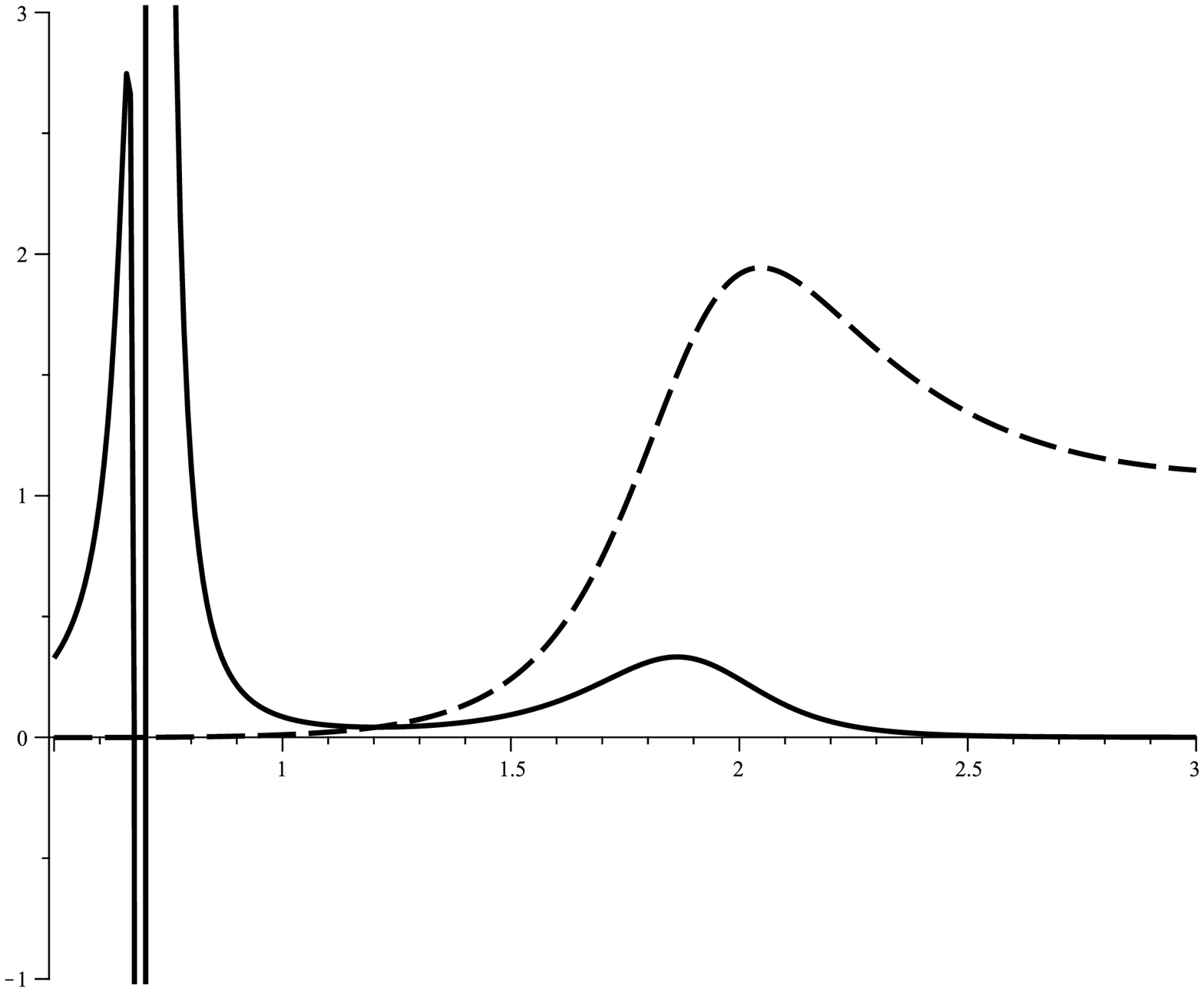} &
\end{array}
$%
\caption{$\mathcal{R}$ (continuous line - Case $I$), $C_{Q}$ (dashed line)
versus $r_{+}$ for $q=1$, $c=c_{1}=c_{2}=c_{3}=2$, $c_{4}=0$, $\protect\alpha%
=0.5$, $d=5$, $k=1$ and $m=0.1$. \newline
up: $P=0.9P_{c}$, "for different scales", \newline
middle: $P=P_{c}$ "for different scales", \newline
down: $P=1.1 P_{c}$, "for different scales".}
\label{Fig12}
\end{figure}

In this paper, we only consider $Case$ $I$ metric for studying GTs in
extended phase space. Considering Eqs. (\ref{TotalM}), (\ref{Heat}), (\ref{P}%
) and (\ref{GTDPV}) with obtained critical pressure in table $1$, we plot
following diagram (Fig. \ref{Fig12}).

%%%%%%%%%%%%%%%%%%%%%%%%%%%%%%%%%%%%%%%%%%%%%%%%%%%%%%%%%%%%%%%%%%%%%%%%%%%%%%%%%%%%
%%%%%%%%%%%%%%%%%%%%%%%%%%%%%%%%%%%%%%%%%%%%%%%%%%%%%%%%%%%%%%%%%%%%%%%%%%%%%%%%%%%

In plotted graphs, a divergency is observed for Ricci scalar which is due to
existence of the root for heat capacity (Fig. \ref{Fig12} left and
down-right panels). For pressure smaller than critical pressure, two
divergencies for heat capacity and Ricci scalar observed (Fig. \ref{Fig12}
up right panel). This behavior is similar to that was observed in Fig. \ref%
{Fig1} middle panel for pressures smaller than critical pressure. In other
words, similar to $T-r_{+} $ diagrams, for pressures smaller than critical
pressure system will have more than one phase transition point. Next, for
pressure being critical pressure only one divergence point exists (Fig. \ref%
{Fig12} down middle). The place of divergency is exactly where critical
horizon radius was found. In other words, in case of considering critical
pressure, the place of phase transition in extended phase space and
divergencies of heat capacity and Ricci scalar coincide with each other
(compare Fig. \ref{Fig12} down middle with Fig. \ref{Fig1} middle). Finally,
for pressures larger than critical pressure no divergence point for heat
capacity and Ricci scalar was observed which is consistent with result of $%
T-r_{+}$ diagram in Fig. \ref{Fig1}.

According to the results, the GTs metric proposed for extended phase space
is providing an effective method which its results are consistent with
thermodynamical concepts. Also, this metric is able to point out phase
transition points which one can obtain through studying $T-r_{+}$ and $%
P-r_{+}$ diagrams. More importantly, it is evident that heat capacity (its
divergencies), GTs (its divergencies) and phase diagrams ($P-r_{+}$, $%
T-r_{+} $ and $G-T$) lead to consistent results.

\section{Heat capacity and critical values in the extended phase space \label%
{HC}}

The final section of this paper is devoted to calculation of critical values
in extended phase space by using heat capacity. In last section, it was seen
that critical values in which phase transition takes place in extended phase
space present themselves as divergencies in heat capacity. Motivated by this
result, in Ref. \cite{PVEinstein} a new method for obtaining critical values
was introduced which is based on denominator of the heat capacity.

In this method, one consider denominator of the heat capacity and rewrite it
with relation between thermodynamical pressure and cosmological constant.
Next, one solve the denominator of the heat capacity with respect to
pressure. Obtained relation for pressure is different from the one that is
obtained by using temperature. If obtained relation for pressure has
maximum(s), then in the place of that maximum (horizon radius and pressure),
phase transition takes place. In other words, instead of mentioned method
for obtaining critical values in section $V$, one can easily study the
maximums of obtained relation for pressure through denominator of the heat
capacity. The advantages of this method are pointed out in Ref. \cite%
{PVEinstein}.

In order to obtain mentioned relation for pressure, one should use
denominator of Eq. (\ref{Heat}) with Eq. (\ref{P}) which leads to
\begin{equation}
P=\frac{\Xi _{3}}{16\pi r_{+}^{2d}\left( 6\alpha ^{\prime }+r_{+}^{2}\right)
},  \label{PNEW}
\end{equation}%
in which%
\begin{eqnarray}
\Xi _{3} &=&\left[ \left(
d_{3}d_{4}d_{5}c_{4}c^{3}-d_{3}c_{2}r_{+}^{2}-2c_{1}r_{+}^{3}\right)
2d_{2}\alpha ^{\prime }cr_{+}^{2d_{2}}+\left(
3d_{4}d_{5}c_{4}c^{2}+c_{3}r_{+}+c_{2}r_{+}^{2}\right)
d_{2}d_{3}cr_{+}^{2d_{1}}\right] m  \notag \\
&&-4q^{2}\left[ 2d_{7/2}\alpha ^{\prime }+d_{5/2}r_{+}^{2}\right]
r_{+}^{4}+cd_{2}\left[ 2d_{5}\alpha ^{\prime 2}+d_{9}\alpha ^{\prime
}r_{+}^{2}+d_{3}r_{+}^{4}\right] r_{+}^{2d_{2}}.  \notag
\end{eqnarray}

\begin{figure}[tbp]
$%
\begin{array}{cc}
\epsfxsize=6cm \epsffile{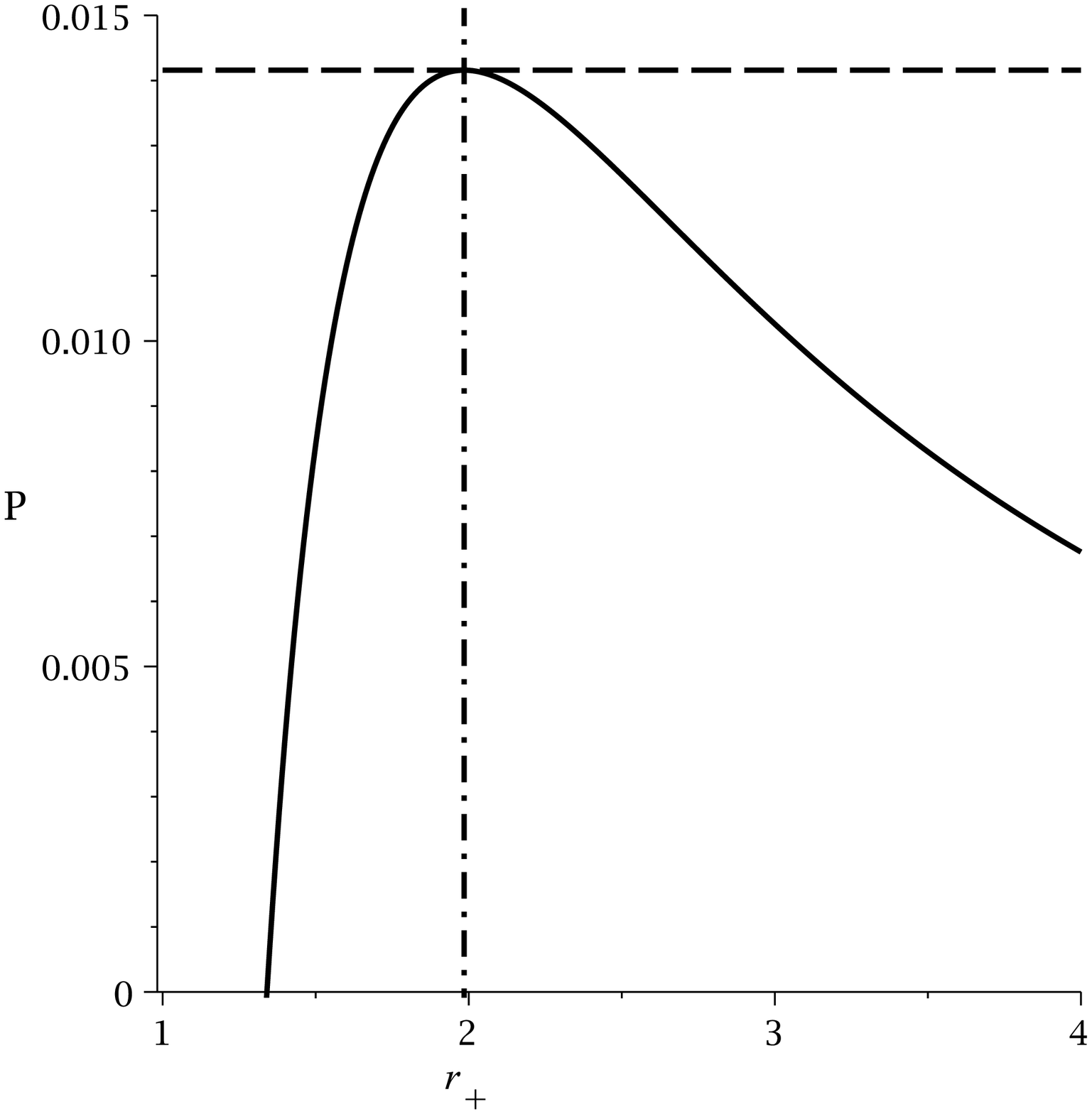} & \epsfxsize=6cm %
\epsffile{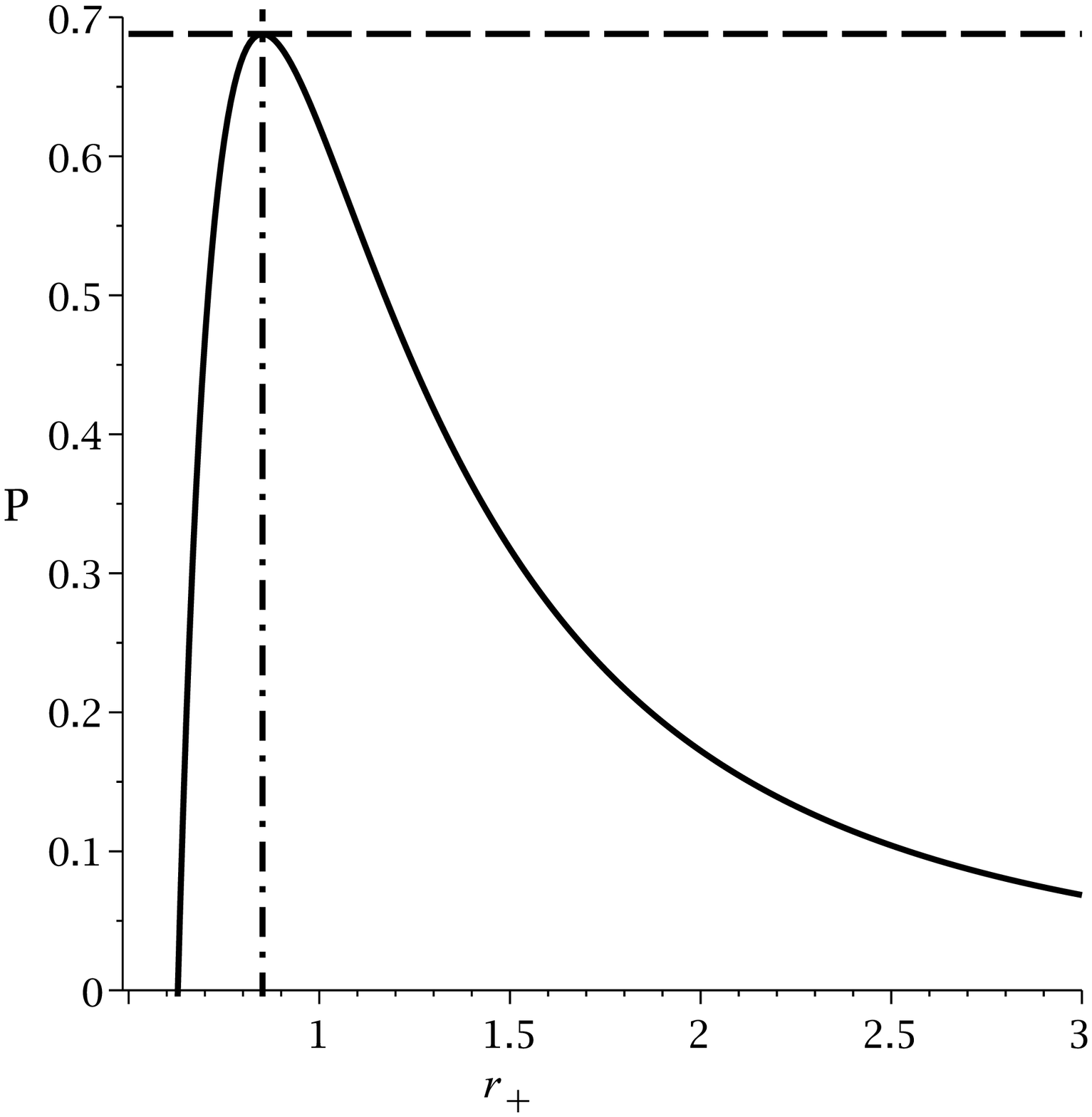}%
\end{array}
$%
\caption{$P$ versus $r_{+}$ diagrams for $q=1$, $c=c_{1}=c_{2}=c_{3}=2$, $%
c_{4}=0$, $d=5$ and $k=1$. \newline
left panel: $\protect\alpha =0.5$ and $m=0.1$ (continues line), $P=0.01415$
(dashed line) and $r_{+}=198493$ (dashed-dotted line)\newline
right panel: $\protect\alpha =0.1$ and $m=0.5$ (continues line), $P=0.68803$
(dashed line) and $r_{+}=0.85114$ (dashed-dotted line). }
\label{Fig13}
\end{figure}

Considering mentioned values in tables $1$ and $2$, we plot following
diagrams for Eq. (\ref{PNEW}), (Fig. \ref{Fig13}).

%%%%%%%%%%%%%%%%%%%%%%%%%%%%%%%%%%%%%%%%%%%%%%%%%%%%%%%%%%%%%%%%%%%%%%%%%%%%%%%%%%%%
%%%%%%%%%%%%%%%%%%%%%%%%%%%%%%%%%%%%%%%%%%%%%%%%%%%%%%%%%%%%%%%%%%%%%%%%%%%%%%%%%%%
It is evident from plotted graphs that the maximum of Eq. (\ref{PNEW}) is
exactly located at the place in which black hole goes under phase transition
(compare Fig. (\ref{Fig13}) with two tables) with same critical pressure. In
other words, the maximums of Eq. (\ref{PNEW}) for different values of
parameters, are representing critical horizon radii and pressures in which
phase transitions take place. Therefore, by employing this method, one is
able to calculate critical horizon radius and pressure at the same time
without going through trouble of obtaining relation for calculating critical
horizon radius and other relations. Another important property of this
method is the consistency of pressure's behavior with thermodynamical
concept. In other words, for pressures smaller than critical pressure, two
values for pressure of phase transition are observed (corresponds to two
divergencies for heat capacity in Fig. \ref{Fig12} and $T-r_{+}$ diagram in
Fig. \ref{Fig1}) and for pressures larger than critical pressure no pressure
for phase transition point is observed.

In conclusion, one can see that this method enables us to obtain critical
values and thermodynamical behavior of the system faster and its properties
are consistent with thermodynamical concepts.

\section{Closing Remarks}

In this paper, we have studied charged massive black holes in Gauss-Bonnet
gravity. We obtained metric function for this gravity and showed that due to
contribution of the massive part, the geometrical structure of black holes
which includes number of the horizons and their places, have been modified.
Therefore, the phenomenology differers completely from usual charged
Gauss-Bonnet black hole. Plotted Carter---Penrose diagrams showed that
although massive part of metric function can change the horizon structure of
black holes, it does not affect the type of singularity and asymptotical
behavior of the solutions. Next, we obtained conserved and thermodynamical
quantities and showed that for obtained values the first law of
thermodynamics is valid.

We also investigated thermal stability of the solutions in context of the
canonical ensemble. We showed that variation of different parameters,
affects the stability conditions of the black holes. Interestingly, it was
found that the variation of massive and GB parameters has opposite effects
on stability and phase transitions of the solutions. Also, it was shown that
dimensionality has strong contribution to type and number of phase
transition which resulted into modification of stability conditions. In
other words, in specific dimensions, the black hole may only enjoy one type
of the phase transition whereas in the other dimensions, it may have two
types of phase transition in several points.

Next, by considering cosmological constant as thermodynamical pressure, we
study phase transition of these black holes in the context of extended phase
space. It was shown that thermodynamical volume is independent of GB and
massive extensions and it only depends on topology of the solutions. Plotted
phase diagrams showed that obtained critical values are the ones in which
phase transition takes place. It was pointed out that interestingly, the
effects of massive and GB parameters are opposite of each other. The
critical temperature and pressure were highly sensitive to variation of
massive parameter. In opposite, the GB parameter, hence strength of
gravitational field put restriction on this rapid growth of temperature
which will be a controlling factor for massive included systems.

In addition, through the use of GTs method, phase transition of these black
holes was investigated in context of heat capacity and extended phase space.
It was found that both employed geometrical metrics for studying
thermodynamical behavior of the system yield consisting results with heat
capacity and extended phase space. In other words, the divergencies of the
Ricci scalar for these metrics coincided with divergencies of the heat
capacity and thermodynamical behavior of the system in context of phase
diagrams. Therefore, these three pictures yield consisting results.

Finally, a new method that was introduced in Ref. \cite{PVEinstein} employed
for calculating critical pressure and horizon radius. It was shown that the
results of this method (using maximum of obtained relation for pressure) and
study conducted in extended phase space were in agreement and obtained
critical pressure and horizon radius were the same.

It will be interesting to study the holographic aspects of obtained
solutions in context of superconductors. Also, it will be worthwhile to
study casual structure and casuality conditions of these systems. In
addition, investigation of the causal structure of massive gravity in
cosmological background by using of interesting Carter--Penrose diagrams is
attractive. Moreover, following the approach of \cite{Alberte,Alberte1}, one
can study the stability and physical singularities of the solutions with
massive degrees of freedom. We left these problems for the future works.

\begin{acknowledgements}
We would like to thank the anonymous referee for useful
suggestions and enlightening comments. We thank Shiraz University
Research Council. This work has been supported financially by the
Research Institute for Astronomy and Astrophysics of Maragha,
Iran.
\end{acknowledgements}

\end{document}